\documentclass[prd,showpacs,altaffilletter,twocolumn,
superscriptaddress,floatfix,nofootinbib,preprintnumbers]{revtex4-1} 

\pdfoutput=1
\usepackage{amsfonts,amsmath,graphicx,bm,siunitx}
\usepackage{dcolumn}
\usepackage[pdftex,
       colorlinks=true,
       pdftitle={The Search for Beauty-fully Bound Tetraquarks Using Lattice Non-Relativistic QCD},
       pdfpagemode=None,
       bookmarksopen=true]{hyperref}

\usepackage{ifpdf}
\usepackage{multirow}
\usepackage[colorlinks=true]{hyperref}
\usepackage{url}
\usepackage[caption=false]{subfig}

\newcommand{\glasgow}{SUPA, School of Physics and Astronomy,
  University of Glasgow, Glasgow, G12 8QQ, UK}

\newcommand{\fermilab}{Fermi National Accelerator Laboratory, Batavia, Illinois, 60510, USA}

\newcommand{\VEC}[1]{{\bf \bm{#1}}} 
\newcommand{\subrm}[1]{{\scriptscriptstyle\mathrm{#1}}}
\newcommand{\csch}{\mathrm{csch} \,}

\def\today{\number\day\space\ifcase\month\or
January\or February\or March\or April\or May\or June\or
July\or August\or September\or October\or November\or December\fi
\space\number\year}
\newcount\mins \newcount\hours
\def\now{\hours=\time \mins=\time
	\divide\hours by60 \multiply\hours by60 \advance\mins by-\hours
	\divide\hours by60 
	\number\hours:\ifnum\mins<10 0\fi\number\mins }

\begin{document}

\title{The Search for Beauty-fully Bound Tetraquarks Using \\Lattice Non-Relativistic QCD}

\author{Ciaran \surname{Hughes}} 
\email[]{chughes@fnal.gov}
\affiliation{\fermilab}

\author{Estia \surname{Eichten}}
\email[]{eichten@fnal.gov}
\affiliation{\fermilab}

\author{Christine~T.~H.~\surname{Davies}} 
\email[]{christine.davies@glasgow.ac.uk}
\affiliation{\glasgow}


\preprint{FERMILAB-PUB-17-381-T}

\begin{abstract}

  Motivated by multiple phenomenological considerations, we perform the first search for the existence of a $\bar{b}\bar{b}bb$ tetraquark bound state with a mass below the lowest non-interacting bottomonium-pair threshold using the first-principles lattice non-relativistic QCD methodology. We use a full $S$-wave colour/spin basis for the $\bar{b}\bar{b}bb$ operators in the three $0^{++}$, $1^{+-}$ and $2^{++}$ channels. We employ four gluon field ensembles at multiple lattice spacing values ranging from $a = 0.06 - 0.12$ fm, all of which include $u$, $d$, $s$ and $c$ quarks in the sea, and one ensemble which has physical light-quark masses. Additionally, we perform novel exploratory work with the objective of highlighting any signal of a near threshold tetraquark, if it existed, by adding an auxiliary potential into the QCD interactions. With our results we find no evidence of a QCD bound tetraquark below the lowest non-interacting thresholds in the channels studied. 
\end{abstract}

\maketitle



\section{Introduction}

Tetraquarks were first considered theoretically decades ago in the context of light-quark physics in order to explain, amongst other experimental features, the $a_0(980)$ and $f_0(980)$ broad resonances \cite{JaffeI}.\footnote{Recent lattice studies of the scattering amplitude pole do indicate that these states are in fact resonances as opposed to cusp effects, etc. \cite{Dudek:a0, Briceno:sigma}}  More recently, there has been exciting experimental evidence indicating the potential existence of tetraquark candidates amongst the so-called XYZ states - states whose behaviour differs from predictions of the heavy quark-antiquark potential model.  The observed XYZ states apparently contain two heavy quarks, $(c\bar c)$ or $(b\bar b$), and two light quarks \cite{PDG:2016}.  The dynamics of these systems involves both the short distance and long distance behaviour of QCD and hence theoretical predictions are difficult.  Consequently, many competing phenomenological models currently exist for these states \cite{QWG:2010}. Lattice QCD studies of the observed XYZ states are also difficult because these states are high up in the spectrum as well as being in the threshold region for strong decays into two heavy flavour mesons. While there are theoretical arguments that some tetraquark states with doubly heavy flavor (e.g., $bb\bar u\bar d$, $bb\bar u \bar s$ and $bb\bar d \bar s$) should be bound and stable against all strong decays \cite{QQqq:EQ}, no general arguments exist for tetraquarks with heavy quark-antiquark content such as $Q\bar Q' q \bar q'$ states.  

A tetraquark system composed of four heavy quarks is a much cleaner system to study theoretically as long-distance effects from light-quarks are expected not to be appreciable, as opposed to systems which are a mixture of heavy and light quarks. In the limit of very heavy quarks perturbative QCD single-gluon exchange will dominate \cite{QQQQ:Voloshin} and so the dynamics are relatively simple. This makes these systems particularly useful to study in order to shed light on the aforementioned XYZ states. In fact, there is a multitude of phenomenological models (with a quark mass ranging from the bottom to the very heavy limit) which predict the existence of a $\bar{Q}\bar{Q}QQ$ bound tetraquark \cite{4b:Schrod,4b:Bai,4b:ColMag,4b:SumRule16,4b:SumRule17,4b:Rosner,Anwar:2017, Alfredo:String}. However, these are not calculations from first-principles and have an unquantifiable systematic error associated with the choice of four-body potential.  In reality, the heaviest possible tetraquark system in nature would be a $\bar{b}\bar{b}bb$ tetraquark. For this, non-perturbative QCD cannot be ignored, making a first-principles lattice QCD study essential. If such a bound $\bar{b}\bar{b}bb$ tetraquark did exist, how it would be observed at the LHC has already been addressed \cite{Zhen:2017, Vega-Morales:2017}. 

Given these pressing theoretical motivations, in this work we perform the first lattice QCD study of the $\bar{b}\bar{b}bb$ system. The sole objective of this exploratory work is to determine if the dynamics of QCD generates enough binding force between the $\bar{b}\bar{b}bb$ to produce a tetraquark state with a mass below the lowest non-interacting bottomonium-pair threshold, ensuring it is stable against simple strong decays. Searching for such a bound $\bar{b}\bar{b}bb$ tetraquark candidate is particularly well-suited to the first-principles lattice QCD methodology because this state, if it existed, would be the ground-state in the $\bar{b}\bar{b}bb$ system. This means it should be relatively easy to identify. Further, $\bar{b}\bar{b}bb$ annihilation effects are strongly suppressed by the heavy quark mass, as in the bottomonium system, and so can be ignored.

This paper is organised as follows: in Section \ref{sec:CtmOp} the interpolating operators used in this study are discussed, in Section \ref{sec:CompSetup} the computational methodology is given, while the majority of the results are presented in Section \ref{sec:QCD4b}. In Section \ref{sec:HO} we explore a novel method of adding an auxiliary potential into QCD with the objective of highlighting a possible tetraquark signal. We then discuss our conclusions in Section \ref{sec:Conclusions}. 


\section{Infinite-Volume Continuum Eigenstates, Operators and Two-Point Correlators}
\label{sec:CtmOp}

The QCD Fock space contains all colour-singlet single-particle states such as the conventional mesons $|\eta_b(\VEC{k})\rangle$, $|\Upsilon(\VEC{k})\rangle$ etc.~and, if a $\bar{b}\bar{b}bb$ bound state also exists, a tetraquark state $|T^{4b}(\VEC{k})\rangle$. In addition, there are also the two-particle states which can be labelled by appropriate quantum numbers as $|\VEC{P}\subrm{tot},J^{PC}; |\VEC{k}\subrm{rel}|, J_1^{P_1C_1},J_2^{P_2C_2},L\subrm{rel}\rangle$ where $\VEC{P}\subrm{tot}$ ($J^{PC}$) is the total (angular) momentum of the two-particle system, with $J_i^{P_iC_i}$ the quantum numbers of the individual particles and $\VEC{k}\subrm{rel}$ ($L\subrm{rel}$) the relative (orbital angular) momentum between the two particles.

The sole motivation of this work is to search for a possible $\bar{b}\bar{b}bb$ tetraquark candidate within QCD that couples to a bottomonium-pair and which lies below the lowest threshold. The bottomonium mesons we study can be classified as $J^{PC} = {^{2S+1}L_J}$. As any orbital angular momenta is expected to raise the internal kinetic energy of the state (and hence its rest mass) we focus on two-body $S$-wave systems ($L=0$) with no orbital angular momentum between them ($L\subrm{rel}=0$).

With the quantum numbers of the $\Upsilon/\eta_b$ being $J^{PC}=1^{--}/0^{-+}$, the $S$-wave $2\eta_b$ and $2\Upsilon$ can have (through the addition of angular momenta) a quantum number of $0^{++}$, while the $\Upsilon\eta_b$ has $1^{+-}$ and the $2\Upsilon$ can also be in a $2^{++}$ configuration. We now want to construct a full basis of $S$-wave colour/spin interpolating operators that has overlap with these quantum numbers. To do so, we start by forming all possible colour combinations that the 2$\bar{b}$ and 2$b$ can be in. These are specified in Table \ref{tab:ColourCombos}.
\begin{table}[t]
  \caption{ The colour representations of the different quark combinations. Note that, as described in the text, once the colour representation of the (anti-) diquark is chosen, the Pauli-exclusion principle enforces certain spin combinations in $S$-wave. Also given are the $SU(3)$ colour contractions needed for the $\bar{b}\bar{b}bb$ operators.}
\label{tab:ColourCombos}
\begin{center}
  \begin{tabular}{l |c |c |c |c |c}
    \hline \hline 
                 &  $b$ & $\bar{b}$ &  $\bar{b}b$ & $\bar{b}\bar{b}$ & $bb$  \\ \hline
    Colour Irrep & $3_c$ & $\bar{3}_c$ & $1_c, 8_c$ & $3_c, \bar{6}_c$ & $\bar{3}_c, 6_c$ \\ \hline \hline
    \multicolumn{1}{l}{$\mathcal{G}^{1}_{efg}\mathcal{G}^{1}_{ef'g'}$} & \multicolumn{5}{|l}{$\delta_{fg}\delta_{f'g'}$} \\\hline
    \multicolumn{1}{l}{$\mathcal{G}^{8}_{efg}\mathcal{G}^{8}_{ef'g'}$} & \multicolumn{5}{|l}{$2\delta_{fg'}\delta_{f'g}-2\delta_{fg}\delta_{f'g'}/3$}\\ \hline
    \multicolumn{1}{l}{$\mathcal{G}^{3}_{efg}\mathcal{G}^{3}_{ef'g'}$} & \multicolumn{5}{|l}{$(\delta_{ff'}\delta_{gg'}-\delta_{fg'}\delta_{gf'})/2$} \\\hline
    \multicolumn{1}{l}{$\mathcal{G}^{6}_{efg}\mathcal{G}^{6}_{ef'g'}$} & \multicolumn{5}{|l}{$(\delta_{ff'}\delta_{gg'}+\delta_{fg'}\delta_{gf'})/2$} \\
    \hline\hline
  \end{tabular}
\end{center}
\end{table}

We can construct meson interpolating operators as
\begin{align}
\mathcal{O}^{1(8)}_{M}(t,\VEC{x}) & = \mathcal{G}^{1(8)}_{efg}\bar{b}_f \Gamma_M b_g(t,\VEC{x}) \label{eqn:MesonOp}
\end{align}
where $\Gamma_M = i\gamma^5, \gamma^k$ projects onto the quantum numbers of the $\eta_b$ and $\Upsilon$ respectively, and $\mathcal{G}^{1(8)}_{efg}$ is the colour projection onto the singlet (octet). In addition, it is also possible to construct a (anti-) diquark operator as
\begin{align}
\mathcal{O}^{\bar{3}(6)}_{D}(t,\VEC{x}) & = \mathcal{G}^{\bar{3}(6)}_{efg}\bar{b}^{\hat{C}}_f \Gamma_D b_g(t,\VEC{x}) \label{eqn:DiquarkOp} \\
\mathcal{O}^{{3}(\bar{6})}_{A}(t,\VEC{x}) & = \mathcal{G}^{{3}(\bar{6})}_{efg}\bar{b}_f \Gamma_A b^{\hat{C}}_g(t,\VEC{x}) \label{eqn:AntiDiquarkOp}
\end{align}
where $(b^{\hat{C}})_{\alpha}=C_{\alpha\beta}\bar{b}_{\beta}$ is the charge-conjugated field with $C=-i\gamma^0\gamma^2$.\footnote{$\gamma^0 =$ diag$(1,-1)$ in the convention used by NRQCD.} As the two quarks have the same flavour, the Pauli-exclusion principle applies and the wavefunction has to be completely anti-symmetric. With our choice to focus on $S$-wave combinations of particles, the spatial wave-function must be symmetric. As the colour (triplet) sextet has a (anti-) symmetric colour wavefunction, this forces the spin-wavefunction to be in a (triplet) singlet with ($\Gamma=\gamma^k$) $\Gamma=i\gamma^5$. 

With these building blocks, we can form four classes of $\bar{b}\bar{b}bb$ colour-singlets by contracting the colour factors $\mathcal{G}$ in any irreducible representation (irrep) with its conjugate colour factors, i.e., $1_c\times 1_c$, $8_c\times {8}_c$\footnote{Since the $8_c$ irrep of $SU(3)$ is the adjoint, it is similar to its conjugate $\bar{8}_c$.}, $3_c\times \bar{3}_c$ and $6_c\times \bar{6}_c$. These $SU(3)$ invariant colour contractions are given in Table \ref{tab:ColourCombos}. After doing this, we need to project the operators onto a specific angular momentum $J^{P}$ by using the standard $SO(3)$ Clebsch-Gordan coefficients (using a spherical basis of spin-matrices \cite{Dudek:Excited}) as
\begin{align}
\mathcal{O}^{J,m}_{(P,Q)} (t,\VEC{x}) & =  \sum_{m_1,m_2}\langle J,m | J_1,m_1,J_2,m_2 \rangle \nonumber \\  & \hspace{1.3cm} \times \mathcal{O}^{J_1,m_1}_{P}(t,\VEC{x}+\VEC{r})\mathcal{O}^{J_2,m_2}_{Q}(t,\VEC{x}) \label{eqn:ClebschGordan}
\end{align}
with $(P,Q)$ describing the blocks this configuration is built from, i.e, $(\eta_b,\eta_b)$, $(\Upsilon,\Upsilon)$, $(D,A)$, etc. We also allow the possibility of the two blocks being separated by a distance $\VEC{r}$. For the $\VEC{r}=\VEC{0}$ case, the operators project onto a definite total angular momentum $J$. For the $\VEC{r}\ne \VEC{0}$ case, one can Taylor expand $\mathcal{O}^{J_1,m_1}_{P}(t,\VEC{x}+\VEC{r})$ around $\VEC{r}=\VEC{0}$ to notice that the operator projects onto a superposition of quantum numbers. Consequently, it is possible to utilise this to further search for the lowest ground state of the four quark system. When dealing with the diquark components, to project onto a definite value of charge-conjugation in Eq.~(\ref{eqn:ClebschGordan}) one can form the linear combination $\mathcal{O}^{1,m_1}_{D}\mathcal{O}^{1,m_2}_{A} \pm \mathcal{O}^{1,m_2}_{D}\mathcal{O}^{1,m_1}_{A}$. 

In fact, not all of these colour combinations are independent. Fierz relations constrain the number of independent colour-spin operators that are possible. For the local operators in $S$-wave, the relations between the two-meson and diquark-antidiquark states are given in Table \ref{tab:Fierz}. 
\begin{table}[t]
  \caption{Fierz relations in the $\bar{b}\bar{b}bb$ system relating the two-meson and the diquark-antidiquark bilinears.}
\label{tab:Fierz}
\begin{center}
  \begin{tabular}{l|c|c}
    \hline \hline
    $J^{PC}$ & Diquark-AntiDiquark & Two-Meson \\\hline
    $0^{++}$ & $\bar{3}_c\times 3_c$ & $-\frac{1}{2}|0;\Upsilon\Upsilon\rangle + \frac{\sqrt{3}}{2}|0;\eta_b\eta_b\rangle$ \\
    $0^{++}$ & $6_c\times \bar{6}_c$ & $\frac{\sqrt{3}}{2}|0;\Upsilon\Upsilon\rangle + \frac{1}{2}|0;\eta_b\eta_b\rangle$ \\
    $1^{+-}$ & $\bar{3}_c\times 3_c$ & $\frac{1}{\sqrt{2}}\left(|1;\Upsilon\eta_b\rangle + |1;\eta_b\Upsilon\rangle\right)$ \\
    $2^{++}$ & $\bar{3}_c\times 3_c$ & $|2;\Upsilon\Upsilon\rangle$\\\hline \hline
  \end{tabular}
\end{center}
\end{table}

The simplest quantity that can be calculated on the lattice in order to extract particle masses is the Euclidean two-point correlator. This is defined as
  \vspace{-0.1cm}
\begin{align}
C_{i,j}^{J^{PC}}(t,\VEC{P}\subrm{tot}=\VEC{0}) \hspace{-0.0cm} = \hspace{-0.07cm} \int d^3x \langle \mathcal{O}^{J,m_i}_i (t,\VEC{x}) \mathcal{O}^{J,m_j}_j (0,\VEC{0})^{\dagger}\rangle  \label{eqn:2pt}
\end{align}
where we choose to project to zero spatial-momentum and $i,j$ label potentially different operators at the source and sink with the same $J^{PC}$, e.g., $i = (\eta_b,\eta_b)$, $j=(\Upsilon,\Upsilon)$. The single-particle contributions to the correlator are determined by inserting a complete set of single-particle states in the Hilbert-space formalism into Eq.~(\ref{eqn:2pt}) to yield
\begin{align}
  C_{i,j}^{J^{PC}}(t,\VEC{P}\subrm{tot}=0) = \sum_n Z_n^i Z_n^{j,*} e^{-E_nt} \label{eqn:2ptFitSingle}
\end{align}
with $Z_n^i = \langle 0 | \mathcal{O}^{J,m_i}_i | n \rangle $ the non-perturbative overlap of the operator to the eigenstate $|n\rangle$ and $E_n|n\rangle = H|n\rangle$ the energy eigenvalue. Note that all states $|n\rangle$ with the same quantum numbers contribute to this correlator, e.g., for the bottomonium $0^{-+}$ pseudoscalar correlator the $|\eta_b\rangle$ as well as all radial excitations contribute. 

The two-particle contributions to the correlator are slightly more complicated. In this case, as derived in Appendix \ref{app:Fit}, the non-relativistic two-particle states give a contribution to the correlator that is
\begin{align}
  &  C_{i,j}^{J^{PC}}(t,\VEC{P}\subrm{tot}=0)  =  \left(\frac{\mu_r}{2\pi t}\right)^{\frac{3}{2}}\sum_{X_2} e^{-(M^S_1+M^S_2)t}\nonumber \\
  & \hspace{1.2cm} \times \left\{Z^0_{X_2} +  Z^{2}_{X_2}  \frac{3}{(t\mu_r)} + Z^{4}_{X_2}  \frac{15}{(t\mu_r)^{2}} + \cdots \right\} \label{eqn:2ptFitTwo}
\end{align}
where the sum is over all distinct two-particle states $X_2$ with quantum numbers $J^{PC}$ and $\VEC{P}\subrm{tot}=0$, $M_i^S$ ($M_i^K$) is the static (kinetic) mass of the particles $-$ as defined in Eq.~(\ref{eqn:NRExpansion}) $-$, $\mu_r = M_1^KM_2^K/(M_1^K+M_2^K)$ is the reduced mass and $Z_{X_2}^{2l}$ are non-perturbative coefficients. 

\begin{figure*}[th]
  \hspace*{\fill}
  \subfloat[Direct$1$]{\label{fig:D1}
    \includegraphics[width=0.30\textwidth]{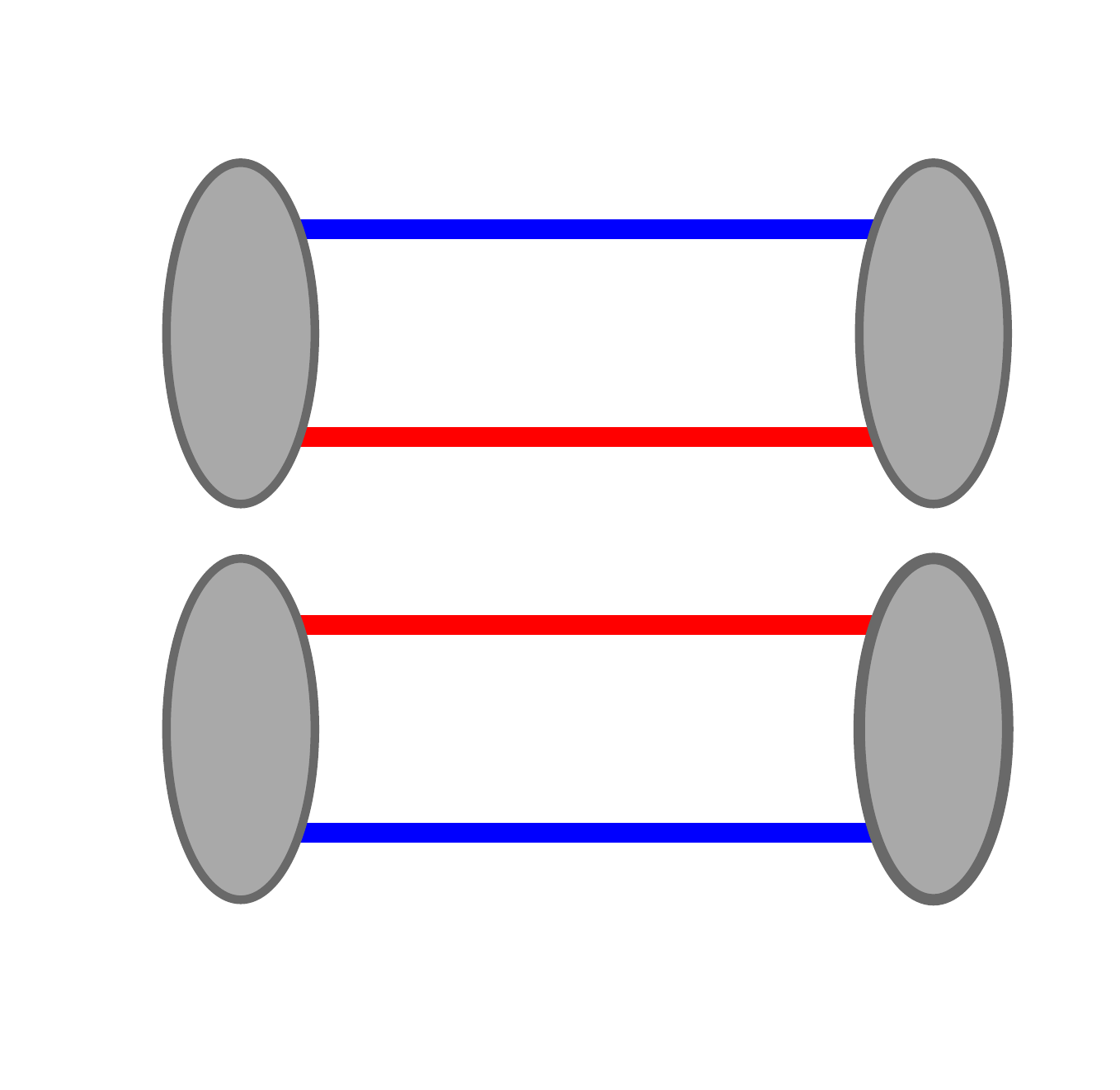}}
  \hfill
  \subfloat[Xchange$2$]{\label{fig:X2}
    \includegraphics[width=0.30\textwidth]{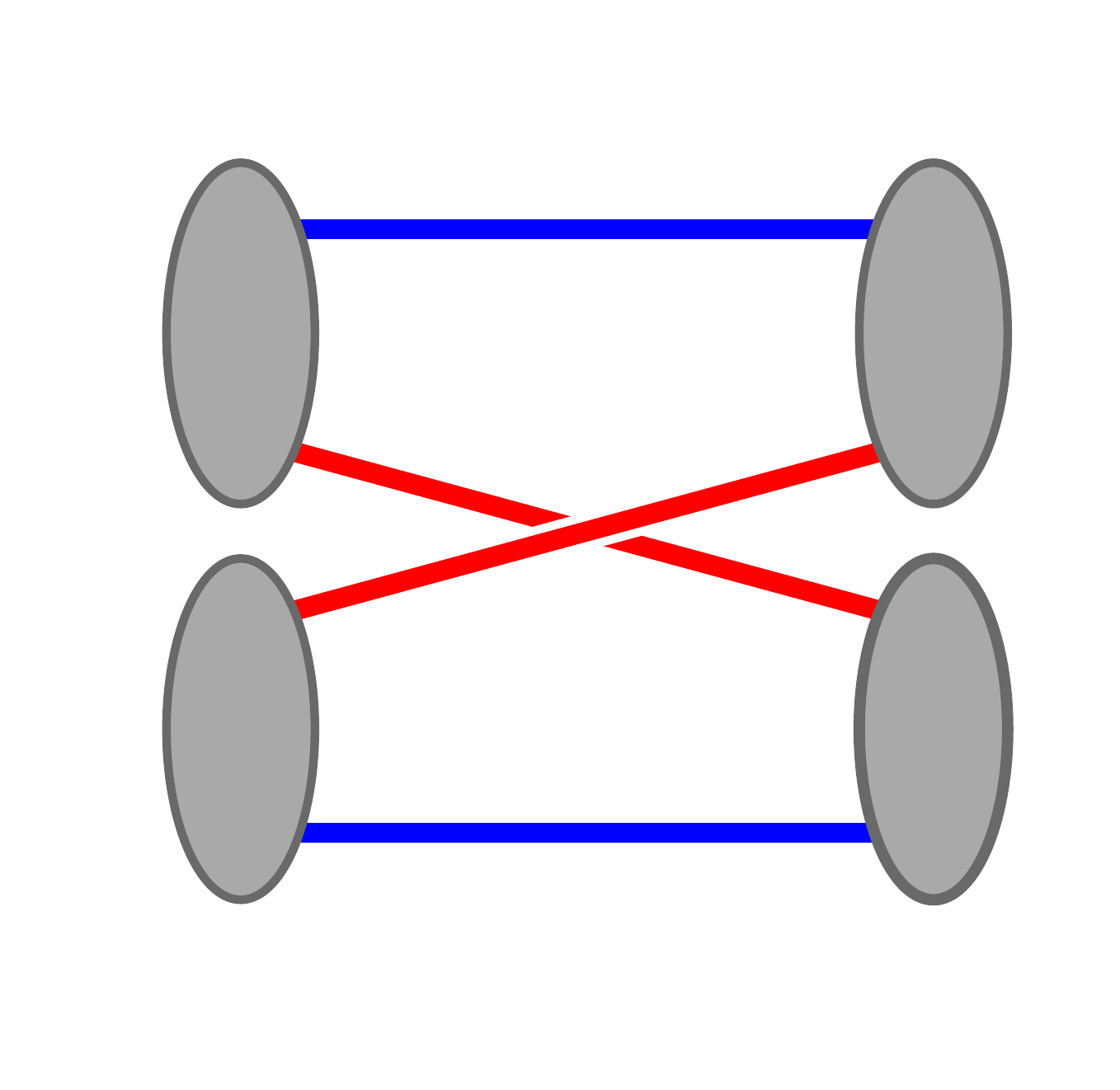}}
  \hspace*{\fill}
  \\
  \hspace*{\fill}
  \subfloat[Direct$3$]{\label{fig:D3}
    \includegraphics[width=0.30\textwidth]{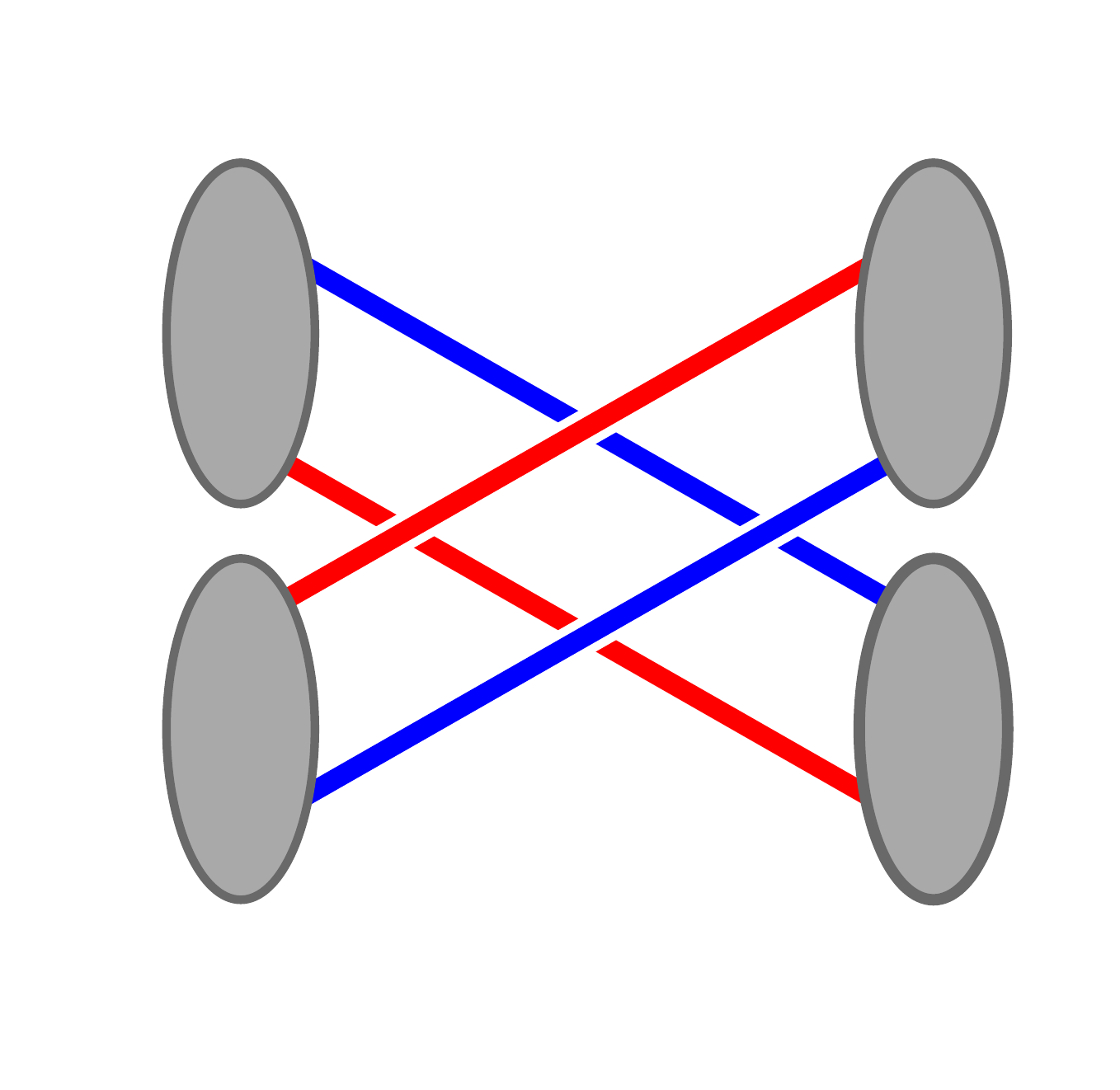}}
  \hfill
  \subfloat[Xchange$4$]{\label{fig:X4}
    \includegraphics[width=0.30\textwidth]{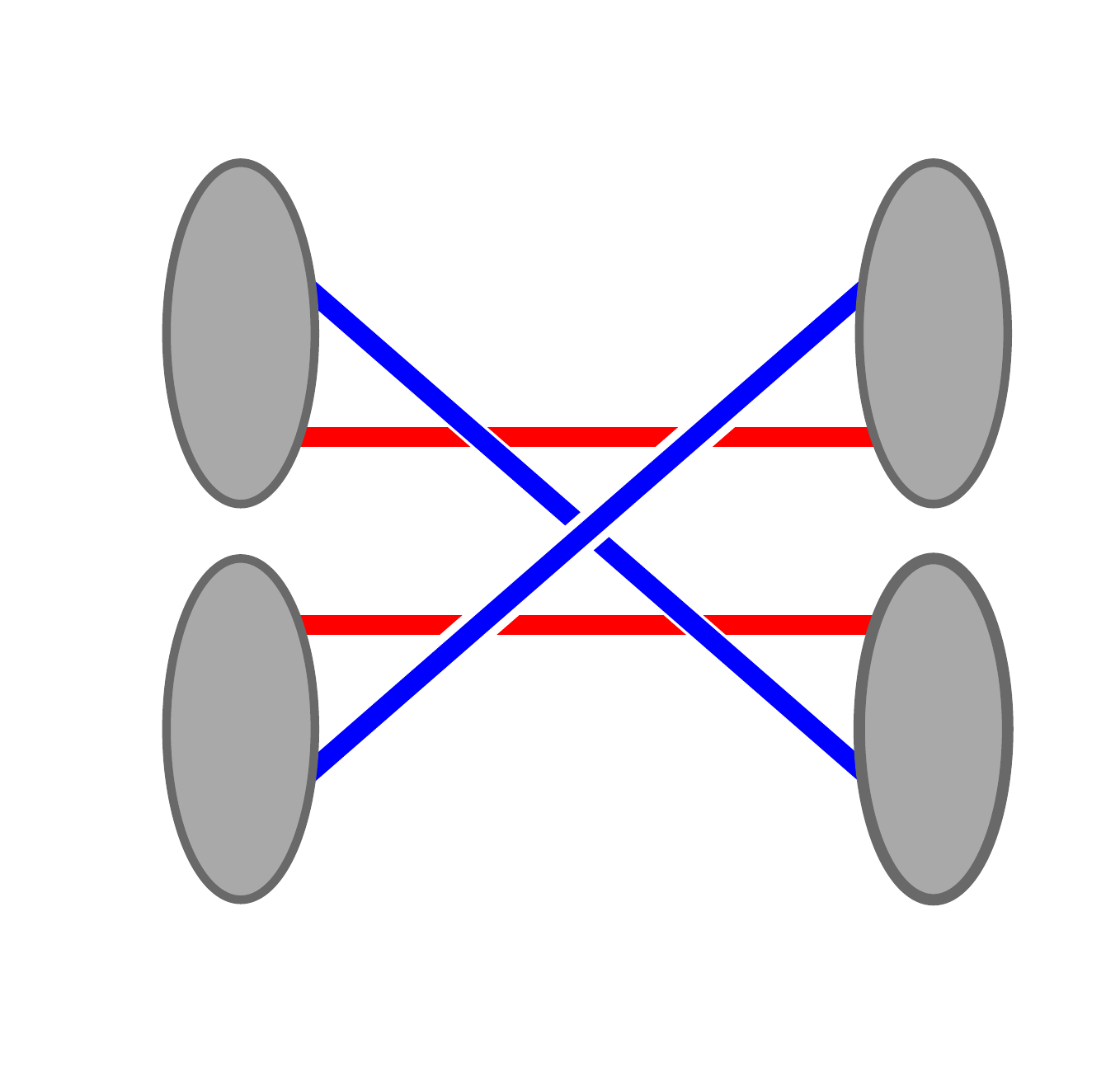}}  
  \hspace*{\fill}
  \caption{There are four connected Wick contractions for the two-meson type correlator when the quarks have the same flavour. The grey region represents a colour neutral meson, the blue line a quark and the red line an antiquark. We call these the (a) Direct$1$ contraction where each meson propagates to itself. (b) Xchange$2$  where an anti-quark is exchanged between the meson pair, (c) Direct$3$ where each meson propagates to the other, and (d) Xchange$4$ where a quark is exchanged between the meson pair. (colour online)  }
  \label{fig:WickTwoMeson}
\end{figure*}

Energies of states can be extracted using the above functional form once the correlator has been computed. Examining Eq.~(\ref{eqn:2pt}) in the path-integral formalism we can perform the connected Wick contractions\footnote{Annihilation diagrams are suppressed by powers of the heavy quark mass \cite{Bodwin:Annihilation} and are expected to be negligible.} so that the correlator can be written as an integral over the gluon-fields with the integrand consisting of products of $b$-quark propagators. For each two-meson type operator, e.g., $\mathcal{O}^{1_c}_{\eta_b}\mathcal{O}^{1_c}_{\eta_b}$, as all quarks have the same flavour there are four connected Wick contractions. These are shown diagrammatically in Figure \ref{fig:WickTwoMeson}. 

\begin{figure*}[t]
  \hspace*{\fill}
  \subfloat[]{\label{fig:Diq1}
    \includegraphics[width=0.30\textwidth]{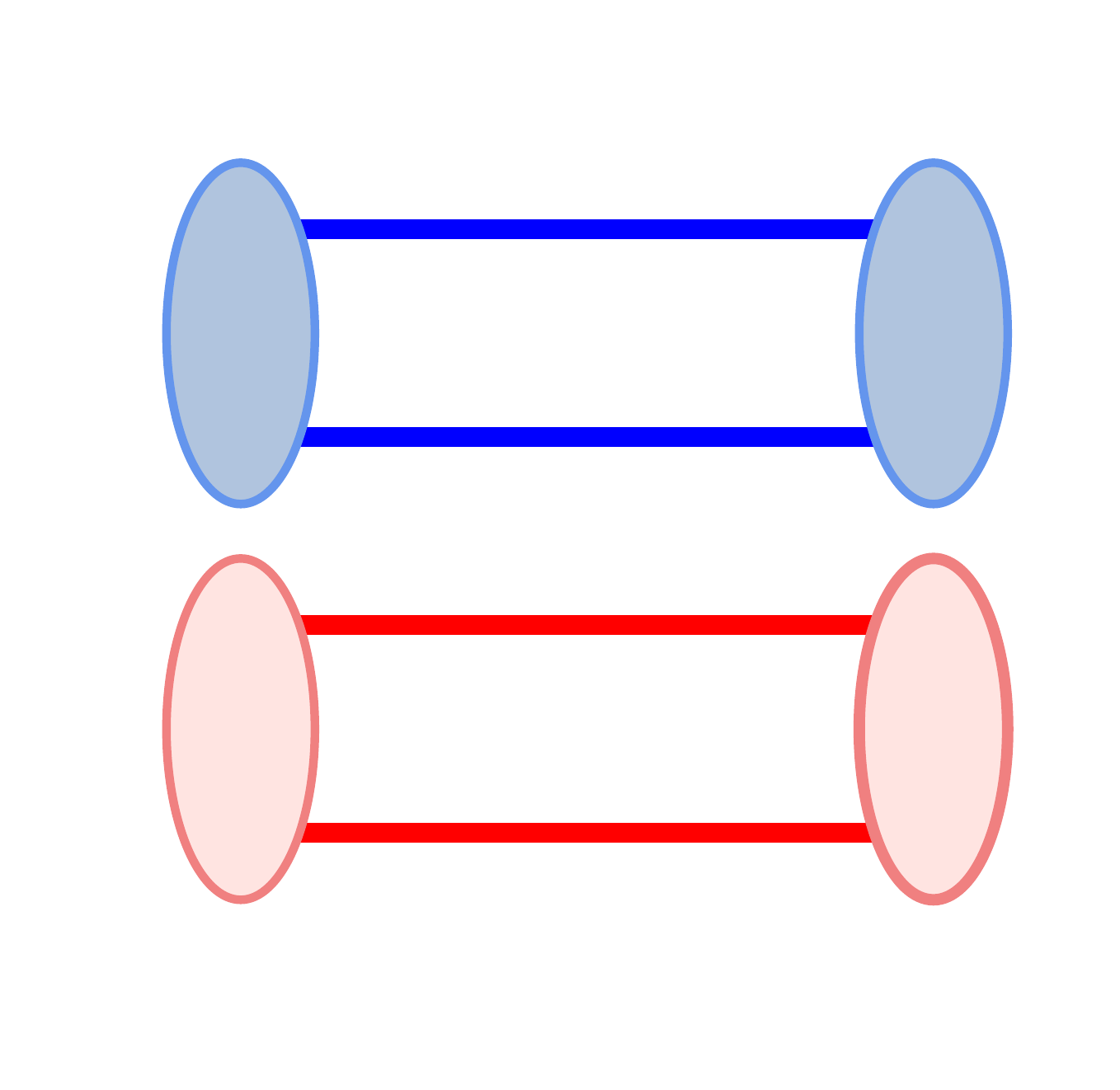}} 
  \hfill
  \subfloat[]{\label{fig:Diq2}
    \includegraphics[width=0.30\textwidth]{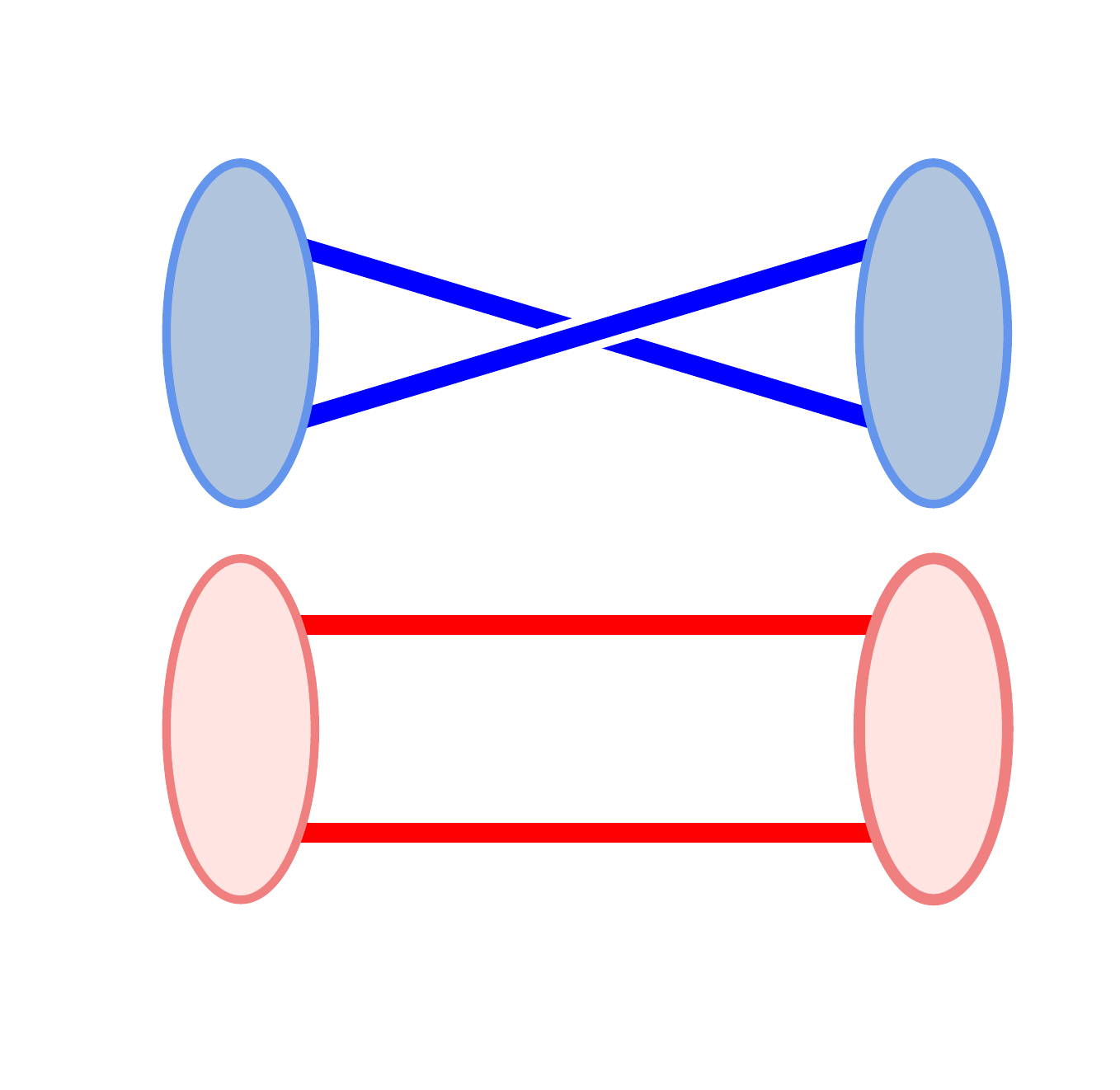}} 
  \hspace*{\fill}
  \\
  \hspace*{\fill}
  \subfloat[]{\label{fig:Diq3}
    \includegraphics[width=0.30\textwidth]{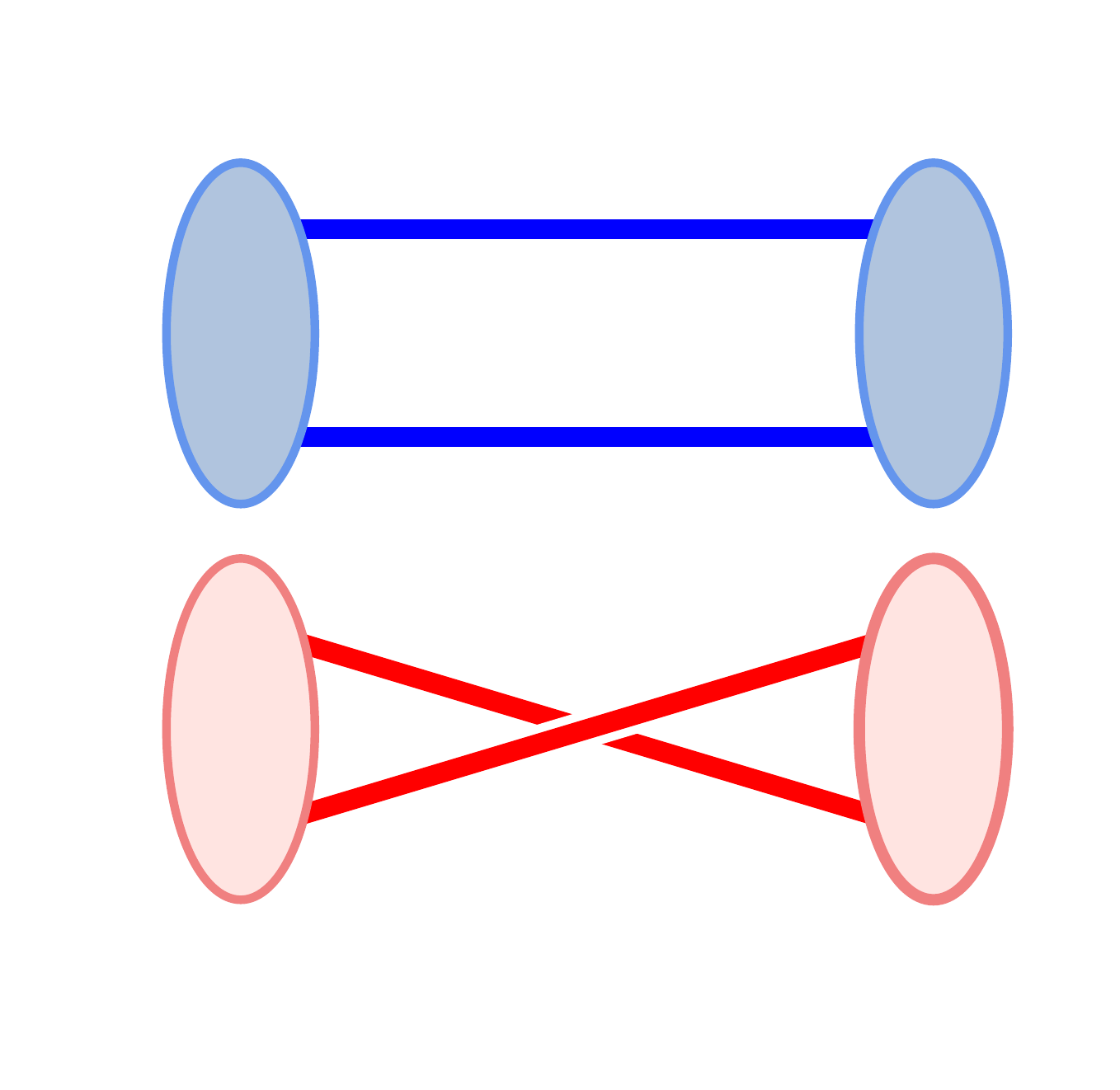}}
  \hfill
  \subfloat[]{\label{fig:Diq4}
    \includegraphics[width=0.30\textwidth]{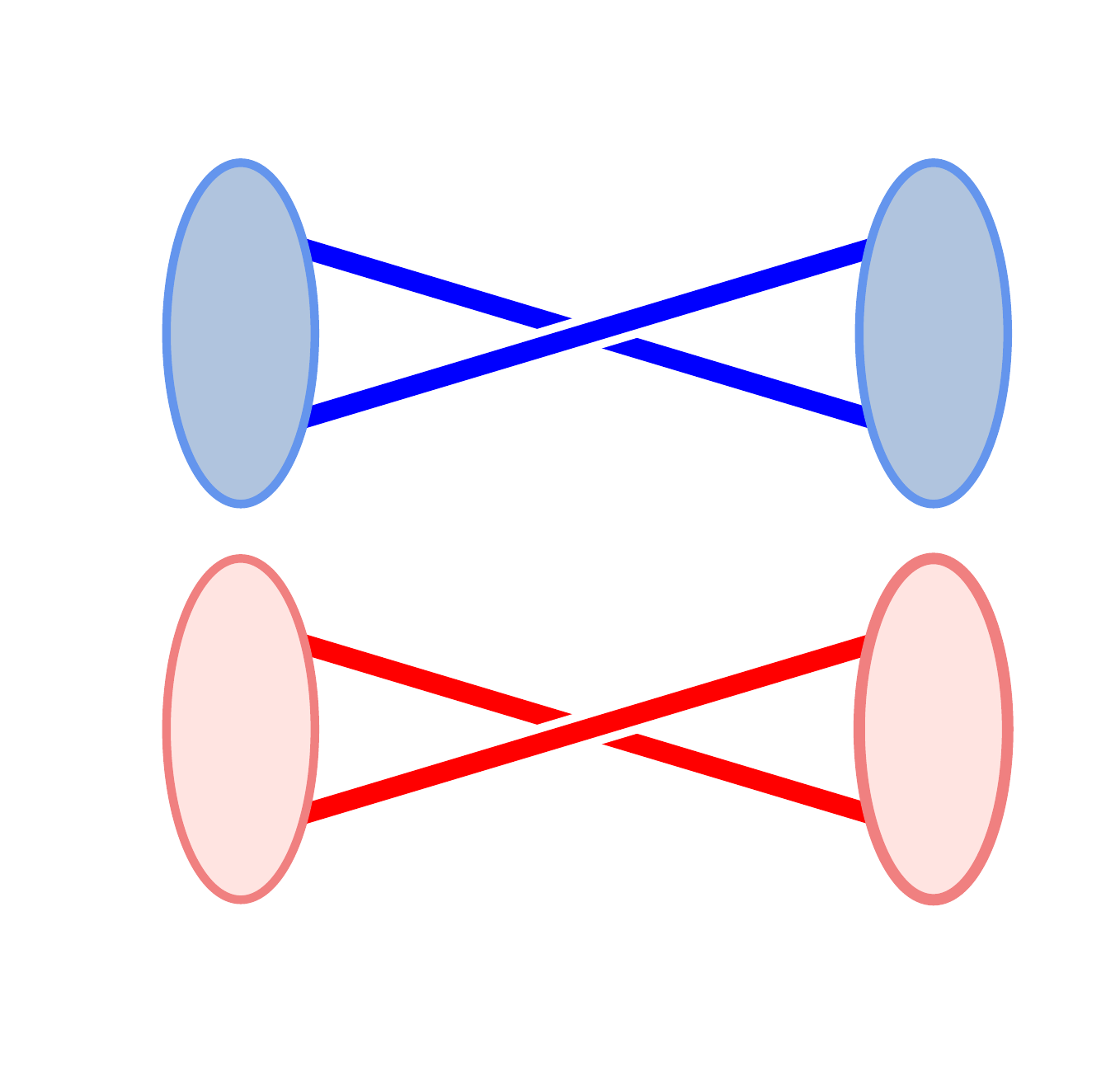}}  
  \hspace*{\fill}
  \caption{There are four connected Wick contractions for the diquark-antidiquark type correlator when the quarks have the same flavour. The blue shaded region represents a diquark, the red shaded region the antidiquark, a blue line a quark and the red line an antiquark. The uncrossing of the lines in Figure \ref{fig:Diq2} to produce Figure \ref{fig:Diq1} gives a $\pm$, as discussed in the text, which enforces the Pauli-exclusion principle. (colour online) }
  \label{fig:WickDiquark}
\end{figure*}

The first Wick contraction for the two-meson correlator, called Direct$1$ and shown in Figure ~\ref{fig:D1}, has the expression
\begin{widetext}
  \vspace{-0.6cm}
\begin{align}
\mathcal{G}^R_{Ade}\mathcal{G}^R_{Ad'e'}\mathcal{G}^R_{Bgh}\mathcal{G}^R_{Bg'h'}\sum_{\VEC{x}} \text{Tr} \left[\Gamma_{M_1} K^{-1}(t,\VEC{x};0,\VEC{z})_{e'g}\Gamma_{M_1}^{\dagger}K^{-1}(t,\VEC{x};0,\VEC{z})_{hd'}^{\dagger}\right] \text{Tr} \left[\Gamma_{M_2} K^{-1}(t,\VEC{x}';0,\VEC{z}')_{eg'}\Gamma_{M_2}^{\dagger}K^{-1}(t,\VEC{x}';0,\VEC{z}')_{h'd}^{\dagger}\right] \label{eqn:D1}
\end{align}
  \vspace{-0.3cm}
\end{widetext}
while the second, called Xchange$2$, is given by
\begin{widetext}
  \vspace{-0.6cm}
\begin{align}
-\mathcal{G}^R_{Ade}\mathcal{G}^R_{Ad'e'}\mathcal{G}^R_{Bgh}\mathcal{G}^R_{Bg'h'}\sum_{\VEC{x}} \text{Tr} \left[\Gamma_{M_1} K^{-1}(t,\VEC{x};0,\VEC{z})_{e'g}\Gamma_{M_1}^{\dagger}K^{-1}(t,\VEC{x}';0,\VEC{z})_{hd}^{\dagger}\Gamma_{M_2} K^{-1}(t,\VEC{x}';0,\VEC{z}')_{eg'}\Gamma_{M_2}^{\dagger}K^{-1}(t,\VEC{x};0,\VEC{z}')_{h'd'}^{\dagger}\right] \label{eqn:X2}
\end{align}
  \vspace{-0.3cm}
\end{widetext}
where $\VEC{x}' = \VEC{x} + \VEC{r}$. The other diagrams, Direct$3$ and Xchange$4$, have similar expressions. For the diquark-antidiquark type operators, e.g., $\mathcal{O}^{6_c}_{D}\mathcal{O}^{\bar{6}_c}_{A}$, there are also four Wick contractions which can be combined into one expression as
\begin{widetext}
  \vspace{-0.5cm}
  \begin{align}
  & C^{J^{PC}}(t,\VEC{P}\subrm{tot}=0) = \left[1 \pm sgn(C\Gamma_D)^T \pm sgn(C\Gamma_A)^T  + sgn(C\Gamma_D)^Tsgn(C\Gamma_A)^T\right]\mathcal{G}^R_{Ade}\mathcal{G}^R_{Ad'e'}\mathcal{G}^R_{Bgh}\mathcal{G}^R_{Bg'h'} \nonumber \\
  & \hspace{0.8cm}\times \sum_{\VEC{x}} \text{Tr} \left[C\Gamma_{D} K^{-1}(t,\VEC{x};0,\VEC{z})_{eg'}\Gamma_{D}^{\dagger}CK^{-1}(t,\VEC{x};0,\VEC{z})_{dh'}^{T}\right] \text{Tr} \left[\Gamma_{A} C K^{-1}(t,\VEC{x}';0,\VEC{z}')^*_{e'g}C \Gamma_{A}^{\dagger}K^{-1}(t,\VEC{x}';0,\VEC{z}')_{hd'}^{\dagger}\right] \label{eqn:WickDiquark}
  \end{align}
    \vspace{-0.3cm}
\end{widetext}
where the sign function is defined by $sgn(X)^T=\pm 1$ if $X^T=\pm X$. The $\pm$ in Eq.~(\ref{eqn:WickDiquark}) corresponds to the ${3}$ or ${6}$ colour representation. It is this prefactor with the signs which enforces the Pauli exclusion principal: the sum cancels for spin combinations that do not make the wavefunction overall anti-symmetric. The spin-triplet/singlet configurations we consider here obey $(C\gamma^k)^T = + (C\gamma^k)$ and $(C\gamma^5)^T = - (C\gamma^5)$. Diagrammatically the four connected Wick contractions contributing to the diquark correlator are shown in Figure \ref{fig:WickDiquark}.

To calculate the two-point correlators described above within the first-principles Feynman path-integral approach to QCD needs the methodology of lattice QCD. We now discuss our lattice QCD approach.

\section{Lattice QCD Methodology}
\label{sec:CompSetup}

\subsection{Second Generation $N_f = 2+1+1$ Gluon Ensembles}
\label{sec:Ensembles}

Our lattice calculation uses gauge field configurations generated by the MILC collaboration \cite{MILC:Configs}. For the gauge fields, they used the tadpole-improved L\"{u}scher-Weisz gauge action correct to $\mathcal{O}(\alpha_s a^2)$ \cite{Hart:GluonImprovement} and include $2+1+1$ flavours in the sea, the up and down quarks (treated as two degenerate light quarks with mass $m_l$), the strange quark, and the charm quark. The sea-quarks are included using the Highly Improved Staggered Quark formulation \cite{HISQAction}. 

Four ensembles are chosen in this study. As one might expect roughly double the discretisation errors in a $2\bar{b}b$ system relative to the $\bar{b}b$ one, to ensure that the heavy quark potential is accurately represented (where short distance details may be important for a compact tetraquark candidate) we utilise three ensembles that span relatively fine lattice spacings ranging from $a = 0.06-0.12$ fm. Details of the ensembles are given in Table \ref{tab:GluonEnsembles}. Due to the computational expense, most of the ensembles use heavier $m_l$ than in the real world. However, to test $m_l$ dependence, we use one ensemble (Set $2$ in Table \ref{tab:GluonEnsembles}) that has physical $am_l/am_s$. Additionally, the ensembles have been fixed to Coulomb gauge to allow non-gauge invariant non-local operators to be used (as constructed in Eq.~(\ref{eqn:ClebschGordan})).

\subsection{$b$-quarks Using iNRQCD}
\label{subsec:NRQCD}

A non-relativistic effective field theory is appropriate for physical systems where the relative velocity of the constituent particles inside the bound state is much smaller than one (in Planck units). When applied to the strong force, this framework is called non-relativistic QCD (NRQCD).  It is well known that $b$-quarks are very nonrelativistic inside their low-lying bound states, where $v^2\approx 0.1$ \cite{Lepage:ImprovedNRQCD}.

For lattice NRQCD, the continuum NRQCD action is discretised onto the lattice \cite{Lepage:ImprovedNRQCD} with operators included to a predetermined level in $v^2$. Here we use an action accurate through $\mathcal{O}(v^4)$ with additional spin-dependent terms at $\mathcal{O}(v^6)$.\footnote{The spin-independent  $\mathcal{O}(v^6)$ terms are subleading effects for the $\bar{b}\bar{b}bb$ energies relevant to this study.} Operators are also added to correct for discretisation effects. We make the further systematic improvement here, introduced in \cite{Dowdall:Upsilon}, to include coefficients of $\mathcal{O}(v^4)$ operators that have been matched to continuum QCD through $\mathcal{O}(\alpha_s)$. We call this improved-NRQCD (iNRQCD). This action has already been used to successfully determine bottomonium $S$, $P$ and $D$ wave mass splittings \cite{Dowdall:Upsilon, Daldrop:Dwave}, precise hyperfine splittings \cite{Dowdall:Hyperfine, Dowdall:Hl}, $B$ meson decay constants \cite{Dowdall:BMeson}, $\Upsilon$ and $\Upsilon'$ leptonic widths \cite{Brian:LeptonicWidth}, $B$, $D$ meson mass splittings \cite{Dowdall:Hl} and hindered M$1$ radiative decays between bottomonium states \cite{Hughes:M1}.

\begin{table}[t]
  \caption{Details of the gauge ensembles used in this
    study. $\beta$ is the gauge coupling. $a$ (fm) is the lattice spacing \cite{Dowdall:Upsilon,Chakraborty:2014aca}, $am_q$ are the sea
    quark masses, $N_s \times N_T$ gives the spatial and temporal
    extent of the lattices in lattice units and $n_{\text{cfg}}$  is the number of
    configurations used for each ensemble. We use $16$ time sources on each configuration to increase statistics. Ensembles $1$ and $2$ are referred to as ``coarse'',  $3$ as ``fine,'' and $4$ as 
    ``superfine''. }
\label{tab:GluonEnsembles}
\begin{center}
  \begin{tabular}{l l l l l l l l }
    \hline \hline 
    Set&  $\beta$ & $a$ (fm) &  $am_l$ &$am_s$ & $am_c$ & $N_s
    \times N_T$ & $n_{\text{cfg}}$ \\ \hline
    $1$ & $6.00$ &$0.1219(9)$ & $0.0102$ & $0.0509$ & $0.635$ &$24 \times
    64$ & $1052$ \\ \hline 
    $2$ & $6.00$ & $0.1189(9)$ & $0.00184$ & $0.0507$ & $0.628$ &$48 \times
    64$ & $1000$ \\ \hline
    $3$ & $6.30$ &$0.0884(6)$ & $0.0074$ & $0.037$ & $0.440$ &$32 \times
    96$ & $1008$ \\ \hline
    $4$ & $6.72$ & $0.0592(3)$ & $0.0048$ & $0.024$ & $0.286$ &$48 \times
    144$ & $400$ \\  \hline \hline
  \end{tabular}
\end{center}
\end{table}

The iNRQCD Hamiltonian evolution equations can be written as
\begin{align}
G({\VEC{x}},t+1)  & = e^{-aH}  G({\VEC{x}},t) \nonumber \\
G({\VEC{x}},t_{\text{src}})  & = \delta_{\VEC{x},\VEC{x_0}} \label{eqn:NRQCDSrc}
\end{align}
with 
\begin{align}
e^{-aH} & = \left( 1 - \frac{a\delta H|_{t+1}}{2} \right) \left( 1
  - \frac{aH_0|_{t+1}}{2n} \right)^n U_t^{\dagger}(x) \nonumber \\
&\hspace{0.7cm}  \times \left( 1 -  \frac{aH_0|_t}{2n} \right)^n \left( 1 - \frac{a\delta H|_t}{2}
  \right) \label{eqn:NRQCDGreensFunction}
\end{align}
\begin{align}
aH_0  & = -\frac{\Delta^{(2)}}{2am_b}, \nonumber \\
a\delta H  & =  a\delta H_{v^4} + a\delta H_{v^6}; \nonumber \\
a\delta H_{v^4} & = -c_1 \frac{(\Delta^{(2)})^2}{8(am_b)^3} 
+c_2\frac{i}{8(am_b)^2}\left( {\VEC{\nabla \cdot \tilde{E}}} -
  {\VEC{\tilde{E} \cdot \nabla  }} \right)  \nonumber \\ &
-c_3\frac{1}{8(am_b)^2} {\VEC{\sigma \cdot \left( \tilde{\nabla}
      \times \tilde{E} - \tilde{E} \times \tilde{\nabla} \right) }}
\nonumber \\ &
-c_4\frac{1}{2am_b} {\VEC{ \sigma \cdot \tilde{B}}} 
+c_5 \frac{\Delta^{(4)}}{24am_b}
-c_6 \frac{(\Delta^{(2)})^2}{16n(am_b)^2} \nonumber 
\end{align}
\begin{align}
a\delta H_{v^6} & = 
-c_7\frac{1}{8(am_b)^3} {\VEC{ \left\{ \Delta^{(2)} , \sigma \cdot
      \tilde{B} \right\} }}  \nonumber \\ & 
-c_8\frac{3}{64(am_b)^4} {\VEC{ \left\{ \Delta^{(2)}, \sigma \cdot \left( \tilde{\nabla}
      \times \tilde{E} - \tilde{E} \times \tilde{\nabla}
    \right)\right\} }}  \nonumber \\ &
-c_9 \frac{i}{8(am_b)^3} {\VEC{\sigma \cdot \tilde{E} \times
    \tilde{E} }} ~. \label{eqn:LNRQCDv6} 
\end{align}
The parameter $n$ is used to prevent instabilities at large
momentum from the kinetic energy operator. A value of $n=4$ is
chosen for all $am_b$ values. We evaluate the propagator using local sources Eq.~(\ref{eqn:NRQCDSrc}). Here, $am_b$ is the bare $b$ quark mass, $\nabla$ is the symmetric lattice derivative, with $\tilde{\nabla}$ the improved version, and
$\Delta^{(2)}$, $\Delta^{(4)}$ are the lattice discretisations of 
$\Sigma_i D_i ^2$, $\Sigma_i D_i^4$ respectively. ${\VEC{\tilde{E}}}$,
${\VEC{\tilde{B}}}$ are the improved chromoelectric and chromomagnetic
fields, details of which can be found in \cite{Dowdall:Upsilon}. Each of these fields, as
well as the covariant derivatives, must be tadpole-improved. We take the mean trace of the gluon field in
Landau gauge, $u_{0L} = \langle \frac{1}{3}\text{Tr}\,U_{\mu}(x)\rangle$, as the
tadpole parameter, calculated in \cite{Dowdall:Upsilon, Dowdall:BMeson}. The matching coefficients $c_1$, $c_2$, $c_4$, $c_5$, $c_6$ in the above Hamiltonian have been computed perturbatively to one-loop  \cite{Hammant:2013, Dowdall:Upsilon}.  $c_3$ was found to be close to the tree-level value of one non-perturbatively \cite{Dowdall:Upsilon} and we set it, as well as the rest of the matching coefficients, to their tree-level values.  The quark mass $am_b$ is tuned fully nonperturbatively in iNRQCD \cite{Dowdall:Hyperfine} using the spin-averaged kinetic mass of the $\Upsilon$ and $\eta_b$ (which is less sensitive to spin-dependent terms in the action). The above Hamiltonian neglects the four-fermion operators which appear at $\mathcal{O}(\alpha_s^2v^3)$, as well as other operators which appear at higher order in the non-relativistic expansion. Simple power-counting estimates \cite{Dowdall:Upsilon, Dowdall:Hyperfine} would lead us to expect contributions of order a few percent (or a few MeV) at most to binding energies from these terms. In the case where a tetraquark bound state is observed then we can estimate the systematic effect from neglecting these contributions.

The parameters used in this study are summarised in Table \ref{tab:NRQCDParams}. There, $c_1$, $c_5$ and $c_6$ are the correct values for an $\mathcal{O}(v^4)$ iNRQCD action \cite{Dowdall:Upsilon}. For Set $4$, all parameters are those for the $\mathcal{O}(v^4)$ action. The small changes to these coefficients in going to an $\mathcal{O}(v^6)$ iNRQCD action (which are similar in magnitude to the two-loop corrections) are not appreciable for the purpose of this work: whether or not a tetraquark candidate exists below the lowest bottomonium-pair threshold. All other parameters listed in Table \ref{tab:NRQCDParams} are taken from \cite{Dowdall:Hyperfine, Hughes:M1}. 
\begin{table}[t]
  \caption{Parameters used for the valence quarks. $am_b$ is the bare
    $b$-quark mass in lattice units, $u_{0L}$ is the tadpole
    parameter and the $c_i$ are coefficients of terms in the NRQCD
    Hamiltonian (see Eq.~\ref{eqn:LNRQCDv6}). Details of their calculation can be found in \cite{Hammant:2013, Dowdall:Upsilon}. $c_3,c_7,c_8$ and $c_9$ are included at tree-level.  \label{tab:NRQCDParams}}
\begin{center}
  \begin{tabular}{l l l l l l l }
    \hline \hline 
    Set&  $am_b$ & $u_{0L}$ &  $c_1$, $c_6$ & $c_2$ &  $c_4$ & $c_5$  \\ \hline
    $1$ & $2.73$ & $0.8346$ & $1.31$ & $1.02$ &  $1.19$ & $1.16$\\ \hline
    $2$ & $2.66$ & $0.8350$ & $1.31$ & $1.02$ & $1.19$ & $1.16$ \\ \hline
    $3$ & $1.95$ & $0.8525$ & $1.21$ & $1.29$ & $1.18$ & $1.12$ \\ \hline
    $4$ & $1.22$ & $0.8709$ & $1.15$ & $1.00$ & $1.12$ & $1.10$ \\ \hline\hline 
  \end{tabular}
\end{center}
\end{table}

Within iNRQCD the single-particle energy-eigenstates can be decomposed in the standard non-relativistic expansion as
\begin{align}
  E(\VEC{P}) = M^S + \frac{|\VEC{P}|^2}{ 2M^K} + \ldots \label{eqn:NRExpansion}
\end{align}
with $M^S$, $M^K$ the static and kinetic masses respectively \cite{Dowdall:Upsilon}. Because the quark mass term is removed from the iNRQCD Hamiltonian, the static mass is unphysical, differing by a constant shift from the physical (kinetic) mass. This means that only static mass differences determined fully non-perturbatively can be compared to experimental results. However, this constant shift can be calculated in lattice perturbation theory if required \cite{Morningstar:NRQCD}, or by using an additional experimental input. The kinetic mass does not suffer this problem as it acquires the quark mass contributions from the quark kinetic terms \cite{Dowdall:Upsilon}. Also, within iNRQCD the Dirac field $\Psi$ can be written in terms of the quark $\psi$ and anti-quark $\chi$ as $\Psi = (\psi, \chi)^T$. The propagator
is then 
\begin{equation}
K^{-1}(x|y) =
 \begin{pmatrix}
  G_{\psi}(x|y) & 0 \\
  0 & -G_{\chi}(x|y) 
 \end{pmatrix}
\label{eqn:DiracMatrixNRQCD}
\end{equation}
where $G_{\psi}(x|y)$ is the two-spinor component quark propagator and
$G_{\chi}(x|y)$ is the two-spinor component anti-quark
propagator. Taken together, we can now compute the $b$-quark propagator via the iNRQCD evolution equations on the gluon ensembles listed in Table \ref{tab:GluonEnsembles}.  The last piece needed to calculate the two-point correlators, and hence energies, are the discretised finite-volume versions of the interpolating operators. 

\subsection{Discrete Finite-Volume Operators}
\label{subsec:Ops}

Together, the isotropic discretisation and the periodic finite-volume break the infinite-volume continuum $SO(3)$ symmetry of NRQCD to the octahedral group, $O_h$ \cite{Moore:2005}. Thus, while the operators constructed in Sec.~\ref{sec:CtmOp} have well-defined $J^{PC}$ quantum numbers associated with $SO(3)$, we need to construct operators which transform within the irreps of the $O_h$ symmetry group (relevant for lattice calculation). This can be achieved by the method of subduced representations, where an operator with a specific $J^{PC}$ can be taken to a specific lattice irrep\footnote{The conserved quantum numbers of a symmetry group are determined using the little group, which for $SO(3)$ and $O_h$  depend on the momenta type \cite{Thomas:Helicity}. In this study we focus on states at rest.} $\Lambda^{PC}$ by using the subduction coefficients found in Appendix A of \cite{Dudek:Excited}. At rest, each of our $J^{PC}=0^{++}$ and $J^{PC}=1^{+-}$ operators trivially subduce into a single lattice irrep labelled by $A_1^{++}$ and $T_1^{+-}$ respectively. However, the $J^{PC}= 2^{++}$ case is slightly more complicated and subduces into two lattice irreps labelled by $T_2^{++}$ and $E^{++}$ (which are three- and two-dimensional). We construct both the $T_2/E$ operators as $\sqrt{2}\mathcal{O}^{[2]}_{T_2/E} = \mathcal{O}^{J=2,m=2} \pm \mathcal{O}^{J=2,m=-2}$, which are correctly subduced from the $J=2^{++}$ operators defined in Sec.~\ref{sec:CtmOp}. In principle, each lattice irrep allows mixing between different $J$ states, e.g., the $A_1$ irrep contains not only the $J=0$ states but also the $J=4$ \cite{Moore:2005}. However in practice since we are only looking for the ground state of the $\bar{b}\bar{b}bb$ correlators we are not sensitive to these mixing effects.

A complete list of $\bar{b}\bar{b}bb$ interpolating operators used to produce the correlator data herein is given in Table \ref{tab:Ops}. In fact, this is an over-constrained set due to the Fierz identities (shown in Table \ref{tab:Fierz}) which relate the two-meson and diquark-antidiquark correlators. We include this over constrained system and ensure that we reproduce the Fierz relations to numerical precision, performing a non-trivial check on our data. Additionally, we also reproduce the relations between the $8_c\times {8}_c$ colour combination and the others \cite{Tetraquark:Symmetry} on a subset of the data. 

It has been found \cite{CalLat:2NScattering} that separating each hadron within the two-hadron interpolating operator by a specific distance $\VEC{r}$ can significantly aid in the extraction of the (ground) state energy. In this direction, we use three different spatial configurations of the $\bar{b}\bar{b}bb$ correlators where the individual building blocks are separated by a distance of $r_x = 0,1$ or $2$ lattice units in the $x$-direction\footnote{The $\eta_b$ and $\Upsilon$ energies used to determine the non-interacting  $2\eta_b$ and $2\Upsilon$ thresholds (needed to determine if a state exists below them) are found from locally smeared meson correlators only.}.

Finally, the subduced lattice interpolating operators are defined in terms of the Dirac fields as in Eq.~(\ref{eqn:MesonOp}), and the correlators are defined analogously, as in Eq.~(\ref{eqn:WickDiquark}). We can then use the decomposition of $K^{-1}(x|y)$ given in Eq.~(\ref{eqn:DiracMatrixNRQCD}) with suitable boundary conditions to write the correlator in terms of the iNRQCD quark propagator $G_{\psi}(x|y)$. Due to our use of iNRQCD, there are no backward propagating valence anti-quarks in our calculation. Consequently, the appreciable finite-temporal effects seen in relativistic two-meson lattice correlators\footnote{These arise in lattice two-meson calculations when a relativistic formulation of valence quark is used due to one of the mesons propagating forward in time while the other propagates around the temporal boundary backwards in time \cite{Dudek:SDPhase}.} \cite{Dudek:SDPhase}  do not arise in our calculation, simplifying the analysis. With this methodology, it is now possible to compute the lowest energy levels associated with the $\bar{b}\bar{b}bb$ system. 

\begin{table}[t]
  \caption{ The $\bar{b}\bar{b}bb$ interpolating operators used in this study. Operators in each column are subduced from the infinite-volume continuum quantum numbers $J^{PC}$ given in the first row. The superscript on each operator denotes the lattice irrep of that operator and the subscript denotes the building blocks of the operator, as explained in the text. We generate each operator with three different spatial configurations as shown in Eq.~(\ref{eqn:ClebschGordan}): where the building blocks are separated by a distance $r_x = 0,1$ or $2$ lattice units in the $x$-direction. }
\label{tab:Ops}
\begin{center}
  \begin{tabular}{l r | l | l}
    \hline\hline
    \multicolumn{2}{c|}{$0^{++}$} & $1^{+-}$ & $2^{++}$ \\ \hline
    source & sink & source/sink & source/sink \\ \hline 
    $\mathcal{O}^{A_1}_{(\eta_b,\eta_b)}$ & $\mathcal{O}^{A_1}_{(\eta_b,\eta_b)}$ & $\mathcal{O}^{T_1}_{(\Upsilon,\eta_b)}$ & $\mathcal{O}^{T_2}_{(\Upsilon,\Upsilon)}$ \\
    $\mathcal{O}^{A_1}_{(\eta_b,\eta_b)}$ & $\mathcal{O}^{A_1}_{(\Upsilon,\Upsilon)}$ & $\mathcal{O}^{T_1}_{(D_{\bar{3}_c},A_{3_c})}$ & $\mathcal{O}^{E}_{(\Upsilon,\Upsilon)}$\\
    $\mathcal{O}^{A_1}_{(\Upsilon,\Upsilon)}$ & $\mathcal{O}^{A_1}_{(\eta_b,\eta_b)}$ &  & $\mathcal{O}^{T_2}_{(D_{\bar{3}_c},A_{3_c})}$  \\
    $\mathcal{O}^{A_1}_{(\Upsilon,\Upsilon)}$ & $\mathcal{O}^{A_1}_{(\Upsilon,\Upsilon)}$ &  & $\mathcal{O}^{E}_{(D_{\bar{3}_c},A_{3_c})}$  \\
    $\mathcal{O}^{A_1}_{(D_{\bar{3}_c},A_{3_c})}$ & $\mathcal{O}^{A_1}_{(D_{\bar{3}_c},A_{3_c})}$ &  & \\
    $\mathcal{O}^{A_1}_{(D_{{6}_c},A_{\bar{6}_c})}$ & $\mathcal{O}^{A_1}_{(D_{{6}_c},A_{\bar{6}_c})}$ &  & \\
    \hline \hline
  \end{tabular}
\end{center}
\end{table}


\section{The Low-Lying Energy Eigenstates of the $0^{++}$, $1^{+-}$ and $2^{++}$ $\bar{b}\bar{b}bb$ System}
\label{sec:QCD4b}

\begin{figure}[t]
  \centering
  \includegraphics[width=0.49\textwidth]{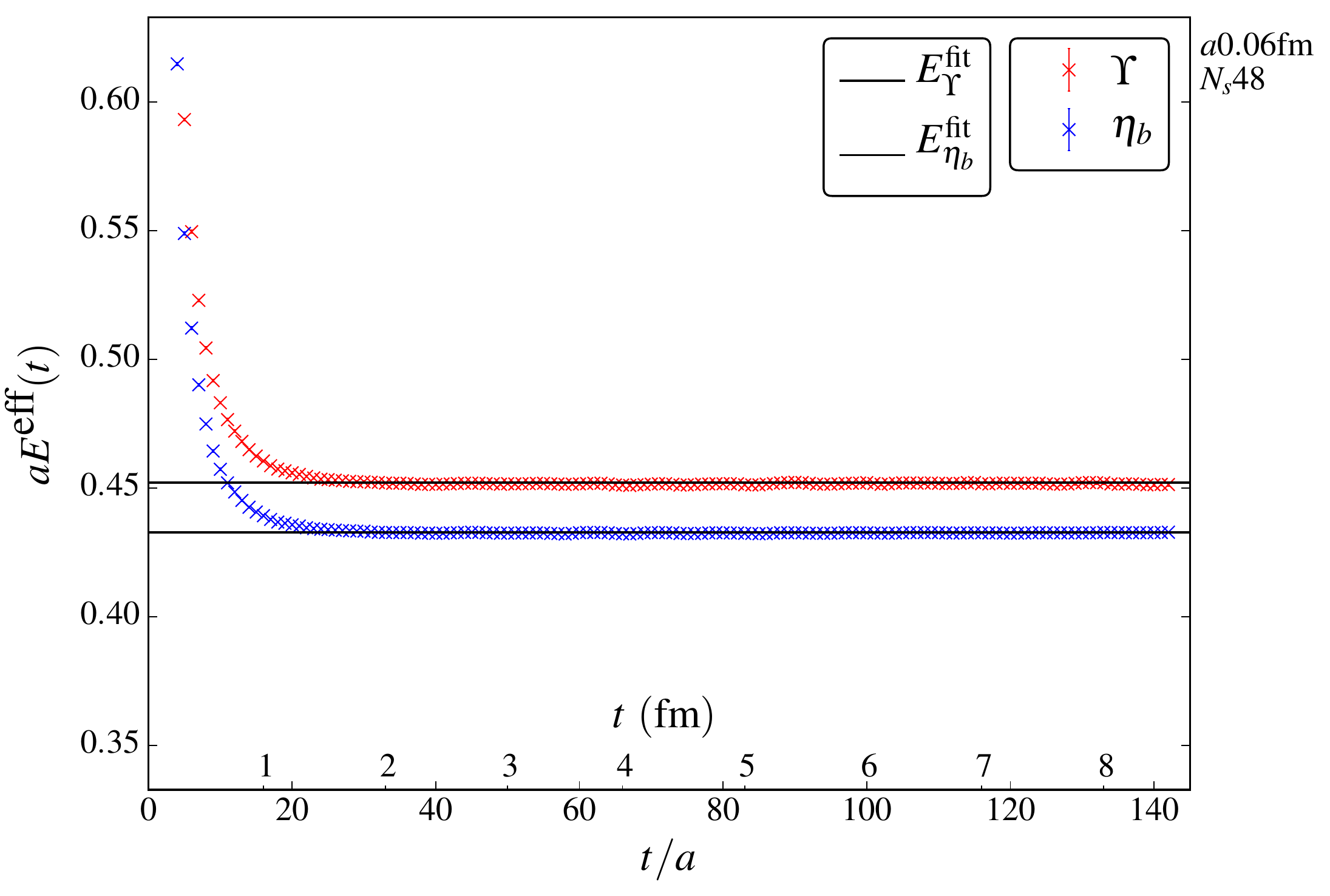}
  \caption{The effective mass plot for the $\eta_b$ and $\Upsilon$ on the ``superfine'' ensemble (Set $4$ listed in Table \ref{tab:GluonEnsembles}). The effective mass plots on the other ensembles are qualitatively identical. (colour online)  }
  \label{fig:EeffSingle}
\end{figure}

In order to determine if there is an energy eigenstate below the $2\eta_b$ threshold, we first need to find the non-interacting thresholds on each ensemble listed in Table \ref{tab:GluonEnsembles}. This is achieved by computing the bottomonium $\eta_b$ and $\Upsilon$ two-point correlators as described above, then fitting them to the functional form given in Eq.~(\ref{eqn:2ptFitSingle}) to extract the single particle energies. As in the range of studies listed in Sec.~\ref{subsec:NRQCD}, amongst others, we utilise the well-established Bayesian fitting methodology \cite{Lepage:Fitting} in this work and refer the reader to \cite{Dowdall:Upsilon} for technical details. Although we fit the correlator data in order to extract fit parameters, in the following we display effective mass plots so the reader can visualise the data. The single particle effective mass is constructed as
\begin{align}
  &aE^{\text{eff}}_{J^{PC}} = \log \left( \frac{C^{J^{PC}}_{i,j}(t)}{C^{J^{PC}}_{i,j}(t+1)} \right)  \label{eqn:Eeff}\\
    &= aE_{J^{PC}} + \frac{Z^i_1Z_1^{j,*}}{Z^i_0Z_0^{j,*}} e^{-(E_1-E_0)t}(1 - e^{-(E_1-E_0)}) + \ldots \label{eqn:EeffApprox} \\
  & \xrightarrow[]{t \to \infty } aE_{J^{PC}}. \label{eqn:EeffSingle}
\end{align}
As can be inferred from Eq.~(\ref{eqn:EeffApprox}), the greater the mass gap $E_1-E_0$ or the larger the ground state overlap $Z_0$ then the quicker $aE^{\text{eff}}_{J^{PC}}$ converges to a plateau, which gives $aE_{J^{PC}}$. The $\eta_b$ and $\Upsilon$ effective masses are shown in Figure \ref{fig:EeffSingle}, where the returned fitted ground-state energy from the correlator fits is also shown overlaid in black. The large difference in energy between the $\eta_b'$ ($\Upsilon'$) and the ground state $\eta_b$ ($\Upsilon$) means that the effective mass plots fall rapidly to a plateau given by the ground state energy, indicating that the fit to the correlator data will extract the ground state energy precisely. 

Also evident from the effective mass plot is the constant signal-to-noise in the $\eta_b$ data, as might be expected from straightforward application of the Parisi-Lepage arguments \cite{Parisi:Noise, Lepage:Noise} for noise growth in a system where all the quarks are the same and no $0^{++}$ bound tetraquark exists. The argument specifies that the noise of the two-point correlator should behave like $\exp\{(E_{J^{PC}} - E_{\text{GS}}/2)t\}$ where $E_{J^{PC}}$ is the lowest energy eigenstate of the bottomonium operator $\mathcal{O}_{J^{PC}}$ constructed to have the quantum numbers $J^{PC}$ and $E_{\text{GS}}$ is the lowest energy eigenstate of the mean squared correlator which controls the noise. Thus, it would be surprising if a tetraquark candidate did exist below the $2\eta_b$ threshold from the lattice perspective alone as then $E_{\text{GS}} < 2E_{\eta_b}$ and the noise of the $\eta_b$ data would grow exponentially. However, the lattice calculation still needs to be performed for a conclusive statement to be made about the existence of this tetraquark candidate since the Parisi-Lepage arguments do not allow for raw crossed Wick contractions that would contribute to either the full two-meson or tetraquark correlator. 

Lattice correlators are affected by both the discrete nature of the space-time lattice and separately by its finite volume. Each has a separate but calculable effect on the extracted lattice energies. Corrections in energies due to discretisation effects are proportional to $a^k$, where $k$ depends on the level of improvement. Here systematic discretisation errors are reduced to $\alpha_s^2a^2$ by the improvements made, as for those studies listed in Sec.~\ref{subsec:NRQCD}, and we expect this to be small enough to have little impact. We can assess this from our results with different values of the lattice spacing. 

Finite-volume effects for single-particle energies (arising from the lightest particle in the sea propagating around the spatial boundaries) are known to behave like $\exp(-aM_{\pi}N_s)$ \cite{Luscher:EnergyStable} and are not appreciable for the ensembles used here which have $aM_{\pi}N_s\approx 4$ \cite{Dowdall:2013rya}. In fact, $N_s$ on Set $1$ and $2$ differ by a factor of two, giving a basic test of volume-dependence. However, finite-volume interactions can shift a two-particle energy by an amount that depends on the infinite-volume scattering matrix. Further, these shifts are non-trivial to parameterise (see for example \cite{Luscher:TwoParticle,Resonance:Moving,Resonance:Spin,Resonance:Coupled,Resonance:Moving2,Resonance:Moving3,Resonance:Coupled2}). As the specific purpose of this study is to search for a hypothetical tetraquark bound state below the lowest threshold, we will not attempt to quantify these finite-volume interactions. 

\begin{figure}[t]
  \centering
  \includegraphics[width=0.49\textwidth]{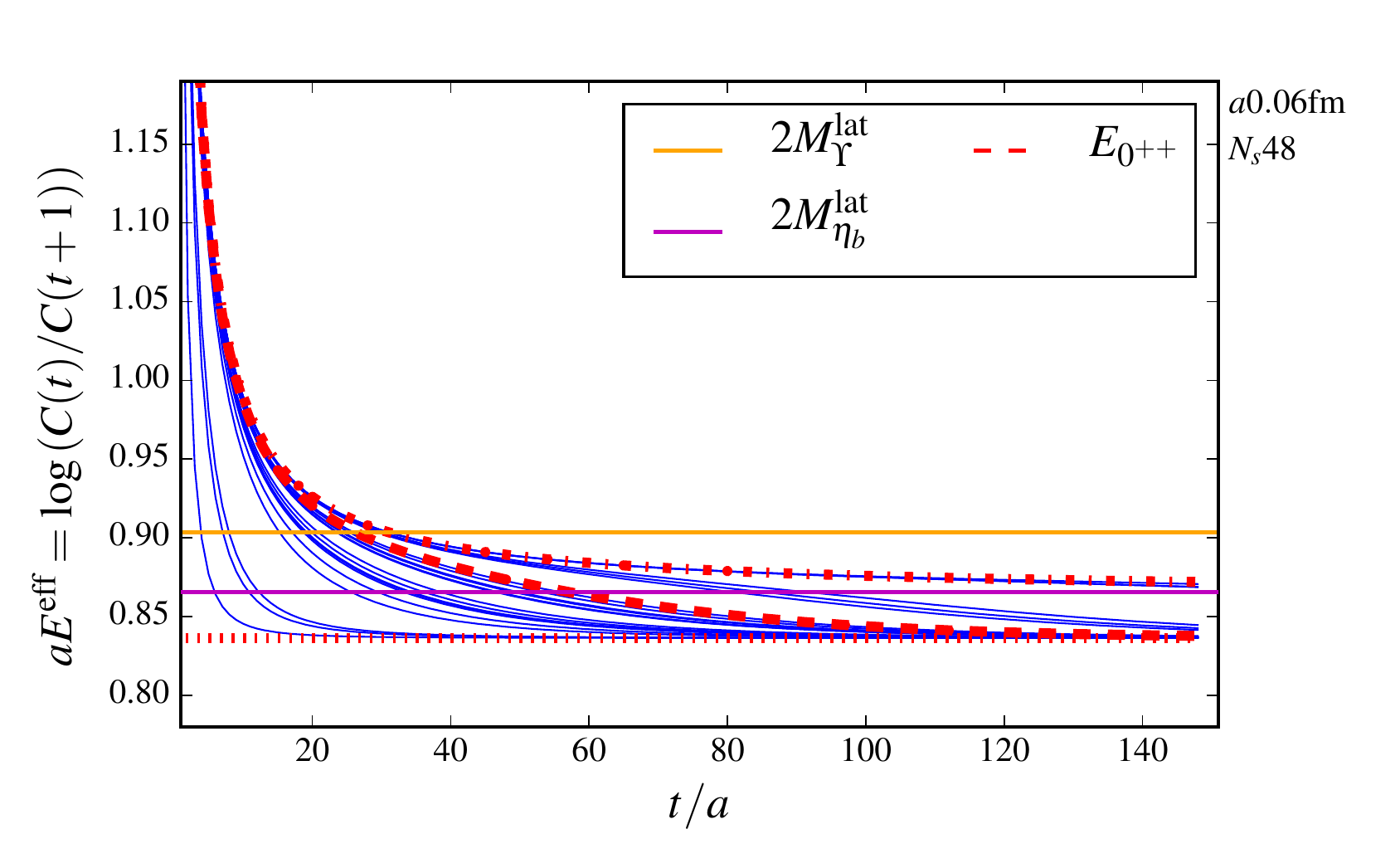}
  \caption{Assuming a tetraquark exists $100$ MeV below the $2\eta_b$ threshold, different normally distributed couplings in $\bar{b}\bar{b}bb$ correlator mock-data produces different effective mass curves on the ``superfine'' ensemble (Set $4$ listed in Table \ref{tab:GluonEnsembles}) as described in the text. (colour online)  }
  \label{fig:EeffMock}
\end{figure}

\begin{figure*}[t]
  \hspace*{\fill}
  \subfloat[$2\eta_b\to 2\eta_b$]{\label{fig:e0pp}
    \includegraphics[width=0.49\textwidth]{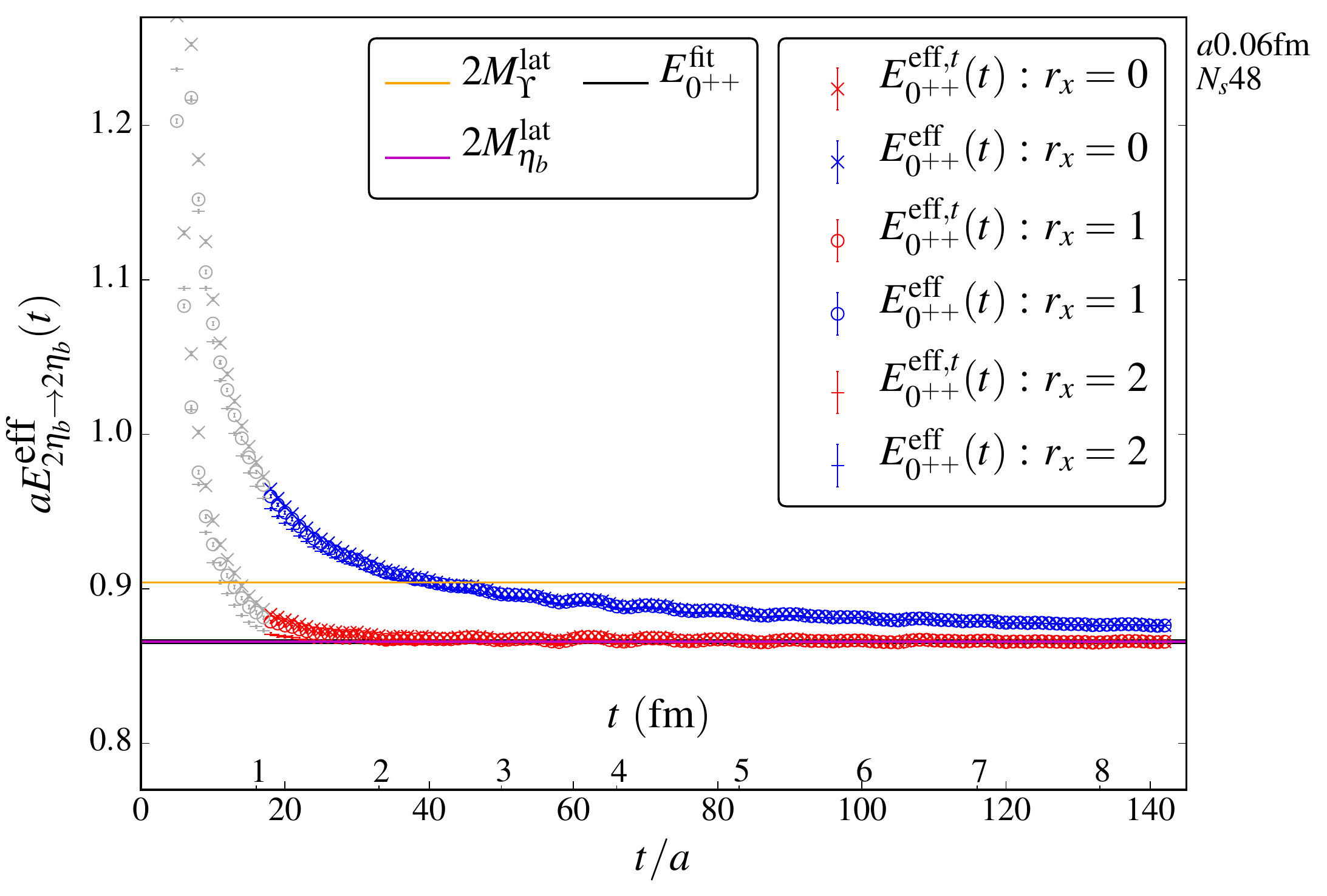}}
  \hfill
  \subfloat[$2\eta_b\to 2\Upsilon$]{\label{fig:p0pp}
    \includegraphics[width=0.49\textwidth]{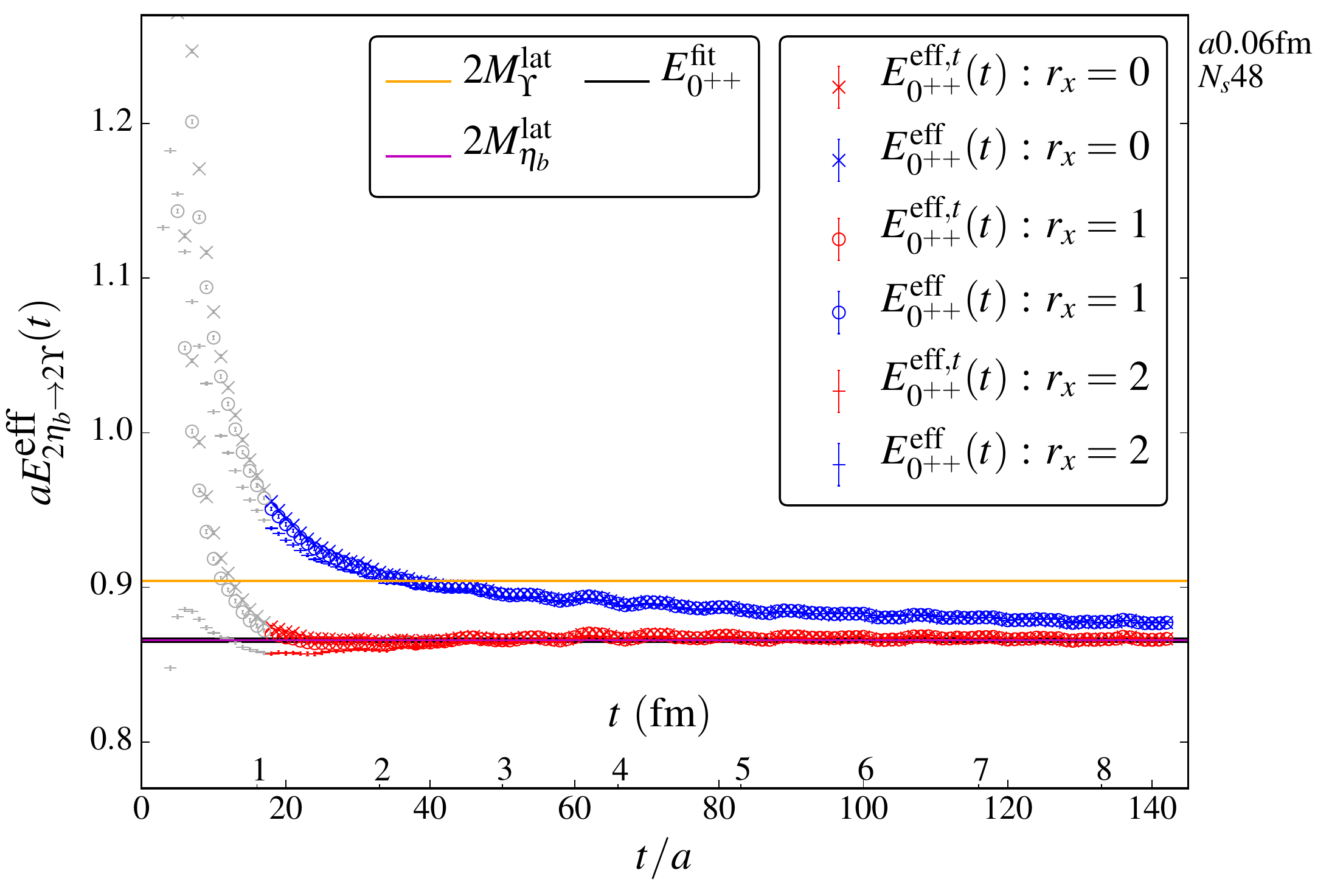}}
  \hspace*{\fill}
  \\
  \hspace*{\fill}
  \subfloat[$2\Upsilon \to 2\eta_b$]{\label{fig:v0pp}
    \includegraphics[width=0.49\textwidth]{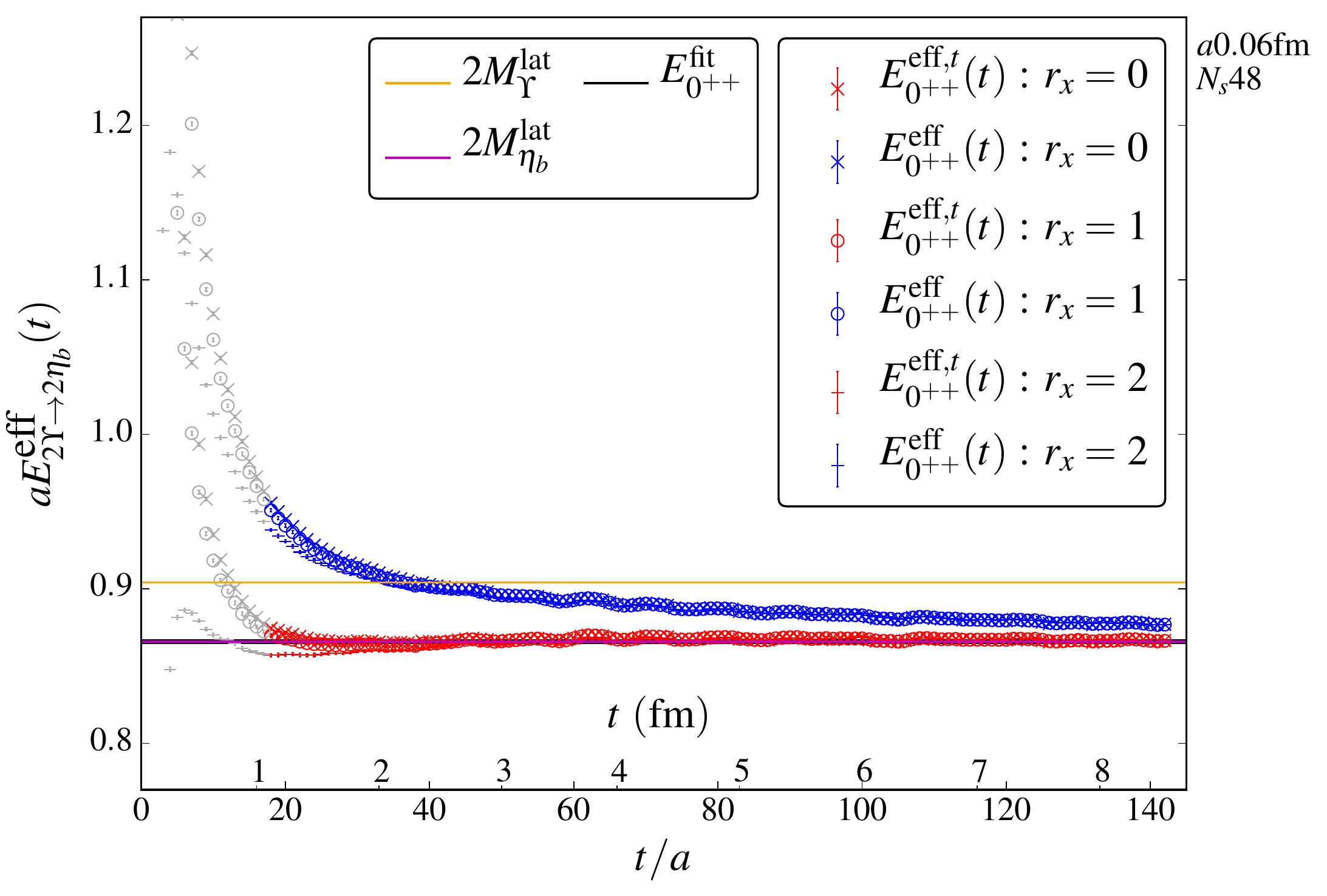}}
  \hfill
  \subfloat[$2\Upsilon \to 2\Upsilon$]{\label{fig:u0pp}
    \includegraphics[width=0.49\textwidth]{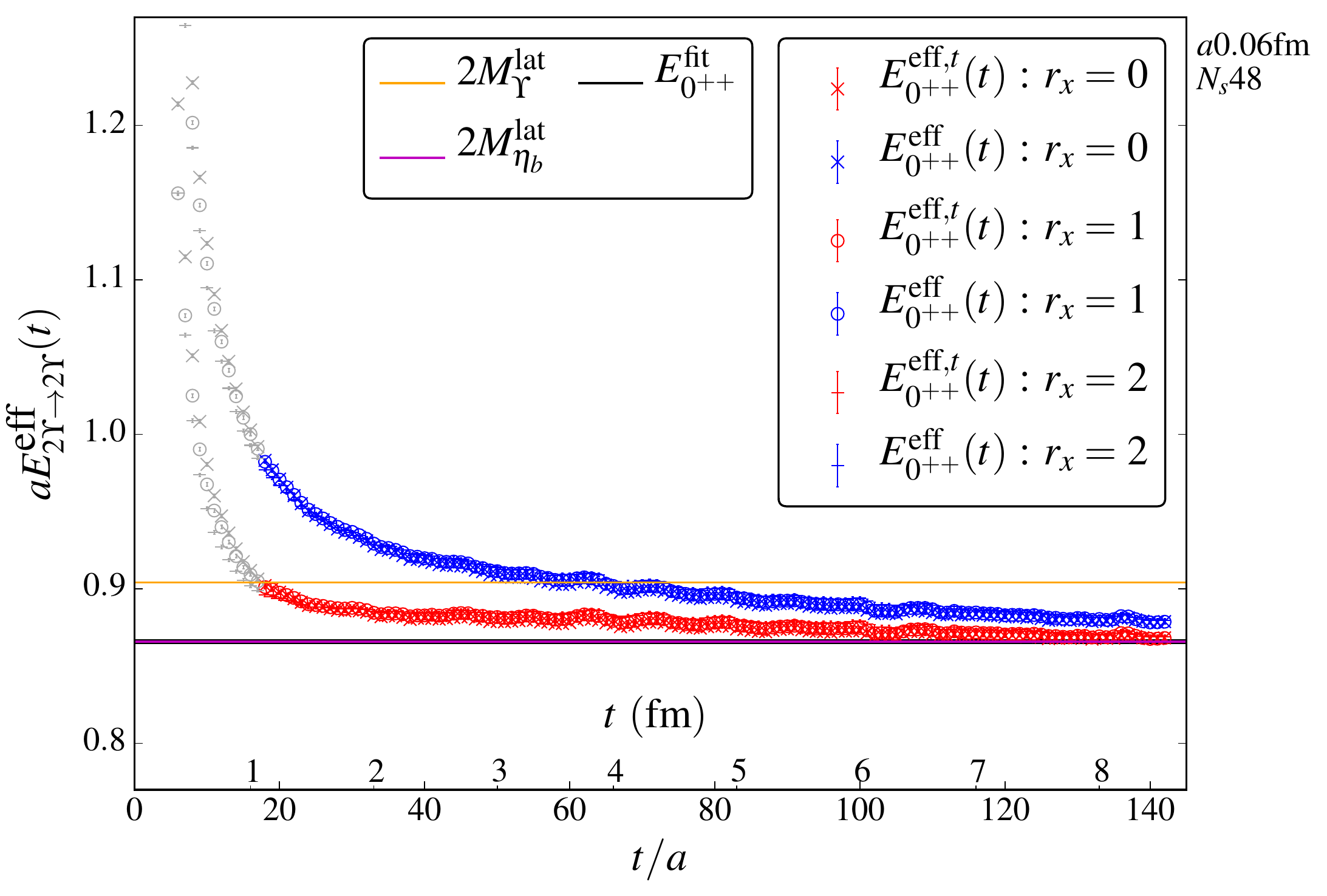}}  
  \hspace*{\fill}
  \caption{The $\bar{b}\bar{b}bb$ effective masses for the $0^{++}$ $(M_1,M_1)\to (M_2,M_2)$ correlators where $M_1$, $M_2$ are the $\eta_b$  or $\Upsilon$. $E^{\textrm{eff}}$ and $E^{\textrm{eff},t}$ are the single- and two-particle effective masses defined in Eq.~(\ref{eqn:EeffSingle}) and Eq.~(\ref{eqn:EeffTwo}) respectively. The mesons are separated by a distance $r_x$ in the $x$-direction when constructing the two-meson interpolating operator as given in Eq.~(\ref{eqn:ClebschGordan}). Gray points are not used when fitting the data (c.f., App.~\ref{app:Fit}). (color online)}
  \label{fig:0ppTwoMeson}
\end{figure*}

Eq.~(\ref{eqn:2ptFitTwo}) describes the non-relativistic two-particle contribution to the correlator after $t=1$ fm (as shown in Appendix \ref{app:Fit}). In this case, because of the additional $1/t^{\frac{3}{2}}$ time-dependence, the effective mass formula for these contributions differs from their single-particle counterparts. Removing the leading order time-dependence yields an effective mass defined as
\begin{align}
  &aE^{\text{eff},t}_{J^{PC}} = \log \left( \frac{t^{\frac{3}{2}}C^{J^{PC}}_{i,j}(t)}{(t+1)^{\frac{3}{2}}C^{J^{PC}}_{i,j}(t+1)} \right). \label{eqn:EeffTwo}
\end{align}

For the $0^{++}$, $1^{+-}$ and $2^{++}$ operators that are constructed, if a stable tetraquark exists below the $2\eta_b$, $\eta_b+\Upsilon$ or $2\Upsilon$ thresholds then it would show up as the ground state of each correlator and hence also in the effective masses. Otherwise, each threshold would be the lowest energy eigenstate. Higher energy states will also appear in each correlator. For example, the $2\Upsilon$ and $\eta_b\eta_b'$ in the $0^{++}$ case, the $\Upsilon\eta_b'$ in the $1^{+-}$ and the $\Upsilon\Upsilon'$ and $\chi_{b0}\chi_{b2}(1P)$ in the $2^{++}$. Of these,  if no tetraquark state is present, only the $2\Upsilon$ in the $0^{++}$ might be noticeable while studying the ground state, as it is $\mathcal{O}(100)$ MeV above the $2\eta_b$. All other excited states have similar energy differences to those appearing in the $\eta_b/\Upsilon$ effective masses shown in Figure \ref{fig:EeffSingle}, which rapidly falls to the ground state. 

\begin{table*}[t]
  \caption{The ground state static masses extracted from the lattice $\bar{b}\bar{b}bb$ correlator data as described in the text.}
\label{tab:Energies}
\begin{center}
  \begin{tabular}{l l l l l l l}
    \hline \hline 
    Set & $aM_{\eta_b}$ & $aM_{\Upsilon}$ & $aM_{0^{++}}$ & $aM_{1^{+-}}$ & $aM_{2_{T_2}^{++}}$ & $aM_{2_E^{++}}$ \\  \hline\hline
    $1$ & $0.25548(13)$ & $0.29180(22)$  & $0.5121(7)$ & $0.5500(12)$ & $0.5840(22)$  & $0.5863(29)$ \\ \hline
    $2$ & $0.25741(19)$ & $0.29365(39)$  & $0.5162(12)$ & $0.5534(21)$ & $0.5921(20)$  & $0.5911(28)$ \\ \hline
    $3$ & $0.26570(8)$ & $0.29288(13)$  & $0.5321(16)$ & $0.5594(11)$ & $0.5899(21)$  & $0.5888(18)$ \\ \hline
    $4$ & $0.43288(8)$ & $0.45209(12)$  & $0.8658(10)$ & $0.8865(23)$ & $0.9083(36)$  & $0.9079(35)$ \\ 
 \hline \hline
  \end{tabular}
\end{center}
\end{table*}

\begin{figure}[t]
  \centering
  \includegraphics[width=0.49\textwidth]{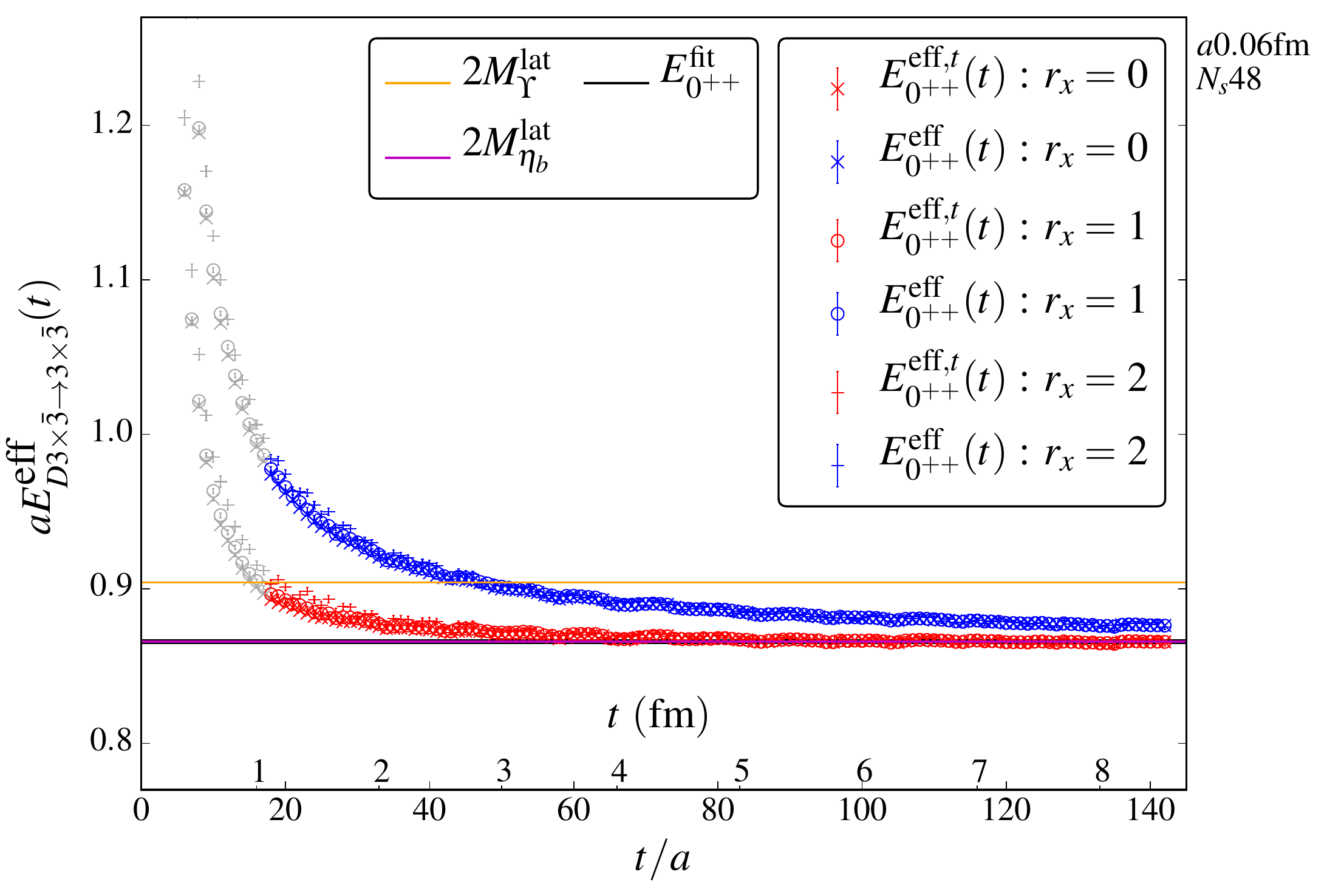}  
  \includegraphics[width=0.49\textwidth]{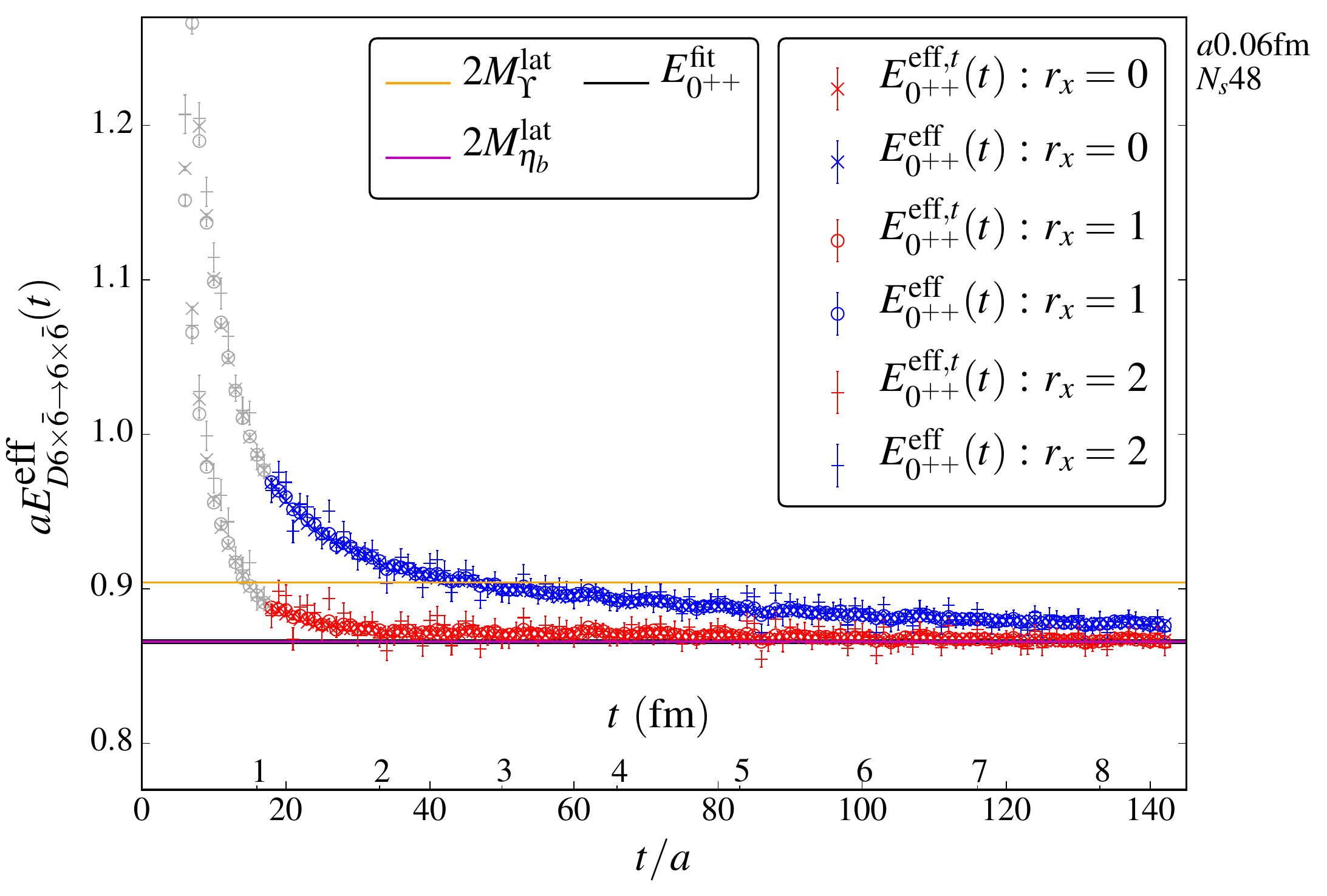}
  \caption{The $\bar{b}\bar{b}bb$ effective masses for the $0^{++}$ diquark-antidiquark $\bar{3}\times \bar{3}$ and ${6}\times \bar{6}$ correlators. $E^{\textrm{eff}}$ and $E^{\textrm{eff},t}$ are the single- and two-particle effective masses defined in Eq.~(\ref{eqn:EeffSingle}) and Eq.~(\ref{eqn:EeffTwo}) respectively. The diquarks are separated by a distance $r_x$ in the $x$-direction when constructing the diquark-antidiquark interpolating operator as given in Eq.~(\ref{eqn:ClebschGordan}). (color online)}
  \label{fig:0ppDiquark}
\end{figure}

It is helpful at this stage to generate mock correlator data to illustrate how we might expect the $\bar{b}\bar{b}bb$ correlator results to behave in the presence or absence of a low-lying tetraquark state (neglecting two-particle finite-volume effects). Using the extracted lattice $\eta_b$ and $\Upsilon$ masses, we can compute the non-interacting $2\eta_b$ and $2\Upsilon$ thresholds on our ensembles. Further, for a fixed value of $Z_{X_2}^0$, we can also compute their leading order two-particle contribution to the correlator from Eq.~(\ref{eqn:2ptFitTwo}). Additionally we can infer the values of the non-interacting $\eta_b\eta_b'$ and $\Upsilon\Upsilon'$ masses on our ensembles using the experimental PDG \cite{PDG:2016} values as input in order to include their two-particle contributions in the mock correlator data also. Further,  in the $0^{++}$ channel, if we assume a tetraquark bound-state exists $100$ MeV below the $2\eta_b$ threshold\footnote{The smallest binding of a below threshold $\bar{b}\bar{b}bb$ tetraquark from the phenomenological studies  (which are shown in Figure \ref{fig:ECompare}) was $108$ MeV \cite{4b:Bai}.}, for a fixed value of the tetraquarks non-perturbative overlap, $Z_{4b}$, this hypothetical state's contribution to the correlator is given by Eq.~(\ref{eqn:2ptFitSingle}). Then, given the effective mass formula defined in Eq.~(\ref{eqn:Eeff}), for each different choice of the non-perturbative coefficients we can generate a separate effective mass curve. In practice, we choose different values of the coefficients from a normal distribution with zero mean and unit variance. Figure \ref{fig:EeffMock} shows such a plot for the ``superfine'' ensemble (Set $4$ in Table \ref{tab:GluonEnsembles}) where the solid blue curves represent different values of the normally distributed coefficients. As the non-perturbative coefficients show up through the ratio $Z_{X_2}^0/Z_{4b}$ in the effective mass formula, analogously to Eq.~(\ref{eqn:EeffApprox}), once the energy difference $E_1 - E_0$ is set the effective mass is only sensitive to the relative size of the of the tetraquark overlap to that of the lowest threshold (once the contribution of excited states has become negligible). With this knowledge, the lower red dotted curve gives mock-data in the situation where there is only a new state present in the correlator data ($Z_{X_2}^0=0$), the middle red dashed curve indicates the case when the tetraquark and two-meson states have the same value of coefficient ($Z_{X_2}^0=Z_{4b}$), while the upper dot-dashed red curve is the mock-data in the case where there is no new state in the data ($Z_{4b}=0$). The blue curves below the middle red one have a larger overlap onto the new state while those above have an increasingly vanishing one. With our over-constrained colour-spin basis of $S$-wave operators at least one operator should have an appreciable overlap onto a tetraquark state below threshold (if it exists) and, as illustrated by the mock-data, give a easy/clean signal similar to the middle red dashed curve where the effective mass drops below the $2\eta_b$ threshold. Even though Figure \ref{fig:EeffMock} is mock-data, the upper red curve with no tetraquark state present looks very like the  real data shown in Figures \ref{fig:0ppTwoMeson} and \ref{fig:0ppDiquark}.

One important point to note is the additional factor of $1/t^{\frac{3}{2}}$ appearing in the two-particle contribution in Eq.~(\ref{eqn:2ptFitTwo}) relative to the single-particle one in Eq.~(\ref{eqn:2ptFitSingle}). This factor suppresses the two-particle contribution relative to the single-particle state, e.g., $t=100$ gives a suppression of the two-particle state of $(0.01)^{1.5}$. This is one reason why the middle red dashed curve has a particularly rapid fall to the new state which lies only $100$ MeV below the $2\eta_b$ (compared to Figure \ref{fig:u0pp} where the $2\Upsilon$ is $\mathcal{O}(100)$ MeV above the $2\eta_b$). Overall this effect would produce an enhancement of the stable tetraquark state if it exists. Further, while we used the ``superfine'' ensemble for illustrative purposes, the other ensembles with larger lattice spacings have similar features but with less temporal resolution due to the larger numerical value of the lattice spacing.

If one is searching for a bound state below threshold then both single- and two-particle contributions appear in the correlator and one may use the  single particle effective mass formula given in Eq.~(\ref{eqn:EeffSingle}) in order to highlight the bound ground state mass (as done in the mock data above). However if no bound state exists in the data, which we do not know {\it{a priori}}, then not removing the $1/t^{\frac{3}{2}}$ dependence from the two-particle contributions has an important effect: an additional factor of $1.5\log(1+1/t)$ is introduced into the effective mass formula in Eq.~(\ref{eqn:EeffSingle}), giving a contamination that vanishes slowly as $t\to \infty$. This would produce a confusing picture of what is actually contributing. The two-particle effective mass formula Eq.~(\ref{eqn:EeffTwo}) removes this contribution. In the results to be reported now, we will overlay both single- and two-particle effective masses on the same plot for the reader's convenience.

We generate the $\bar{b}\bar{b}bb$ correlator data for the operators given in Table \ref{tab:Ops} using the ensembles listed in Table \ref{tab:GluonEnsembles} and fit this data simultaneously with the bottomonium meson data so to include correlations between data sets. All the $\bar{b}\bar{b}bb$ data within a specific irrep and those which are unrelated by a Fierz relation\footnote{Simultaneously fitting data sets related by a Fierz identity would mean the correlation matrix would have a zero eigenvalue and thus not be invertable for use in a least-squares minimisation.} are fit using Eq.~(\ref{eqn:2ptFitTwo}) for the two-particle contributions and Eq.~(\ref{eqn:2ptFitSingle}) for a hypothetical tetraquark state below threshold. The mean of the prior energy of the $2\eta_b$, $\eta_b+\Upsilon$ and $2\Upsilon$ thresholds are roughly estimated based on the effective masses and then given a suitably wide prior width of $100$ MeV while a tetraquark state prior energy is taken to be $250(100)$ MeV below each threshold. As can be seen in Figure \ref{fig:0ppTwoMeson}, since the data plateaus to the non-interacting $2\eta_b$ threshold, no energy eigenstate is found below this threshold and variations of the tetraquark prior energy are insignificant. Similar behaviour is seen with the other quantum numbers. 

\begin{figure}[t]
  \centering
  \includegraphics[width=0.49\textwidth]{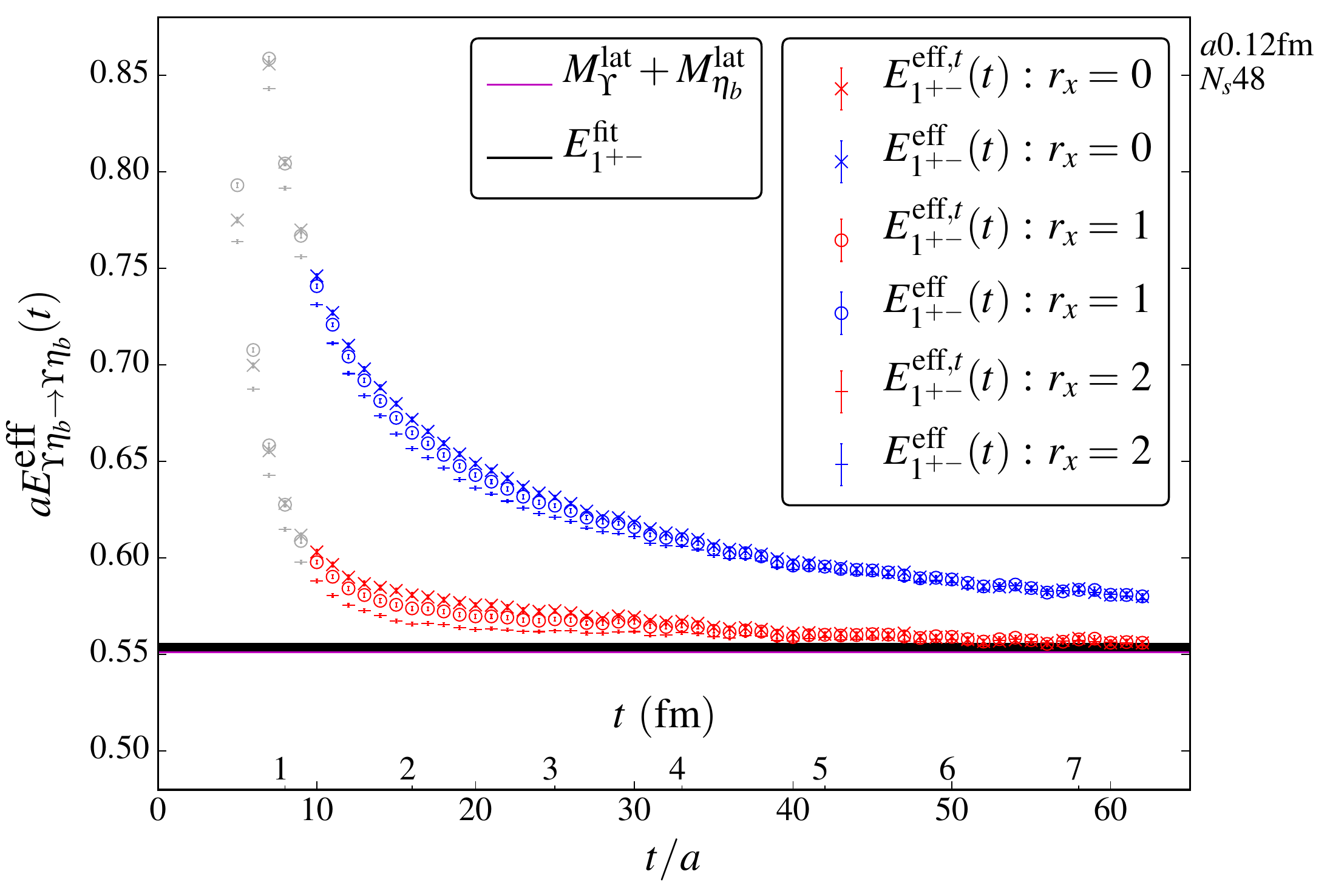}  
  \includegraphics[width=0.49\textwidth]{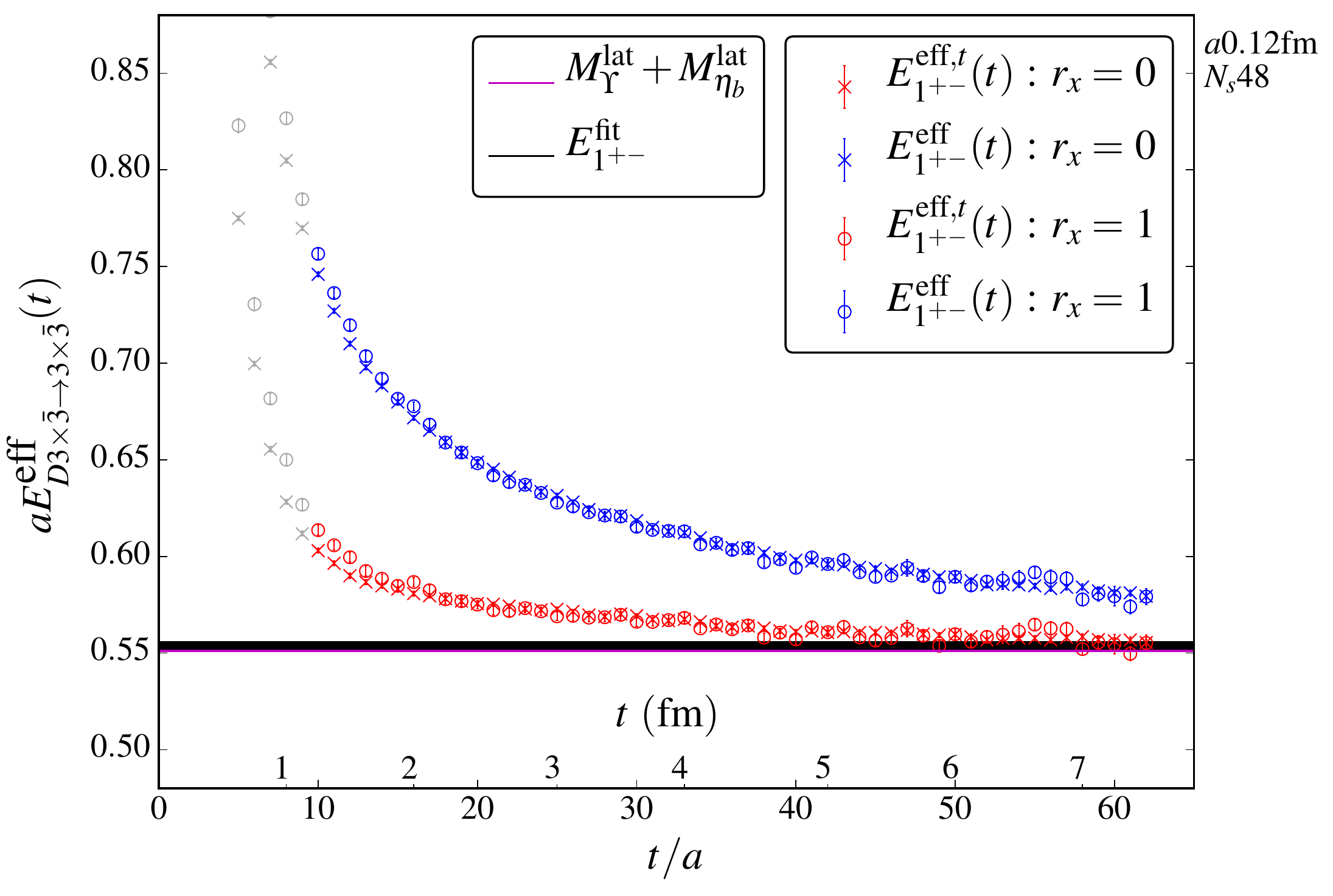}
  \caption{The $\bar{b}\bar{b}bb$ effective masses for the $1^{+-}$ $\Upsilon\eta_b$ and diquark-antidiquark $\bar{3}\times \bar{3}$ correlators. $E^{\textrm{eff}}$ and $E^{\textrm{eff},t}$ are the single- and two-particle effective masses defined in Eq.~(\ref{eqn:EeffSingle}) and Eq.~(\ref{eqn:EeffTwo}) respectively. (color online)}
  \label{fig:1pmTwoDiquark}
\end{figure}

\begin{figure}[t]
  \centering
  \includegraphics[width=0.49\textwidth]{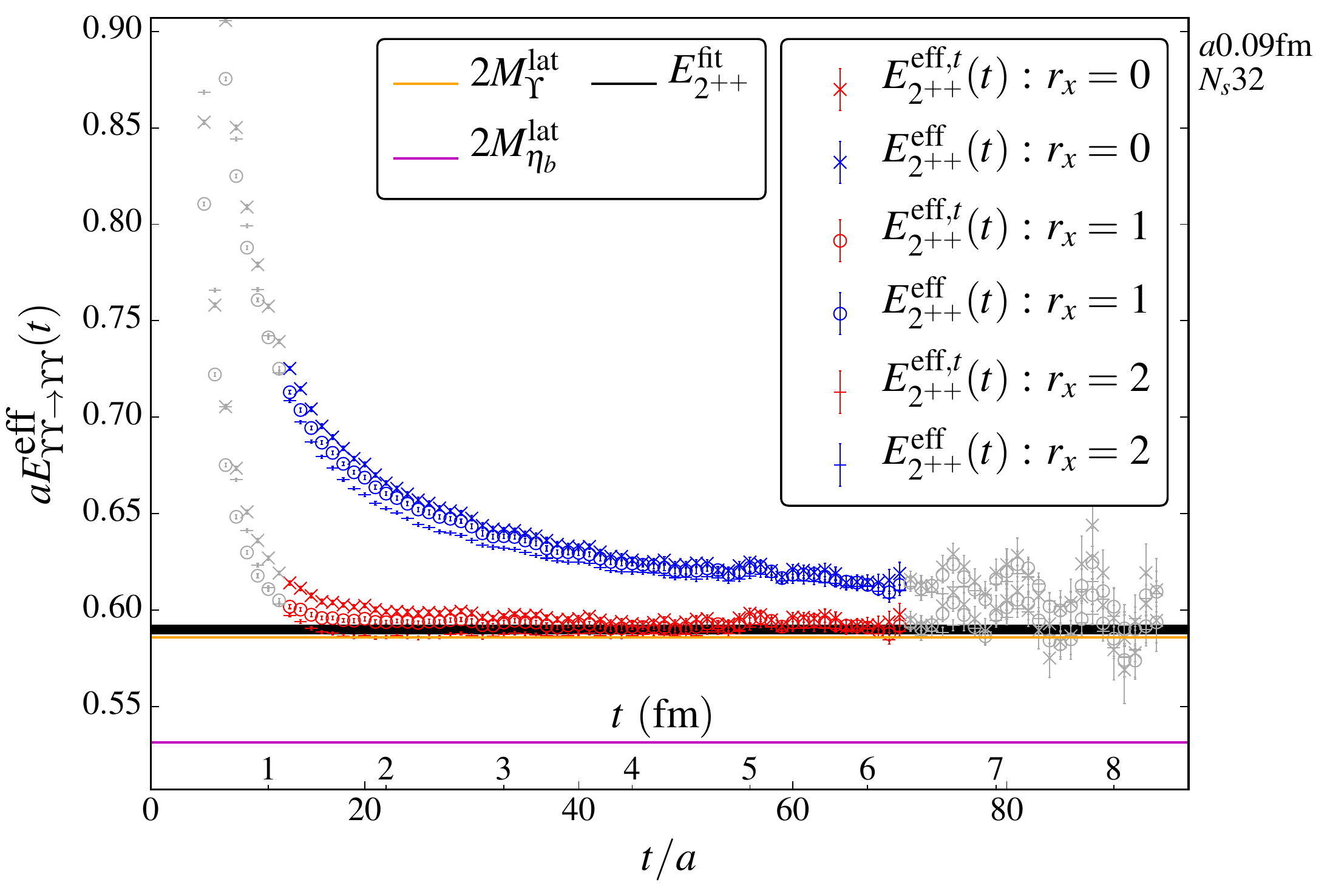}  
  \includegraphics[width=0.49\textwidth]{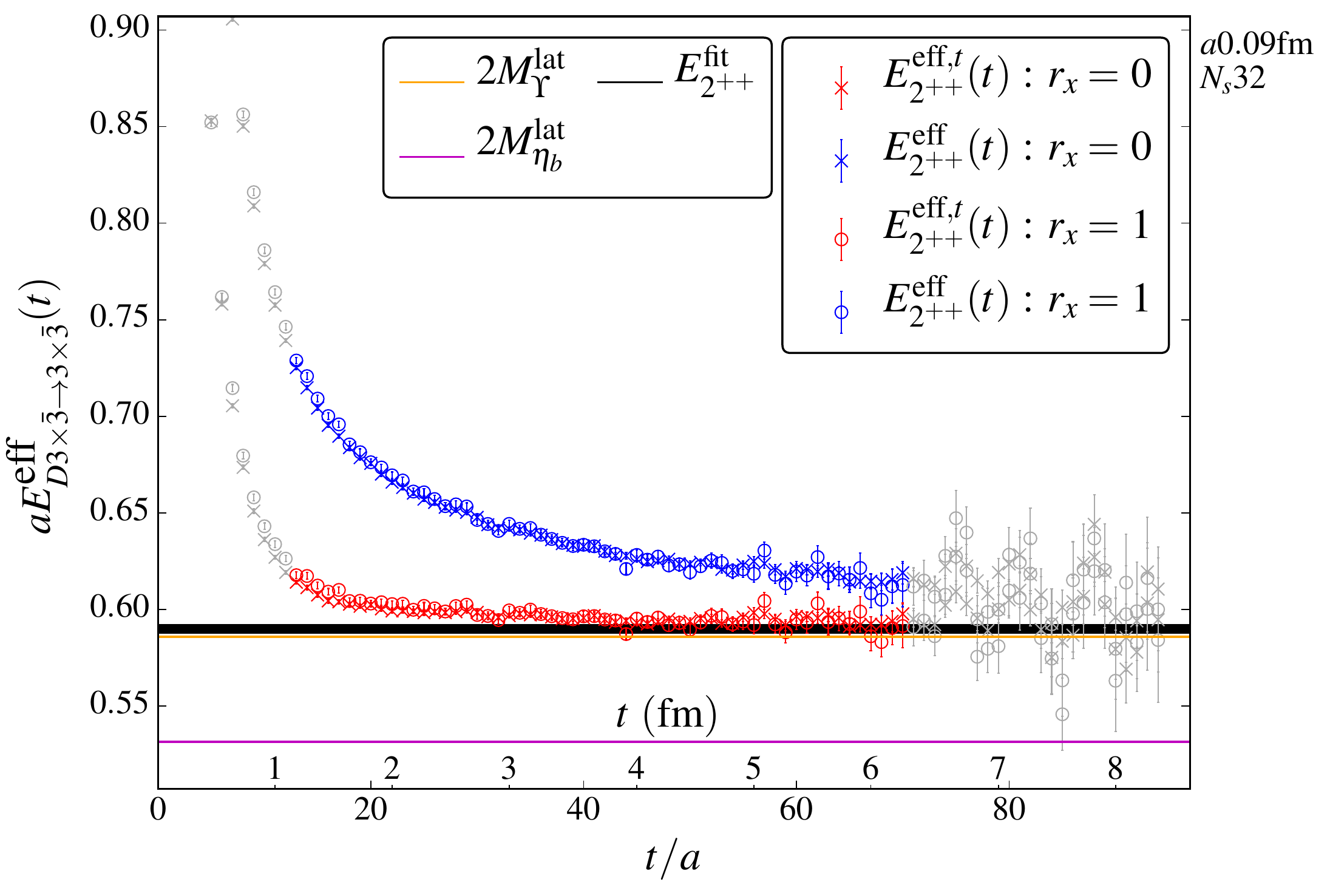}
  \caption{The $\bar{b}\bar{b}bb$ effective masses for the $2^{++}$ $\Upsilon\Upsilon$ and diquark-antidiquark $\bar{3}\times \bar{3}$ correlators subduced into the $T_2$ irrep.  $E^{\textrm{eff}}$ and $E^{\textrm{eff},t}$ are the single- and two-particle effective masses defined in Eq.~(\ref{eqn:EeffSingle}) and Eq.~(\ref{eqn:EeffTwo}) respectively. (color online)}
  \label{fig:2pTTwoDiquark}
\end{figure}

\begin{figure}[t]
  \centering
  \includegraphics[width=0.30\textwidth]{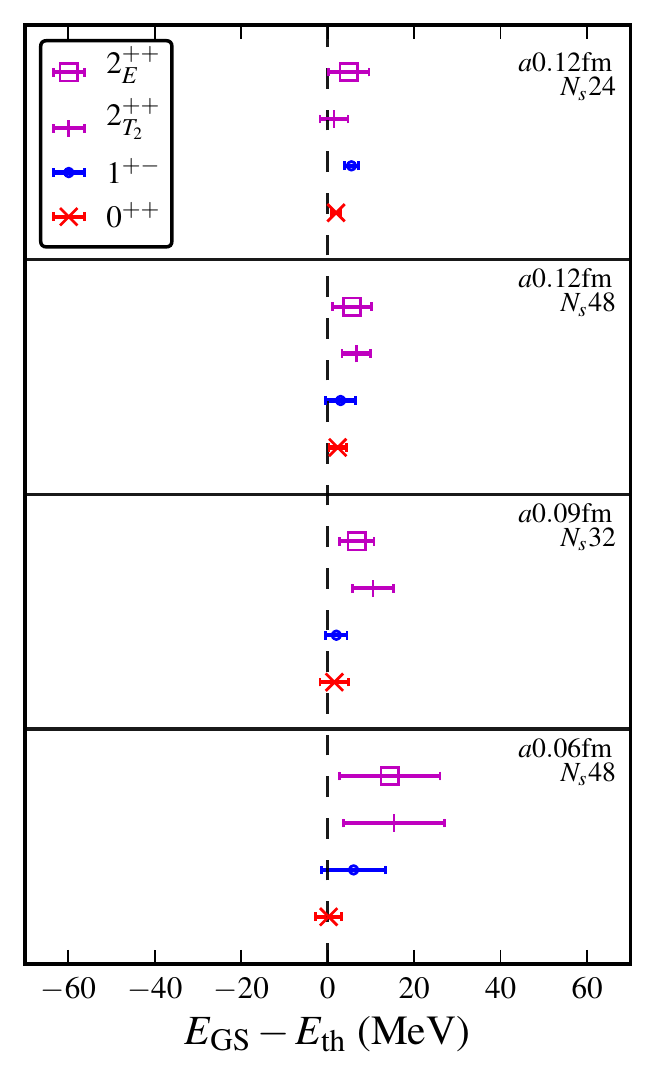}
  \caption{A summary of the $\bar{b}\bar{b}bb$ ground state energies with the lowest non-interacting  bottomonium-pair threshold subtracted, across the different lattice ensembles listed in Table \ref{tab:GluonEnsembles}.  Statistical error only. Note, as shown in Table \ref{tab:GluonEnsembles}, fewer configurations were used on the $a = 0.06$ fm ensemble than on the others. (color online)}
  \label{fig:ESummary}
\end{figure}

Again, while we fit the correlator data in order to extract particle energies, so that the reader can visualise this data we display effective mass plots on the different ensembles. The ``superfine'' ensemble (Set $4$) $0^{++}$ two-meson effective masses are shown in Figures \ref{fig:0ppTwoMeson}, while the $0^{++}$ diquark-antidiquark are given in Figure \ref{fig:0ppDiquark}, the ``physical coarse'' (Set $2$) $1^{+-}$ two-meson and diquark-antidiquark in Figure \ref{fig:1pmTwoDiquark} and the ``fine'' (Set $3$) $2^{++}$ two-meson and diquark-antidiquark correlators subduced into the $T_2$ lattice irrep are shown in Figure \ref{fig:2pTTwoDiquark}. Each plot has the fitted ground state energy overlaid in black for comparison. The $2^{++}$ subduced into the $E$ irrep is similar to the $T_2$ case. Further, the behaviour of the lattice data on all ensembles is qualitatively similar to those shown. The extracted ground state energies in each channel are given in Table \ref{tab:Energies} and a comparison of the energies is shown in graphical form in Figure \ref{fig:ESummary}. 

It should also be noted that the numerical value of the effective mass (shift) plateau in two-hadron correlators has been shown, in certain cases, to be sensitive to the choice of interpolating operators \cite{Iritani:Mirage, Yamazaki:Mirage}. There, the authors found that when the noise growth in the correlator data restricts the study of effective masses to a maximum propagation time of approximately $2$ fm, fake plateaus can appear. These fake plateaus can be a consequence of different choices of source and sink (smeared) operators: this can cause the a negative sign in the $Z_1$ term in the effective mass formula Eq.~(\ref{eqn:EeffApprox}) and a dip below threshold can appear for a short time range which can be misinterpreted as a bound state. Wall-sources were shown to be particularly prone to this behaviour of producing a ``false dip'' and obtaining an appreciably different ``plateau'' than a Gaussian source. Here we only use local quark sources. In addition, the elastic scattering states can also have a dependence on the choice of operator, which can cause a slow decay to the ground state and mimic a slowly varying effective mass that can be mistaken for a plateau over a short time range. As noted in these studies, a necessary check for a real effective mass plateau when using different source and sink operators is the convergence of all data to a single plateau at times larger than approximately $2$ fm. As we separate the operators by $r_x=0,1$ and $2$ lattice units (as described in Section \ref{subsec:Ops}) and propagate to $t>8$ fm, this is a consistency check we satisfy. 

A few notable features of the $\bar{b}\bar{b}bb$ effective mass plots are evident. First and foremost, no value of the effective mass is observed below the lowest non-interacting  bottomonium-pair threshold in any channel, in line with what one would expect if no stable tetraquark candidate existed below threshold. Indeed, the $\bar{b}\bar{b}bb$ effective mass plots are strikingly similar to the upper dot-dashed curve in the mock data in Figure \ref{fig:EeffMock} where no bound tetraquark state is present. Additionally, the $2\eta_b\to 2\eta_b$ effective mass shown in Figure \ref{fig:e0pp} plateaus very early due to the larger overlap onto the $2\eta_b$ threshold, while the $0^{++}$ $2\Upsilon\to 2\Upsilon$ effective mass shown in Figure \ref{fig:u0pp} falls more slowly to the $2\eta_b$ threshold due to the larger overlap onto the nearby $2\Upsilon$ threshold. The cross correlators $2\eta_b\to 2\Upsilon$ show how the different operators converge to a single plateau at a time greater than $t\approx 4$ fm, a necessity for a true plateau as discussed above. The local diquark-antidiquark $0^{++}$ correlator data is a linear combination of the $2\eta_b$ and $2\Upsilon$ two-meson data, related by the Fierz identities given in Table \ref{tab:Fierz}, and the effective masses shown in Figure \ref{fig:0ppDiquark} reflects this. It is empirically observed that separating the diquark from the anti-diquark by too large a distance $r_x$ results in larger noise due to the separation of colour sources. The $1^{+-}$ $\Upsilon\eta_b$ and diquark-antidiquark effective masses are shown in Figure \ref{fig:1pmTwoDiquark}, where the noise starts to increase after $t \approx 7$ fm due to the Parisi-Lepage argument mentioned above with the noise being set by at least the $2\eta_b$ threshold. Based on this, one would also expect the signal-to-noise to be worse for the $2^{++}$ data, which is also evident from the correlator data subduced into the $T_2$ irrep shown in Fig.~\ref{fig:2pTTwoDiquark}. As Set $2$ has physical $m_l/m_s$ corresponding to a pion mass of $\mathcal{O}(131)$ MeV, while the other ensembles have nonphysical $m_l/m_s$ corresponding to pion masses of $\mathcal{O}(300)$ MeV \cite{HISQEnsemble:Params}, no sensitivity to light sea-quarks is observed. This would be expected from the smallness of the Van-der-Waals potential generated by the two-pion exchange between two $1S$ bottomonium mesons \cite{Brambilla:2Pion}. As can be seen in Figures \ref{fig:2pTTwoDiquark} and \ref{fig:ESummary}, the $2^{++}$ ground state obtained from the lattice is slightly higher than that of the non-interacting threshold. However, this is the state which has the largest signal-to-noise, restricting the data to shorter time regions and it is possible that we are sensitive to the same aforementioned issue of a slowly varying fake plateau. Alternatively, this positive shift in the two-particle energy could potentially indicate appreciable infinite-volume continuum scattering arising from finite-volume interactions \cite{Luscher:TwoParticle}, but quantifying these phase-shifts is outside the remit of this study. Regardless, these effects do not indicate that a bound tetraquark state exists in this channel. 

\begin{figure}[t]
  \centering
  \includegraphics[width=0.49\textwidth]{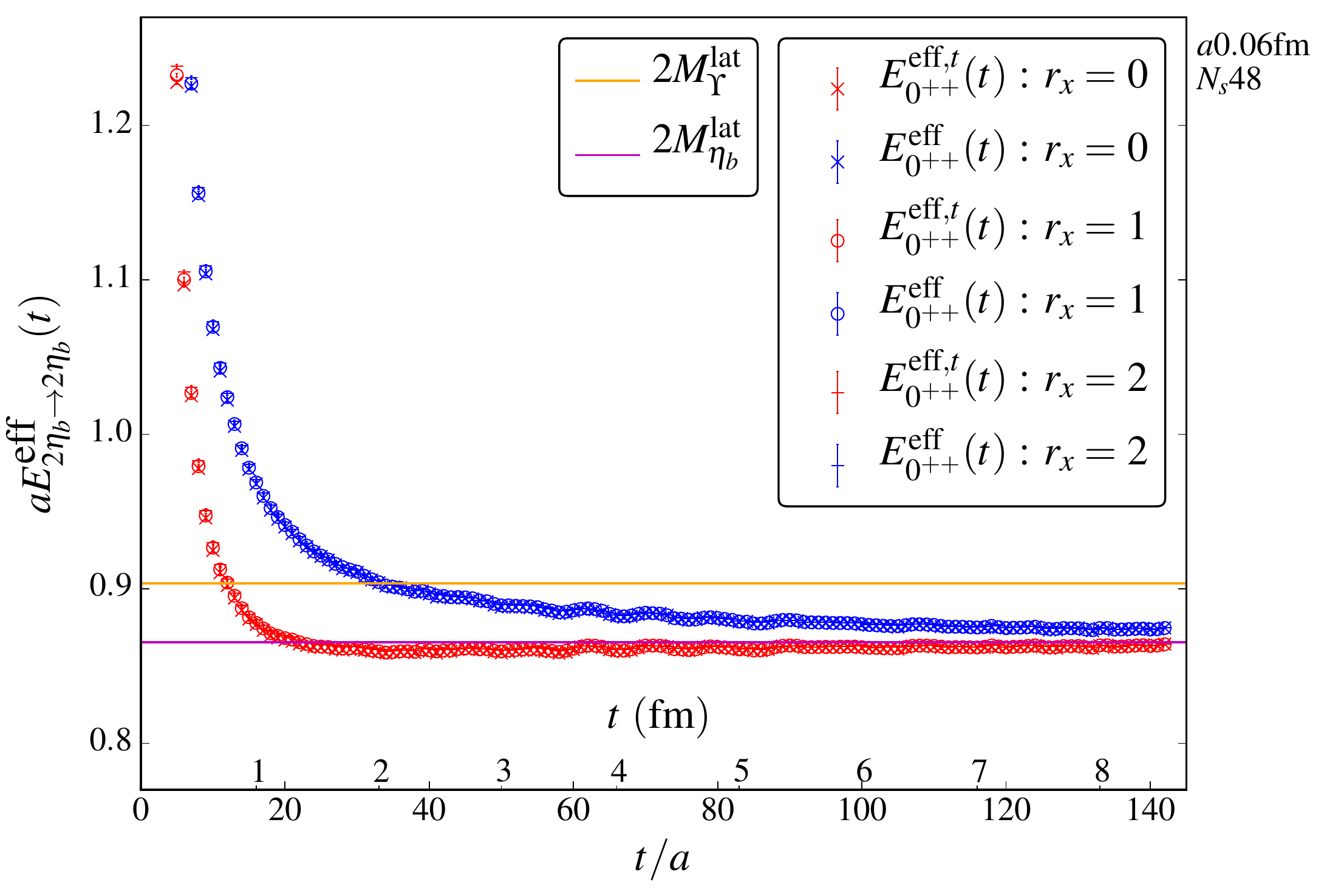}  
  \includegraphics[width=0.49\textwidth]{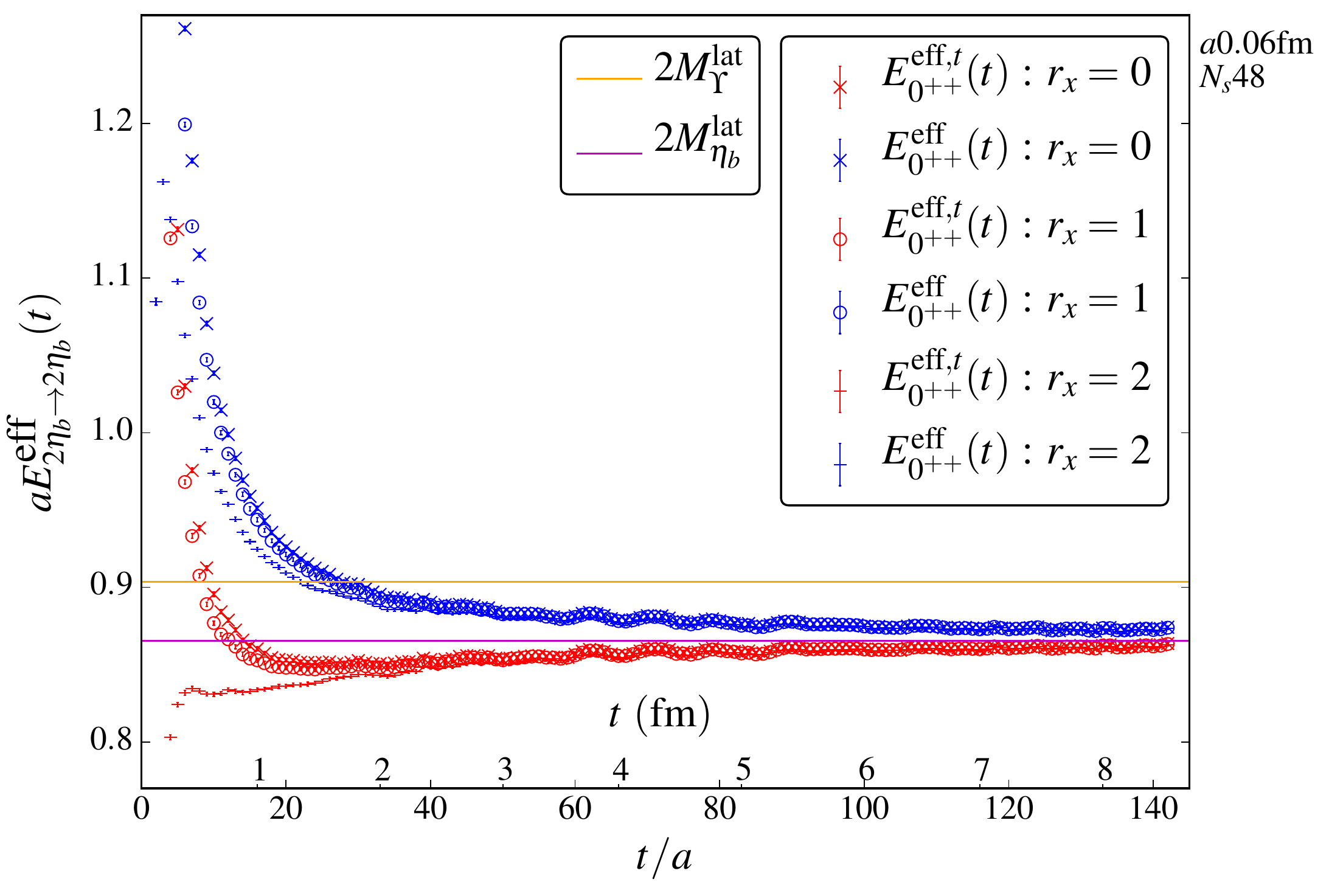}
  \caption{The individual Wick contraction effective masses of the $2\eta_b\to 2\eta_b$ correlators. The upper figure is the Direct$1$ and the lower is the Xchange$2$ contraction (each shown diagrammatically in Figs.~\ref{fig:D1} and \ref{fig:X2}). $E^{\textrm{eff}}$ and $E^{\textrm{eff},t}$ are the single- and two-particle effective masses defined in Eq.~(\ref{eqn:EeffSingle}) and Eq.~(\ref{eqn:EeffTwo}) respectively. (color online) }
  \label{fig:e0ppD1X2}
\end{figure}

\begin{figure}[t]
  \centering
  \includegraphics[width=0.49\textwidth]{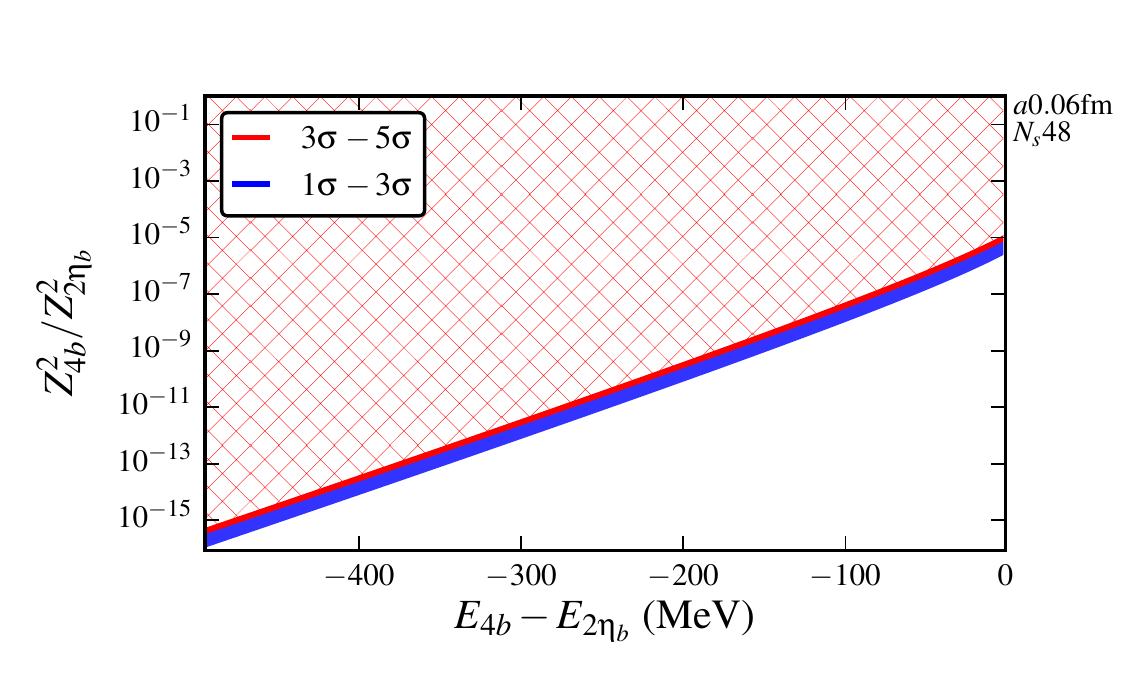}  
  \caption{The excluded region for the ratio of tetraquark/$2\eta_b$ overlaps, $Z_{4b}/Z_{2\eta_b}$,  onto the $\mathcal{O}_{(\eta_b,\eta_b)}$ operator, assuming a tetraquark with mass, $E_{4b}$, lying below the $2\eta_b$ threshold, $E_{2\eta_b}$. The red hashed region is excluded at $5\sigma$ by the data as described in the text. The $1\sigma-3\sigma$ and $3\sigma-5\sigma$ exclusion bands are also shown for reference. (color online) }
  \label{fig:Exclusion}
\end{figure}

For illustration purposes, we also show the effective masses of the individual Wick contractions contributing to the $2\eta_b\to 2\eta_b$  correlator in Figure \ref{fig:e0ppD1X2}. As is evident, in each individual Wick contraction the effective mass drops below the $2\eta_b$ threshold but then rises slowly to threshold. However, importantly, when all Wick contractions are added together to yield the full correlator (shown Figure \ref{fig:e0pp}) the effective mass falls rapidly to threshold from above. This behaviour will be discussed further in Sec.~\ref{sec:Conclusions}. 

After analysing all our data, as hinted by the effective mass plots, there is no indication of a bound tetraquark state below the non-interacting thresholds on any ensemble, as shown in Fig.~\ref{fig:ESummary}. We see no evidence of any change in the ground-state energies (with respect to the thresholds) as we vary lattice spacing or sea quark masses. 

Searching for a new tetraquark candidate has at least a two-dimensional parameter space: the hypothetical state would have an energy and also an overlap onto a specific operator. Using the lattice data presented here, we can determine a relationship between these parameters. Assuming that a tetraquark does exist below the lowest bottomonium-pair threshold in our data, at a certain time $t^*$ the correlator can be modeled with a two-state ansatz. Specifically for the $0^{++}$ channel, given that the tetraquark has an energy $E_{4b}$ and an overlap $Z_{4b}$ onto a particular operator, at a large enough time the only other appreciable contribution will come from the higher $2\eta_b$ threshold which has an overlap $Z_{2\eta_b}$ with the same operator. In this case the correlator is given by
\begin{align}
C(t^*) = Z_{4b}^2e^{-aE_{4b}t^*} + Z_{2\eta_b}^2e^{-aE_{2\eta_b}t^*}\left(\frac{aM_{\eta_b}}{4\pi t^*}\right)^{\frac{3}{2}}. 
\end{align}
Using this ansatz in the effective mass formula Eq.~(\ref{eqn:Eeff}) and rewriting the equation in terms of the non-perturbative overlaps yields the constraint
\begin{align}
&  \frac{Z_{4b}^2}{Z_{2\eta_b}^2}  = \left(\frac{1 - \left(\frac{t^*}{t^*+1}\right)^{\frac{3}{2}}\exp{({aE^{\mathrm{eff}}(t^*)-aE_{2\eta_b}})}}{\exp({aE^{\mathrm{eff}}(t^*)-aE_{4b}}) - 1} \right)  \nonumber \\
  & \hspace{4.0cm} \times e^{-a\Delta E t^*} \left(\frac{aM_{\eta_b}}{4\pi t^*}\right)^{\frac{3}{2}} \label{eqn:QCDConstraint}
\end{align}
with $ a\Delta E = aE_{2\eta_b} - aE_{4b} > 0$. As can be seen, if the tetraquark is not observed by a time $t^*$ then the overlap onto this new state must be (at least) exponentially suppressed with the binding of the tetraquark state, e.g, with $-\Delta E$. 

This point illustrates that if a tetraquark did exist with $E_{4b}<E_{2\eta_b}$ then it is possible that it was not observed in our data because $Z_{4b}\approx 0$ within statistical precision. In this scenario, we can use the constraint Eq.~(\ref{eqn:QCDConstraint}) to estimate an upper bound on the magnitude of the overlaps given that no clear evidence of the tetraquark is observed within our statistical precision. The needed inputs for the constraint include the value of $t^*$ where the two-state ansatz is valid, $aE^{\mathrm{eff}}(t^*)$ from correlator data constructed with a specific operator, as well as $aE_{2\eta_b}=2aM_{\eta_b}$. For the local $\mathcal{O}_{(\eta_b,\eta_b)}$ operator on the $a = 0.06$ fm ensemble, by examining Figure \ref{fig:e0pp},  a choice of ${t}^* = 143$ ensures that the two-state ansatz is valid (given the long plateau at the $2\eta_b$ threshold). Here, $aE^{\mathrm{eff}}(t^*=143)=0.87634(61)$ can also be precisely obtained. The value of $aM_{\eta_b}$, given in Table \ref{tab:Energies}, is found from the $\eta_b$-meson data. Then, using this data in the constraint, for a certain choice of $E_{4b}$, a numerical value of the ratio of overlaps is found such that it is consistent with zero within its small $1\sigma$ statistical error. Any value of the ratio of overlaps larger than this $1\sigma$ error is inconsistent, at this level of confidence, with our data observing a tetraquark at this value of $E_{4b}$.

We use this model to estimate how small the hypothetical tetraquark overlap would need to be so that the tetraquark was not observed within our statistical precision. A $1\sigma$, $3\sigma$ and $5\sigma$ exclusion plot of the parameter space is given in Figure \ref{fig:Exclusion}. As the input data into the constraint has a long propagation time past $t^*>8$ fm and a statistically precise value of $aE^{\mathrm{eff}}(t^*)$ which does not fall below the threshold, a significant amount of parameter space is excluded. It should be understood that this figure is only valid for a particular operator in a certain channel. The given $0^{++}$ channel in Figure \ref{fig:Exclusion} excludes the largest amount of parameter space as it is the most statistically precise. Also Figure \ref{fig:Exclusion} is constructed from data on the ``superfine'' ensemble alone, where discretisation effects are smallest and would not change the quantitative behaviour significantly. 

To conclude this section, we find no evidence of a stable tetraquark candidate below the non-interacting thresholds by studying a full $S$-wave colour-spin basis of QCD operators. In the next section we will perform an exploratory and complementary study of an alternative approach so to ensure the robustness of our conclusions.

\section{NRQCD With A Harmonic Oscillator Potential}
\label{sec:HO}

A stable tetraquark state in the $0^{++}$, $1^{+-}$ and $2^{++}$ channels would overlap with the full basis of $S$-wave colour-spin operators utilised above. In this section, we go one additional step by exploring an alternative approach in order to further investigate the possibility of a tetraquark state. Adding a central confining potential to the quark interactions can produce a more deeply bound tetraquark relative to the threshold as the strength of this interaction is increased (as we will see below). Furthermore, an appropriate choice of additional interaction can reduce the fiducial volume of the lattice and thus thin the allowed discrete momenta states of the two meson degrees of freedom. Adding an external attractive scalar central-potential to the QCD interactions yields these desired effects.

The harmonic oscillator potential is a particularly suitable choice of scalar interaction between quarks. For a particle of mass $m$ at position $\VEC{x}$ away from the centre $\VEC{x_0}$ the potential is just  $\kappa r^2/2 \equiv \kappa|\VEC{x} - \VEC{x_0}|^2/2$.  Defining $\omega = \sqrt{(\kappa/m)}$,  the ground state energy and wavefunction are $E_0 ={3}\omega/2$  and $\psi(r) = C\exp{(-m\omega r^2/2)}$.   Additionally, the separability of the combined QCD and harmonic oscillator potential into total and relative coordinates ensures that solutions of multiquark systems can be split into two parts with the total coordinate piece analytically solvable. This follows from the nature of the harmonic oscillator potential\footnote{Defining  $r = |\VEC{x}_1 -\VEC{x}_2|$ and ${R}_{cm} = ((\VEC{x}_1 + \VEC{x}_2)/2 - \VEC{x}_0)$, we can separate $\kappa[(\VEC{x}_1-\VEC{x}_0)^2 + (\VEC{x}_2 -\VEC{x}_0)^2]/2$ into relative and centre-of-mass  coordinates $[(\kappa/2) r^2 + (2 \kappa) R^2_{cm}]/2$.}. Thus we expect that for small values of $\omega$ the ground state for $2\eta_b$ mesons is approximately $2M_{\eta_b}(\omega) + 3\omega$. The $3\omega$ term comes from two colour-singlet $\eta_b$ mesons in the harmonic oscillator potential and the mass of the $\eta_b$ is shifted slightly from the value at $\omega=0$ because of the additional harmonic oscillator interaction combined with the QCD interactions that bind the two heavy quarks into a $\eta_b$\footnote{For sufficiently small $\omega$ the shift in $\eta_b$  is directly related to the rms radius of the $\eta_b$ state since from perturbation theory it is given by $\langle \eta_b(\omega = 0)| kr^2/2|\eta_b(\omega=0)\rangle$.}. However, for a compact tetraquark state the mass would be $M(\bar{b}\bar{b}bb)(\omega) + {3\omega}/{2}$, as there is only one colour-singlet state in the central harmonic oscillator potential ($3\omega/2$) and if the tetraquark state is also a compact state (on the scale of $\eta_b$ and much less than the effective lattice volume) then its mass will also receive only a modest positive correction due to $\omega$.

\begin{figure}[t]
  \centering
  \includegraphics[width=0.45\textwidth]{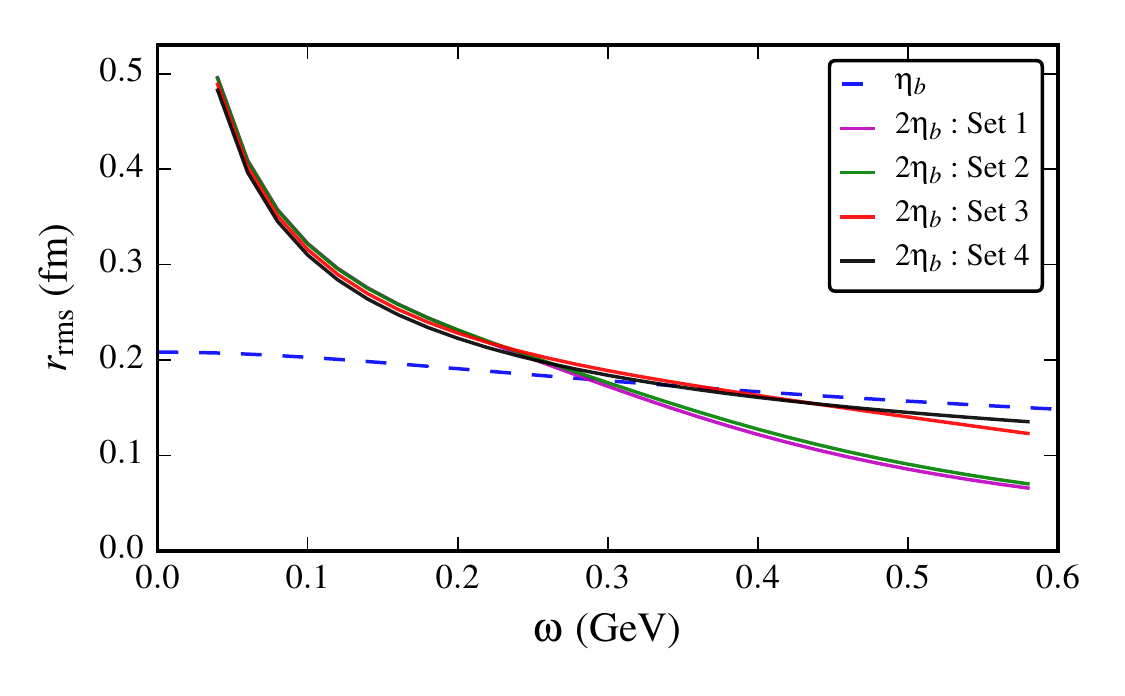}  
  \caption{The root mean square distance $r_{rms}$ of the $\eta_b$ and the $2\eta_b$ as a function of the harmonic oscillator strength calculated from a potential model with the different ensemble parameters listed in Table \ref{tab:GluonEnsembles} as discussed in the text. (color online)}
  \label{fig:rms}
\end{figure}

Hence if there were a tetraquark state near threshold then this additional interaction could drive it further below threshold, giving a much cleaner and distinct signal for the tetraquark candidate in our calculation. As the potential model framework describing the $\eta_b$ has been largely successful, we can use this framework as a general guide for the exploratory non-perturbative lattice calculation when including the harmonic oscillator potential. Of course, we are mainly interested in QCD (and not the harmonic oscillator) and, as such, if we do find a stable tetraquark state when including the harmonic oscillator potential then we must take the $\omega\to 0$ limit. Thus, the objective of this section is to determine if a stable tetraquark state exists when the quarks are exposed to an auxiliary potential (which could push the tetraquark increasingly lower than the threshold) and if it does, will it survive the QCD limit. 

The harmonic oscillator potential is defined as\footnote{In a periodic box of length $L$ the harmonic oscillator is also periodic. }
\begin{align}
\delta H_{HO} &= \frac{m_b\omega^2 }{2} | \VEC{x}-\VEC{x}_0|^2 \label{eqn:HO}
\end{align}
where the quarks are pulled towards the fixed point $\VEC{x}_0$  with a strength $\omega$. We choose $\VEC{x}_0$ to be the same position as the source. Intuitively, as the quarks/$\eta_b$'s start to propagate further away from the source, the harmonic oscillator potential pulls them closer together. In turn, this restricts the quarks/$\eta_b$'s to be in a certain volume. 

First, it is necessary to determine which volumes the quarks/$\eta_b$'s are confined into by the addition of the harmonic oscillator potential. The root mean square distance, $r_{rms}$, gives an indication of this. We determine the $r_{rms}$ of the $\eta_b$ based on solutions of the Schrodinger equation using a Cornell potential\footnote{The potential form is $-\frac{4\alpha_s}{3r} + \frac{r}{b^2}$ with $\alpha_s =0.36$, $b= 2.34\rm~GeV~$ and reduced mass $2.59$ GeV}. The Matthieu equation can describe the behaviour of two free $\eta_b$'s in a harmonic oscillator potential on a periodic box, and the solutions of which can yield $r_{rms}$ for the $2\eta_b$ state. The results from such a calculation are plotted in Figure \ref{fig:rms}.

Based on this, in order to confine the quarks sufficiently so that the two $\eta_b$'s overlap, and also to study the dependence on $\omega$, values of $\omega = (75,150,300,350)$ MeV and $\omega = (75,150, 350,500)$ MeV are chosen for the lattice ensembles called Set $1$ and Set $3$ in Table \ref{tab:GluonEnsembles}. In lattice units, the simulated values of $\kappa /2 = (am_b)(a\omega)^2/2$ are $(0.0029,0.0117, 0.0469, 0.638)$ and $(0.0011,0.0044, 0.0240, 0.0489)$ respectively. 

The harmonic oscillator is implemented through a minor modification of the NRQCD evolution equations via
\begin{align}
  e^{-a\tilde{H}} & = \left( 1 - \frac{a\delta H_{HO}}{2l} \right)^l e^{-aH} \left( 1 - \frac{a\delta H_{HO}}{2l} \right)^l  \label{eqn:NRQCDHOGreensFunction}
\end{align}
where $e^{-aH}$ is the purely NRQCD evolution equation defined in Eq.~(\ref{eqn:NRQCDGreensFunction}). This implementation was chosen so that the evolution equation is still time-reversal symmetric. Here, $l$ is a stability parameter akin to $n$ in Eq.~(\ref{eqn:NRQCDGreensFunction}) which is used to prevent possible numerical instabilities \cite{Lepage:ImprovedNRQCD}. Values of $l=13$ and $10$ were chosen for the calculations on Set $1$ and $3$ respectively. Following these details, we are now able to present results from the non-perturbative lattice calculations. 

\subsection{Numerical Results}
\label{sec:HOSetup}

\begin{figure}[t]
  \centering
  \includegraphics[width=0.51\textwidth]{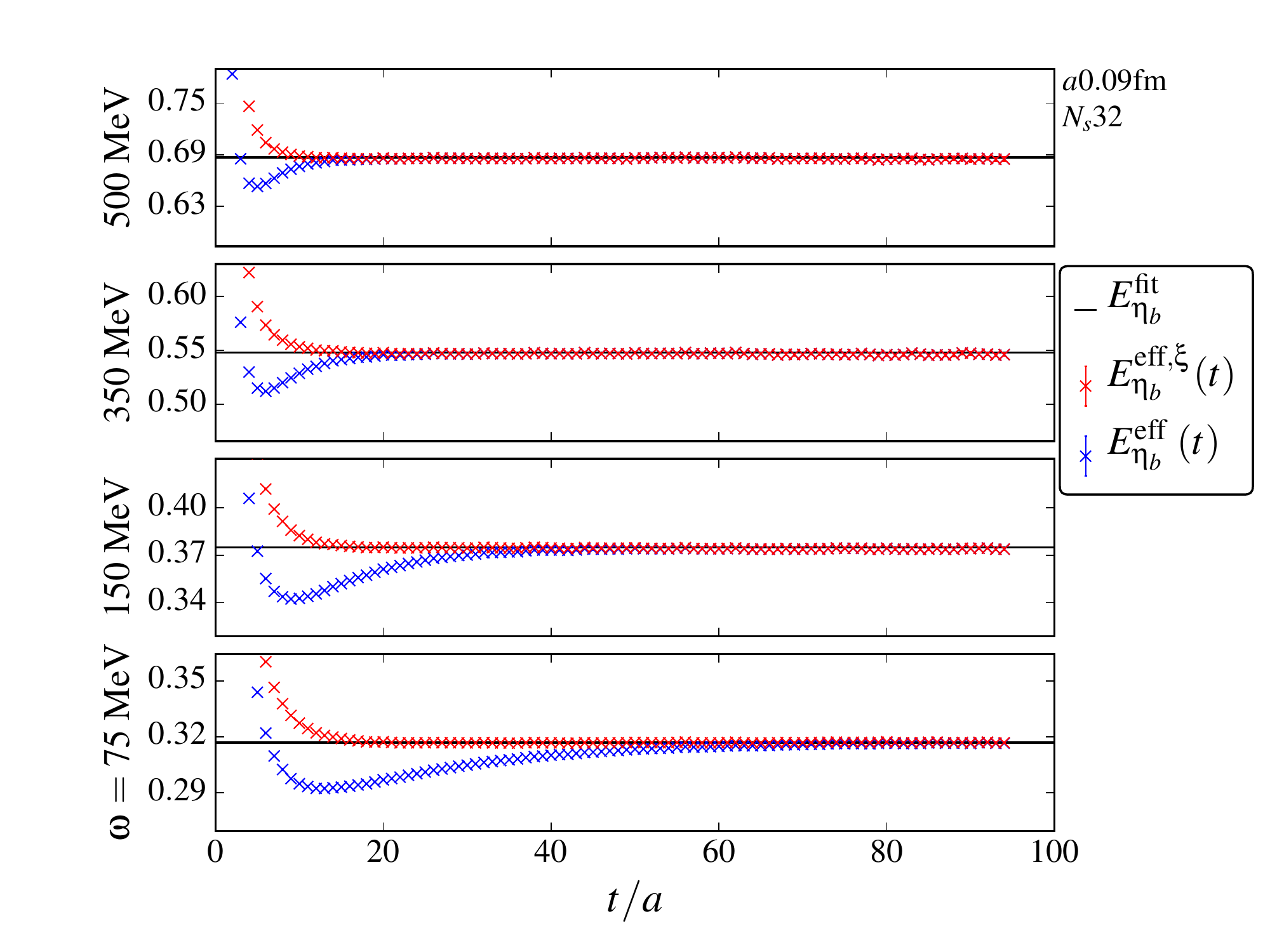}
  \caption{The effective mass plot for the $\eta_b$ when including the harmonic oscillator potential on Set $3$. $E^{\textrm{eff}}$ is given by Eq.~(\ref{eqn:Eeff}) while $E^{\textrm{eff},\xi}$ removes the leading $1+e^{-2\omega t}$ dependence from the correlator (\ref{eqn:2ptHOFitSingle}) to enable a better comparison with the data when no harmonic oscillator potential is included. (colour online) }
  \label{fig:EeffSingleHO}
\end{figure}

All correlator data from Set $1$ and $3$ discussed in Section \ref{sec:QCD4b} was generated again with the inclusion of the harmonic oscillator potential at the four different $\omega$ values given above. The harmonic oscillator alters both the single- and two-particle contributions to the correlator so that they become (as derived in Appendix \ref{app:FitHO}) dependent on $\omega$ as
\begin{align}
  C_{i,j}^{J^{PC}}(t,\omega) & = \sum_n \frac{Z_n^i Z_n^{j,*}}{(1 + e^{-2\omega t})^{\frac{3}{2}} } e^{-(M(\omega)_n + \frac{3}{2}\omega)t} \label{eqn:2ptHOFitSingle} \\
  C_{i,j}^{J^{PC}}(t,\omega) & = \sum_{X_2} Z_{X_2}^i Z_{X_2}^{j,*} \left( \frac{ 2\omega\mu_r\pi^{-1} }{1 - e^{-4\omega t }}\right)^\frac{3}{2} \nonumber \\
  & \hspace{1cm} \times e^{-(M^S_1(\omega)+M^S_2(\omega) + 3\omega)t} + \cdots .   \label{eqn:2ptHOFitTwo}
\end{align}
First, Figure \ref{fig:EeffSingleHO} shows the effective masses, as defined in Eq.~(\ref{eqn:Eeff}), of the $\eta_b$ when including the harmonic oscillator. Also overlaid are the effective masses when removing the $1+e^{-2\omega t}$ dependence to enable a better comparison with the data when no harmonic oscillator is included. As can be seen, the dip in the harmonic oscillator effective masses is from this additional time dependence. Physically, this can be understood to be due to the $b$-quarks travelling non-relativistically and so it takes time for the harmonic oscillator to have an effect.

\begin{figure}[t]
  \centering
  \includegraphics[width=0.49\textwidth]{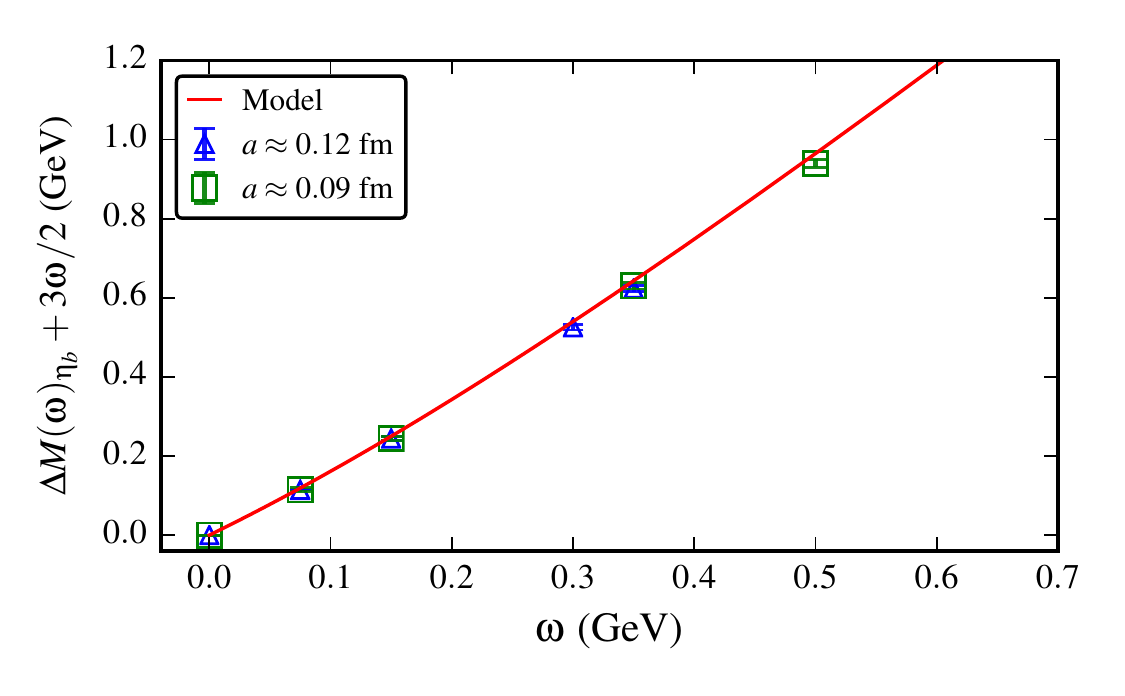}  
  \caption{The lattice $\eta_b$ energy $M(\omega)_{\eta_b} + 3\omega/2$ when including the harmonic oscillator potential with $M(\omega=0)_{\eta_b}$ subtracted compared to the model predictions as discussed in the text. (color online)}
  \label{fig:EHO}
\end{figure}

The $\eta_b$ correlator data when including the harmonic oscillator potential is fit to the functional form given by Eq.~(\ref{eqn:2ptHOFitSingle}) in order to extract the lowest energy eigenstate $M(\omega)_{\eta_b} + 3\omega/2$ from the asymptotic behaviour. We show the fitted result overlaid on the effective mass plot in Figure \ref{fig:EeffSingleHO}. As before, the long plateau indicates that the ground state will be extracted accurately. To compare to the potential model predictions, we subtract the $\eta_b$ mass with no harmonic oscillator included ($M(\omega=0)_{\eta_b}$) and then plot the energy differences against $\omega$, as shown in Figure \ref{fig:EHO}. Good qualitative agreement between the lattice results and the potential model predictions is observed. 

\begin{figure*}[th]
  \hspace*{\fill}
  \subfloat[$2\eta_b\to 2\eta_b$]{\label{fig:e0ppHO}
    \includegraphics[width=0.49\textwidth]{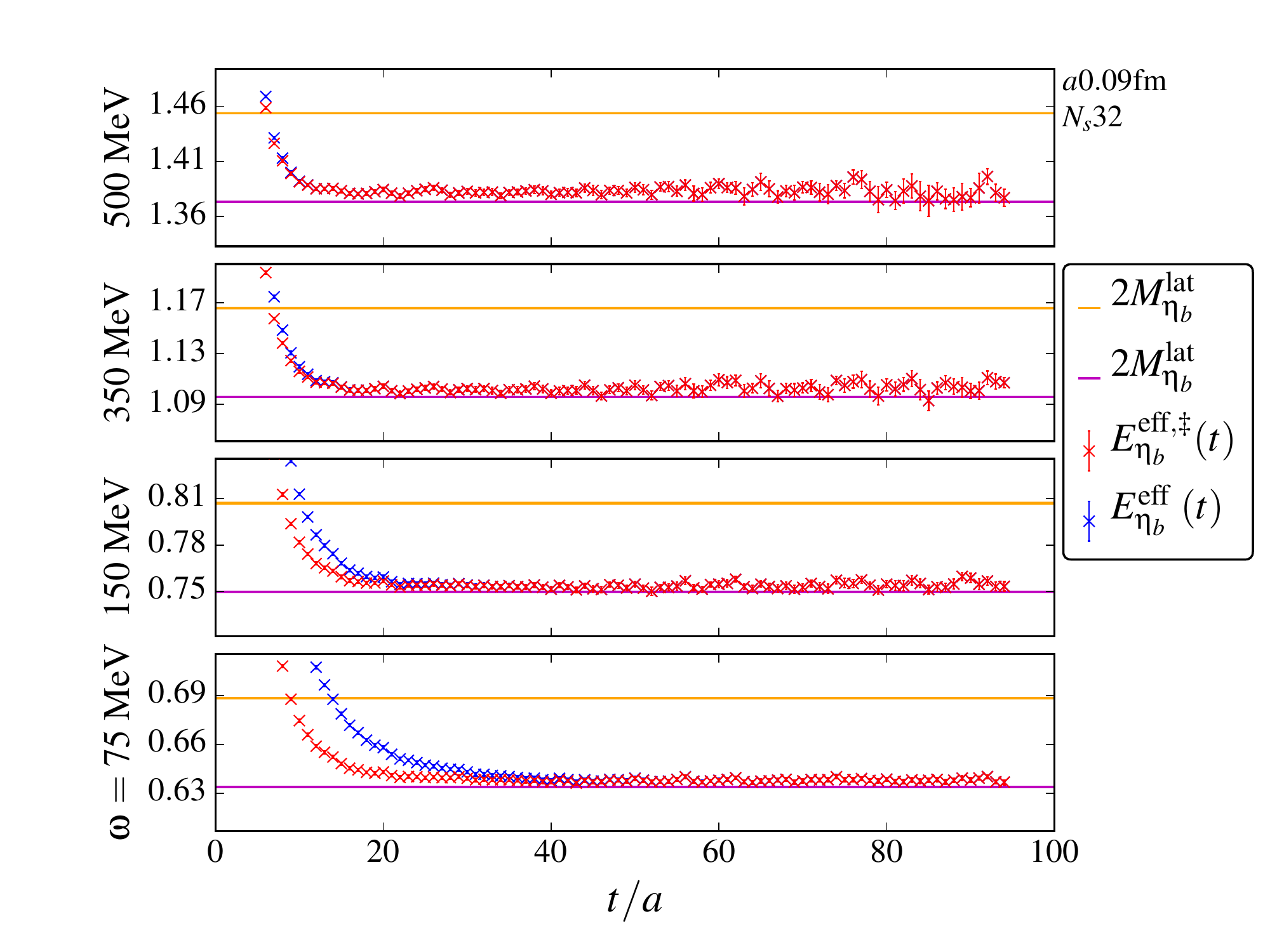}}
  \hfill
  \subfloat[$2\Upsilon\to 2\Upsilon$]{\label{fig:u0ppHO}
    \includegraphics[width=0.49\textwidth]{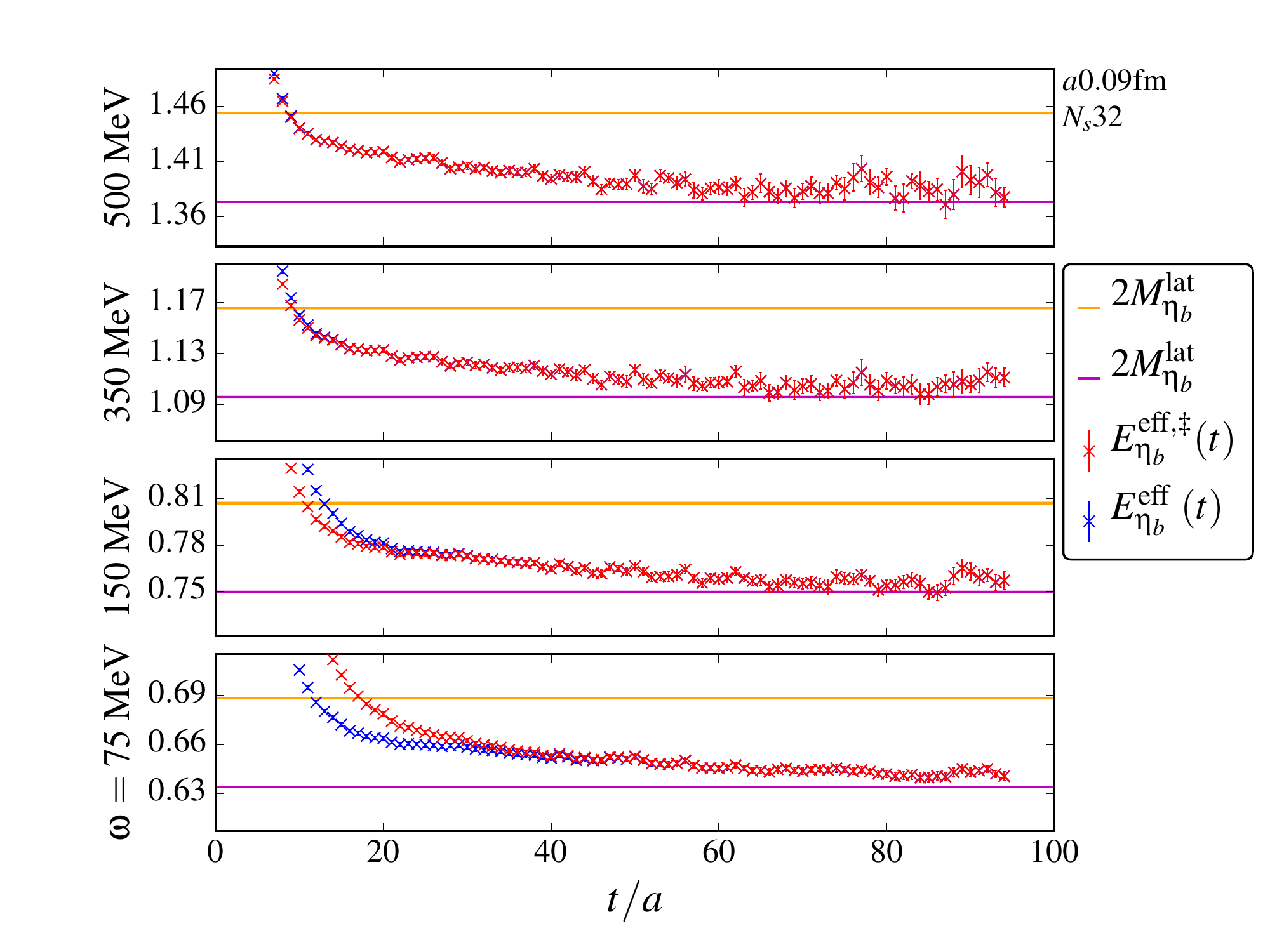}}
  \hspace*{\fill}
  \\
  \hspace*{\fill}
  \subfloat[$3_c\times \bar{3}_c \to 3_c\times \bar{3}_c$]{\label{fig:d0ppHO}
    \includegraphics[width=0.49\textwidth]{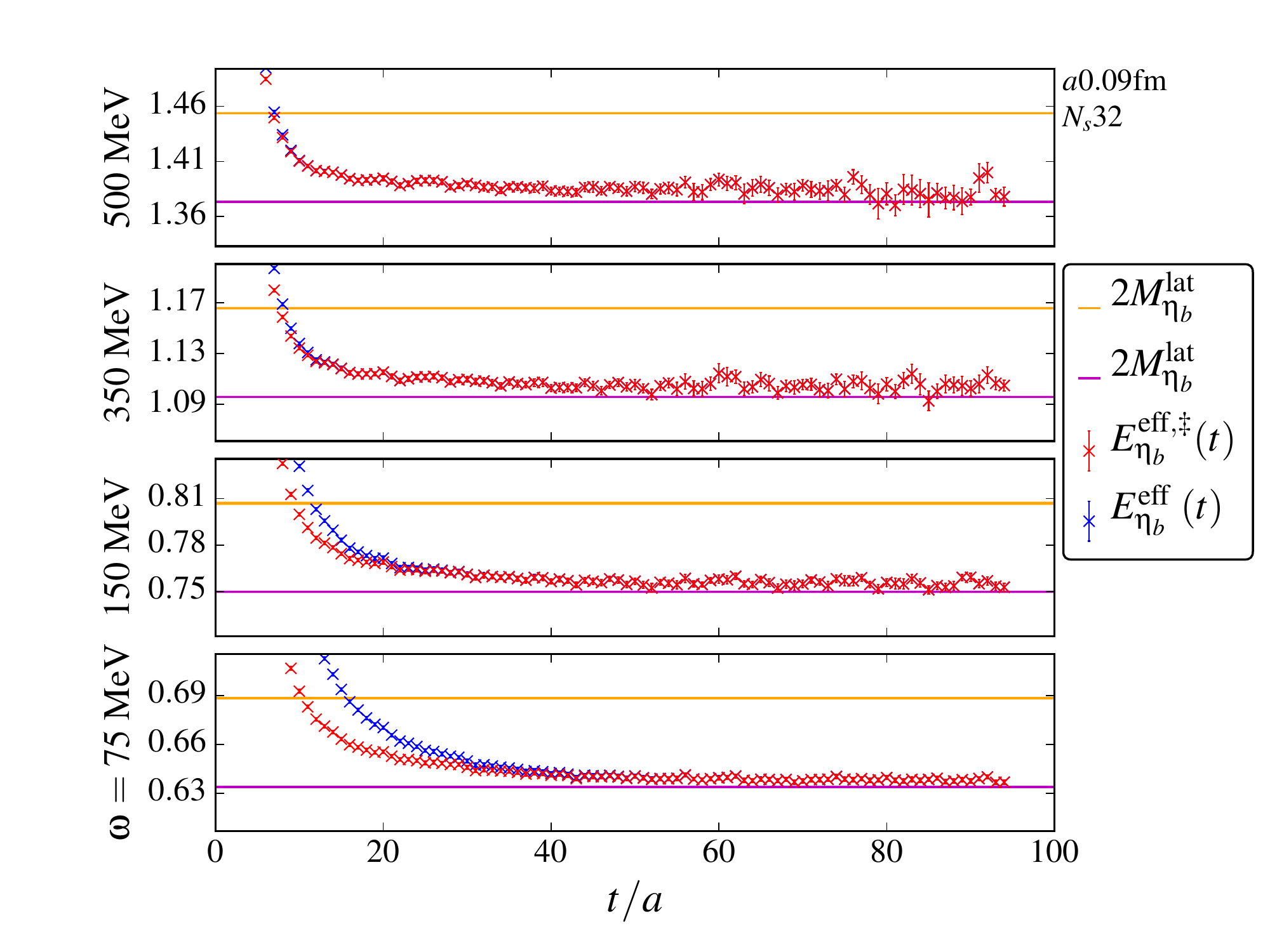}}
  \hfill
  \subfloat[$6_c\times \bar{6}_c \to 6_c\times \bar{6}_c$]{\label{fig:s0ppHO}
    \includegraphics[width=0.49\textwidth]{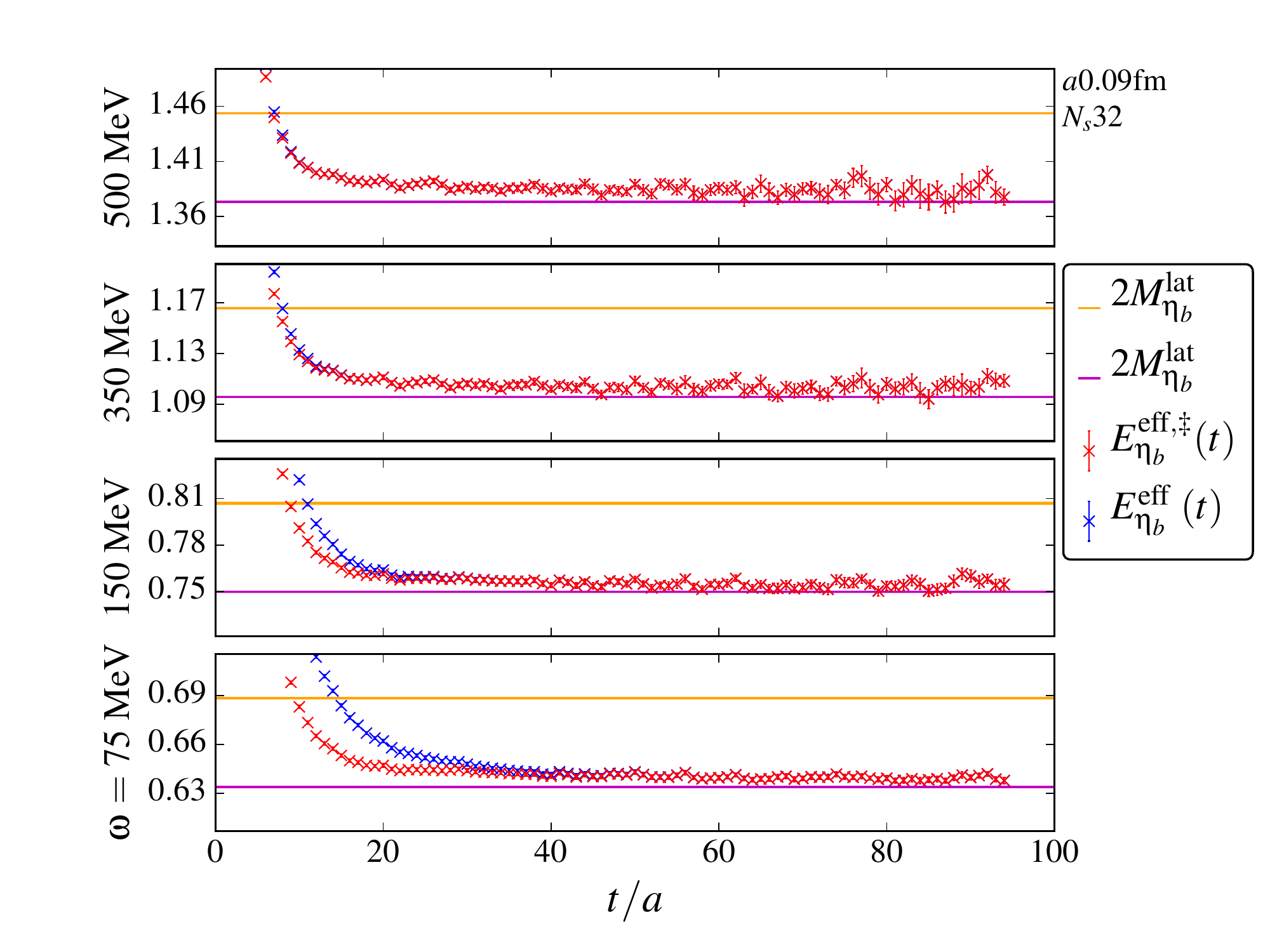}}  
  \hspace*{\fill}
  \caption{The $\bar{b}\bar{b}bb$ effective masses for the $0^{++}$ correlators when including the harmonic oscillator potential on Set $3$. $E^{\textrm{eff}}$ is given by Eq.~(\ref{eqn:Eeff}), while $E^{\textrm{eff},\ddagger}$ removes the leading $1-e^{-4\omega t}$ dependence from the correlator (\ref{eqn:2ptHOFitTwo}) to enable a better comparison with the data when no harmonic oscillator potential is included. (color online)  }
  \label{fig:0ppHO}
\end{figure*}

For the $\bar{b}\bar{b}bb$ system, we show the $0^{++}$ effective masses on Set $3$ in Figure \ref{fig:0ppHO}. It is evident that the $0^{++}$ and the $\eta_b$ data contains more noise when a harmonic oscillator potential is included. While fitting the data to the form in Eq.~(\ref{eqn:2ptHOFitTwo}) can be performed, it is not necessary as the purpose of this exploratory work is to determine if a stable tetraquark exists when $\omega \ne 0$. As can be seen, there is no fall below the $2\eta_b$ threshold for any value of $\omega$. Similar behaviour is seen with the data on Set $1$. 

We show the effective masses for the individual Direct$1$ and Xchange$2$ Wick contractions of the $2\eta_b\to 2\eta_b$ correlator in Figure \ref{fig:e0ppHOD1X2}. As before, the effective masses of the individual Wick contractions drop below the $2\eta_b$ threshold, even though importantly, when added together to yield the full correlator shown in Figure \ref{fig:e0ppHO} the effective mass is always above threshold. As this was also seen in the pure NRQCD data shown in Sec.~\ref{sec:QCD4b}, it may be a problematic feature of models that utilise a phenomenologically motivated four-body potential for this system. 

To conclude this section, despite adding an auxiliary potential into the QCD interactions that should push a near threshold tetraquark candidate increasingly lower we find no indication of any state below the $2\eta_b$ threshold. The conclusions of this section then agree with those of  Sec.~\ref{sec:QCD4b}.

\section{Discussion and Conclusions}
\label{sec:Conclusions}

In this work we have studied the low-lying spectrum of the $\bar{b}\bar{b}bb$ system using the first-principles lattice non-relativistic QCD methodology in order to search for a stable tetraquark state below the lowest non-interacting  bottomonium-pair threshold in three different channels: the $0^{++}$ which couples to the $2\eta_b$ and $2\Upsilon$, the $1^{+-}$ which couples to $\Upsilon\eta_b$ and the $2^{++}$ which couples to $2\Upsilon$. In Section \ref{sec:CompSetup} we describe our numerical methodology. Four gluon ensembles were employed with lattice spacings ranging from $a = 0.06 - 0.12$ fm, and one ensemble which has physical light-quark masses. All ensembles have $u$, $d$, $s$ and $c$ quarks in the sea.

In Sec.~\ref{sec:QCD4b} we presented the majority of the results in this work. Here, we determined the lowest energy eigenstate of the $\bar{b}\bar{b}bb$ system with the quantum numbers $0^{++}$, $1^{+-}$ and $2^{++}$ using an over-constrained $S$-wave colour/spin basis (arising from Fierz relations between the diquark-antidiquark and two-meson systems as shown in Table \ref{tab:Fierz}). We did not observe any state below the lowest non-interacting bottomonium-pair threshold in any channel, as can be seen in Figures \ref{fig:0ppTwoMeson}, \ref{fig:0ppDiquark}, \ref{fig:1pmTwoDiquark} and \ref{fig:2pTTwoDiquark}, and a summary of our results from this section is given in Figure \ref{fig:ESummary}.

In Sec.~\ref{sec:HO}, to ensure the robustness of our conclusions, we performed an exploratory calculation of a novel method which added an auxiliary scalar potential into the QCD interactions with the objective of pushing a near threshold tetraquark increasingly lower than the threshold. This would give a more distinct and cleaner signal for its presence in our calculation. The harmonic oscillator was found to be a suitable central scalar potential. For the $\eta_b$-meson with this potential, we first verified agreement between the non-perturbative lattice calculations and a potential model  (as shown in Figure \ref{fig:EHO}) and then used this potential model as a general guide to choose multiple appropriate values of the potential strength. Despite studying the $\bar{b}\bar{b}bb$ system with this additional scalar potential on the lattice, no indication of the QCD tetraquark was observed. 

\begin{figure}[t]
  \centering
  \includegraphics[width=0.50\textwidth]{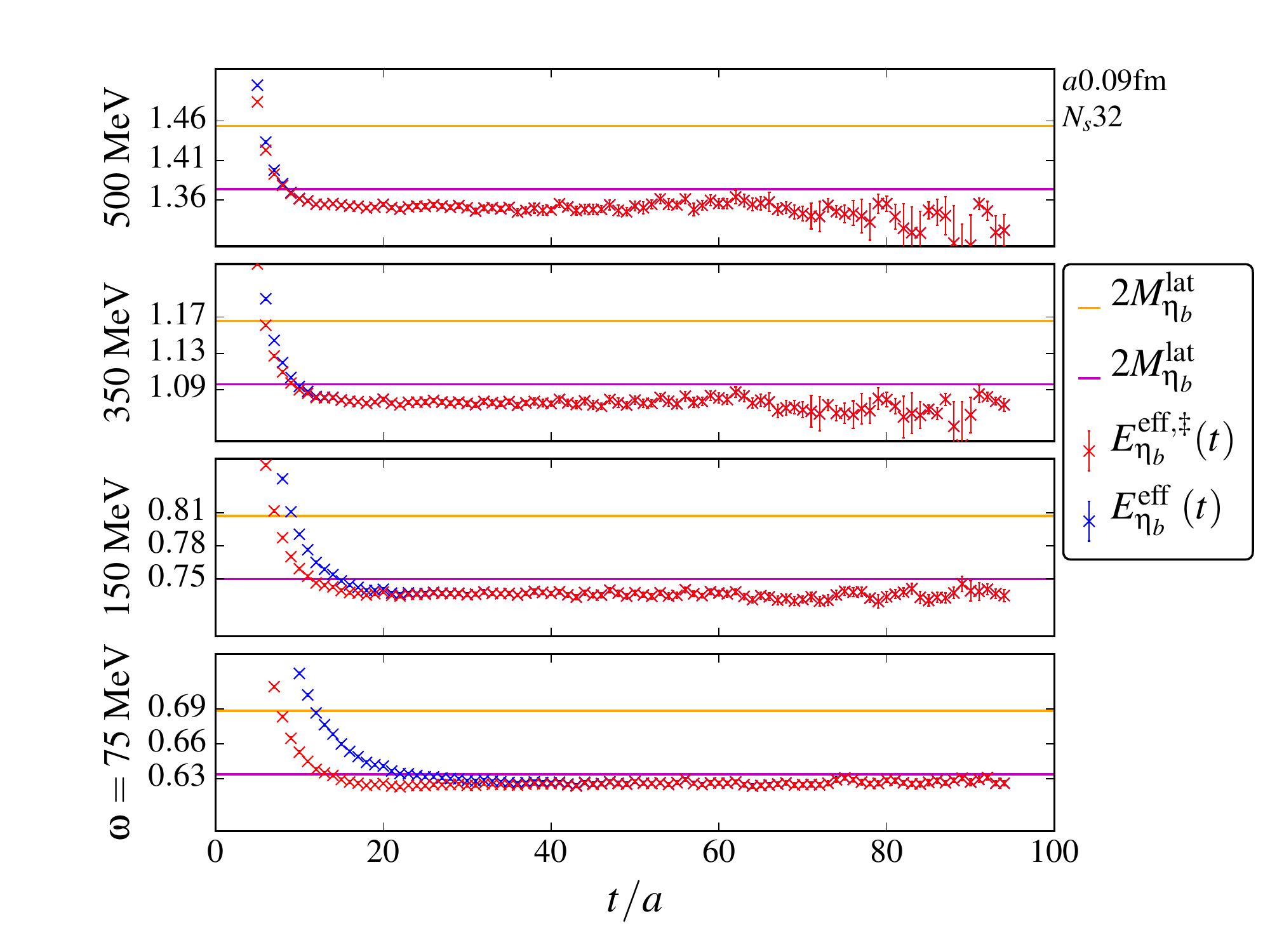}  
  \includegraphics[width=0.50\textwidth]{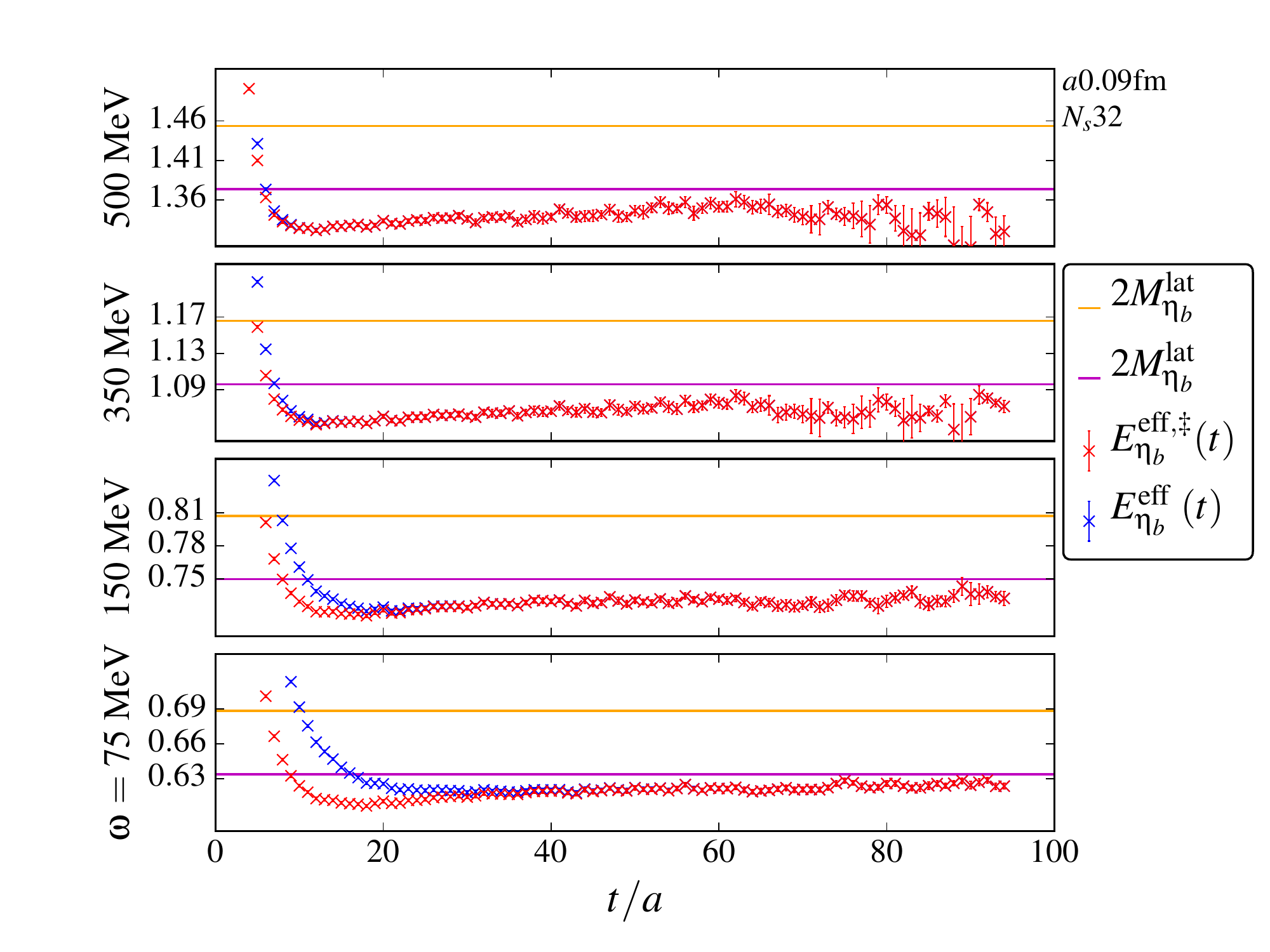}
  \caption{The effective masses for the individual $2\eta_b\to 2\eta_b$ Wick contraction correlator data when including a harmonic oscillator potential on Set $3$. The upper figure is the Direct$1$ and the lower is the Xchange$2$ contraction (each shown diagrammatically in Figs.~\ref{fig:D1} and \ref{fig:X2}). (color online) }
  \label{fig:e0ppHOD1X2}
\end{figure}

\begin{figure}[t]
  \centering
  \includegraphics[width=0.38\textwidth]{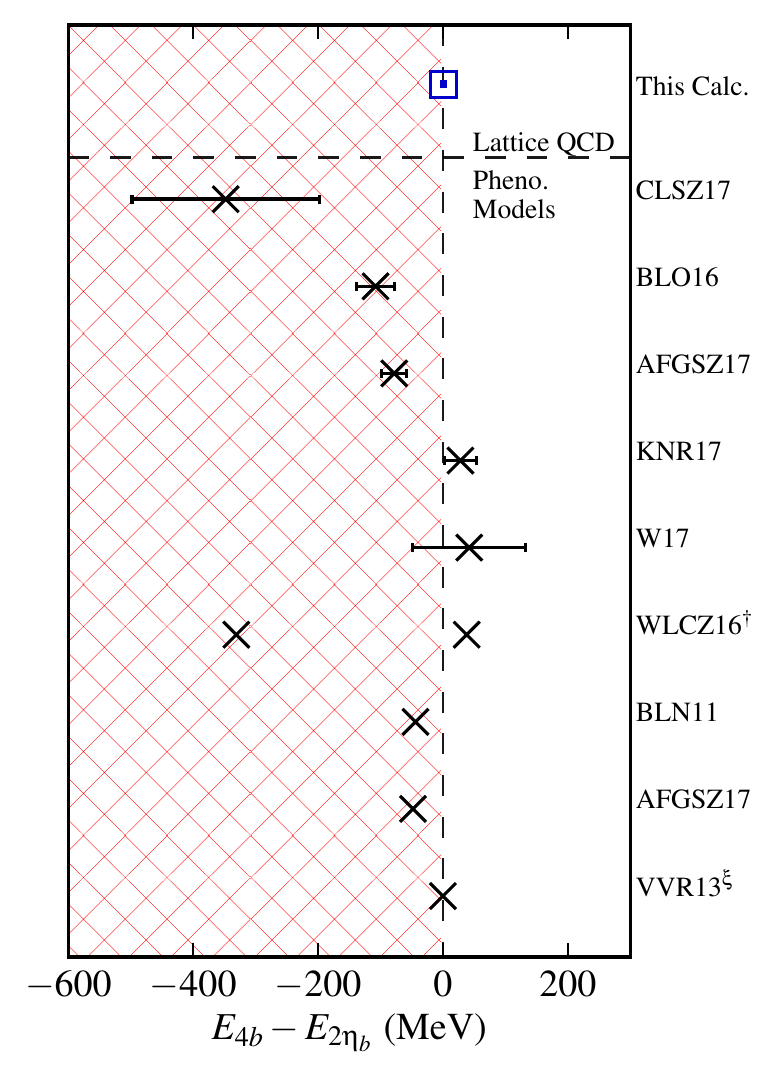}
  \caption{A comparison of our result for the $\bar{b}\bar{b}bb$  ground state energy in the $0^{++}$ channel (stat. error only) and predictions from phenomenological models. The hatched region indicates the exclusion of a bound tetraquark with an energy $E_{4b}$ subject to the value of its non-perturbative overlap as given in Figure \ref{fig:Exclusion}. In this comparison, we take our ground state energy obtained on the ``superfine'' ensemble (Set $4$ in Table \ref{tab:GluonEnsembles}) as a representative because it has the smallest discretisaion effects and the statistical error encompasses the results on the other ensembles as shown in Figure \ref{fig:ESummary}. The $y$-axis labels results from different phenomenological models \cite{4b:Schrod,4b:Bai,4b:ColMag,4b:SumRule16,4b:SumRule17,4b:Rosner,QQQQ:Voloshin} by last initial of authors and year of publication. An error is plotted if given in the reference. The two results from WLCZ$16^{\dagger}$ differ by how the mass scale was set. The result VVR$13^{\xi}$ finds no bound tetraquark and we indicate this by placing their result on threshold. (color online)}
  \label{fig:ECompare}
\end{figure}

This work is the only first-principles study of the low-lying $\bar{b}\bar{b}bb$ spectrum in the literature. However, there are others which utilise different methodologies. For example, \cite{4b:Schrod} predicts the tetraquark mass by solving the two-particle Schrodinger equation with a phenomenologically motivated non-confining potential between the point-like diquark and anti-diquark, and finds a $0^{++}, 1^{+-}$ and $2^{++}$ tetraquark to be bound by $44$, $51$ and $5$ MeV respectively\footnote{A model in which the diquarks are taken to be fundamental particles cannot determine the two-meson threshold from the Schrodinger equation because the diquarks cannot recombine into mesons. Thus, the experimental meson masses \cite{PDG:2016} are used to determine the lowest threshold.}. The authors of \cite{Anwar:2017} used a diquark model including a confining linear potential, but neglected spin effects, and found a $0^{++}$ tetraquark to be bound by $48$ MeV. However, it has also been found that the root-mean-square distance between the diquark and anti-diquark inside the tetraquark in this model is similar in magnitude to the distance between the quarks inside each diquark \cite{4b:Bai}. Consequently, such an approach is internally inconsistent. More recently, \cite{4b:ColMag} used a Hamiltonian including only spin-spin interactions mediated by a one-gluon exchange. Here, all other effects such as the chromoelectric interactions, colour confinement and the $b$-quark mass need to be set separately. The authors set these additional contributions in two ways: by estimating the effects using an effective heavy quark \cite{4b:ColMag} or by using the experimental meson masses as input. In this way, the authors find that the tetraquark state could be either below the $2\eta_b$ threshold or lie inbetween the $2\eta_b$ and $2\Upsilon$ thresholds (where in both cases the thresholds were determined using the experimental meson masses). In \cite{ColMag:1992} the author also uses a model including only chromomagnetic interactions and finds an unbound tetraquark. Within the QCD sum-rules framework, \cite{4b:SumRule16} finds a tetraquark candidate approximately $300$ MeV below the experimental $2\eta_b$ threshold while \cite{4b:SumRule17} finds a tetraquark lying inbetween the experimental $2\eta_b$ and $2\Upsilon$  thresholds. Using phenomenological arguments, \cite{4b:Rosner} also finds a tetraquark candidate lying inbetween the thresholds. Indeed, in the limit of very heavy quarks where the force proceeds through one-gluon exchange containing only colour-Coulomb contributions (safely neglecting spin and long distance effects), the authors of \cite{QQQQ:Voloshin} used a variational methodology to determine that a bound tetraquark exists for a $QQ\bar{Q^\prime}\bar{Q^\prime}$ system when $m_Q / m_{Q^\prime} \lesssim 0.15$ (where both $m_Q$ and $m_{Q^\prime}$ are heavy relative to $\Lambda_{\textrm{QCD}}$). However, if $m_Q / m_{Q^\prime}$ is varied and the tetraquark becomes unbound, then as the free two-meson eigenstate becomes the ground state of the system this numerical methodology has an increasingly slow convergence to a solution \cite{QQQQ:Voloshin}  (being numerically ill-posed) due to a redundant degree of freedom in the minimisation procedure. Indeed \cite{QQQQ:Voloshin} indicates that for $m_Q/m_{Q^\prime} = 1$ the solution is unstable, a hint that no bound tetraquark exists for all identical quarks in the very heavy mass limit. The authors of \cite{Anwar:2017} assume that the $b$-quark is sufficiently heavy so to use the one-gluon exchange with only colour-Coulomb contributions, and by also neglecting the mixing between different colour-components of the $2 \times 2$ potential matrix, finds a tetraquark bound by $78(20)$ MeV (by using the experimental $\eta_b$ mass to determine the threshold). In an orthogonal direction to the above work, the authors of \cite{Alfredo:String} only include a linear string contribution in the one-gluon exchange (neglecting spin effects and the appreciable short-distance Coulomb contributions), and without mixing between the different colour components of the potential matrix, find a bound tetraquark when $m_Q/m_{Q^{\prime}}=1$. However, in subsequent work \cite{Alfredo:StringMix,Alfredo:Summary}, by modelling the aforementioned mixing they concluded that no bound tetraquark exists. Perhaps the most sophisticated non first-principles methodology used to study the four-body $\bar{b}\bar{b}bb$ tetraquark is the diffusion Monte Carlo method utilised by \cite{4b:Bai}. Here, one determines the ground state of a phenomenologically motivated Hamiltonian by solving the Schrodinger equation and examining the stability of $\sum_ne^{-(E_n-E_0)t}\Psi_n(\VEC{x})$ to determine $E_0$, where $E_n$ ($\Psi_n$) is the $n$-th energy-eigenstate (eigenfunction). The authors include both the colour-Coulomb and linear contributions in the gluon exchange but neglect the mixing between the different colour components in the potential matrix, and find a stable tetraquark candidate $108$ MeV below the $2\eta_b$ threshold (determined from the experimental meson mass). Consequently, there is no study in the literature which is not from first-principles that includes all the appreciable effects relevant for the $bb\bar{b}\bar{b}$ system: treating the bottom-quarks as fundamental particles, including both short and long distance effects in the gluon exchange and including the mixing between the different colour components in the $2\times 2$ potential matrix. 

It should be emphasised however that these studies, unlike ours, are not from first-principles and thus have an unquantifiable systematic error associated with the choice of four-body potential. To emphasise this further, thinking of each Wick contraction (shown diagrammatically in Figure \ref{fig:WickTwoMeson}) as a different potential contributing to the QCD dynamics, then only studying a subset of these interactions can change the energies of states. This can lead to the misindentification of a new state below threshold. For example, the effective masses of the individual Wick contractions contributing to the $2\eta_b\to 2\eta_b$  correlator are shown in Figure \ref{fig:e0ppD1X2}. As is evident there, in each individual Wick contraction the effective mass drops below the $2\eta_b$ threshold but rises slowly to threshold even though when all Wick contractions are added together to yield the full correlator (shown Figure \ref{fig:e0pp}) the effective mass falls rapidly to threshold from above. This behaviour is even more pronounced in the data with the additional scalar potential, shown in Figure \ref{fig:e0ppHOD1X2}, possibly indicating that this may be a problematic feature of models that utilise a phenomenologically motivated four-body potential: a subset of the interactions show behaviour that may be misinterpreted as a bound state below threshold, while when all interactions are included no bound state is seen. Particularly, the slow rise to threshold from below could make the diffusion Monte Carlo method practically difficult due to the slowly varying stability condition combined with the fact that a long evolution time (greater than $8$ fm) is necessary.

In conclusion: we find no evidence of a $\bar{b}\bar{b}bb$ tetraquark with a mass below the lowest non-interacting bottomonium-pair thresholds in the $0^{++}$, $1^{+-}$ or $2^{++}$ channels. We give a constraint in Eq.~(\ref{eqn:QCDConstraint}) that future phenomenological models must satisfy if such QCD states are postulated. For the $0^{++}$ channel, we use this constraint to estimate how small the non-perturbative overlap of the hypothetical tetraquark (onto a particular operator) would need to be, relative to the $2\eta_b$, so that it was not observed within our statistical precision. A $1\sigma$, $3\sigma$ and $5\sigma$ exclusion plot of the parameter space is shown in Figure \ref{fig:Exclusion}, and discussed in Sec.~\ref{sec:QCD4b}. As we have propagation times longer than $8$ fm and statistically precise data, we can exclude all but the most finely-tuned parameter space. Our lattice results then rule out the phenomenological models discussed above that predict a tetraquark below the lowest bottomonium-pair thresholds which have a value of non-perturbative overlap that is excluded by Fig.~\ref{fig:Exclusion}. A comparison of these results with ours is shown in Figure \ref{fig:ECompare}. 

Further studies of possible heavy tetraquark channels that include orbital angular momentum either between the mesons in the tetraquark or between the quarks in the meson could be performed with the methodology used here\footnote{It should be noted that although we focused on $S$-wave combinations of quarks, the channels we study also exclude certain combinations of orbital angular momentum from producing a bound tetraquark. For example, the $0^{++}$ overlaps with $2\Upsilon$ in an orbital angular momentum $D$-wave configuration. If this state produced a low-lying bound tetraquark it would also show up in our calculation.}. Similarly one could also study whether stable $\bar{c}\bar{c}cc$, $\bar{b}\bar{c}bc$ or $\bar{b}\bar{b}cc$ tetraquarks exist or not. Additionally, two-hadron systems receive a finite-volume energy shift which depends on the infinite-volume scattering amplitude which is non-trivial to parameterise. Here we do not calculate these finite-volume energy shifts. Doing so in a more extended study would allow statements to be made about the existence of resonant tetraquark states above the lowest thresholds, that likely do exist in nature. Quantifying these shifts would be an exciting avenue for future work.

Finally, recent work based on heavy-quark symmetry \cite{QQqq:EQ} and phenomenological arguments \cite{Karliner:2017qjm} indicates that a $J^P=1^+$ $\bar{b}\bar{b}ud$ tetraquark will be stable in QCD. In fact, by extracting a potential from the lattice in the static heavy-quark limit and solving the Schrodinger equation \cite{Bicudo:bbud2013,Bicudo:bbud2015,Bicudo:bbud2017} also finds binding in this channel. Initial lattice calculations hint that such a state exists \cite{Francis:bbud} but calculations are difficult because of a signal-to-noise problem for heavy-light states \cite{Davies:PrecisionDs}. Lattice QCD calculations in this direction are essential for a conclusive first-principles statement to be made and to give further motivation for a targeted experimental search for these tetraquark configurations of nature.

{\it{The unaveraged correlator data, that have been analysed to produce the results in this work, are publicly available in a SQLite database from any of the authors upon request}}\footnote{The unaveraged correlator data is too large to be hosted for free on a remote server.}. 

\section*{ACKNOWLEDGMENTS}

We would like to thank  William Bardeen and Zhen Liu for the many insightful discussions on tetraquarks, as well as Gavin Cheung who in addition gave guidance on the tetraquark implementation. We are also grateful to the MILC collaboration for the use of their gauge configurations. This manuscript has been authored by Fermi Research Alliance, LLC under Contract No.~DE-AC02-07CH11359 with the U.~S.~Department of Energy, Office of Science, Office of High Energy Physics. The United States Government retains and the publisher, by accepting the article for publication, acknowledges that the United States Government retains a non-exclusive, paid-up, irrevocable, world-wide license to publish or reproduce the published form of this manuscript, or allow others to do so, for United States Government purposes. The results described here were obtained using the Darwin Supercomputer of the University of Cambridge High Performance Computing Service as part of the DiRAC facility jointly funded by STFC, the Large Facilities Capital Fund of BIS and the Universities of Cambridge and Glasgow. 

\appendix
\section{Two-Point Correlator Fit Functions}
\label{app:Fit}

Here we derive the non-relativistic two-particle contribution to the correlator on our ensembles. To begin, the correlator is given in Eq.~(\ref{eqn:2pt}). For clarity, the $i,j$ subscripts are dropped. The completeness relation for a two-hadron system is \cite{QFT:Srednicki}
\begin{align}
I & =  \sum_{X^2}\int \frac{d^3P_{tot}}{(2\pi)^3} \frac{d^3k}{(2\pi)^3}\frac{1}{2E(X^{2})}|X^2_{(\VEC{P}_{tot}, \VEC{k})}\rangle\langle X^2_{(\VEC{P}_{tot}, \VEC{k})}| \label{eqn:2PComplete}
\end{align}
where $|X^2_{(\VEC{P}_{tot}, \VEC{k})}\rangle = |M_1(\VEC{k})M_2(\VEC{P}_{tot}-\VEC{k})\rangle$ is a two-hadron state (with quantum numbers suppressed) and to avoid superfluous notation, we will also set $\VEC{P}_{tot}=0$. A key difference from the one-hadron system is the internal relative momentum, $\VEC{k}$, which contributes an additional three-integral. Substituting the completeness relation Eq.~(\ref{eqn:2PComplete}) into the correlator Eq.~(\ref{eqn:2pt}) and performing the momentum conserving integrals yields
\begin{align}
C(t) & = \sum_{X^2}  \int \frac{d^3k}{(2\pi)^3} {Z_{X^2}{(\VEC{k})}}^2  e^{-E(X^{2})t } \label{eqn:2ptFitTwo1}
\end{align}
where $Z_{X^2}(\VEC{k})$ is a non-perturbative coefficient. 

However, on a discrete finite-volume the above integral over elastic states is replaced by a finite sum with $k_{m} \in (-N_s/2 + 1, \ldots, , N_s/2 )$ in units of $(2\pi/aN_s)$. In turn, for $\VEC{P}_{tot}=0$, Eq.~(\ref{eqn:2ptFitTwo1}) becomes a sum over back-to-back hadronic states which have values of the discrete momenta that are equal in magnitude but opposite in direction. One can expand the two-particle energy using a non-relativistic dispersion relation, appropriate since we are using NRQCD, as
\begin{align}
  E(X^2) & = \sqrt{M_1^2 + |\VEC{k}|^2} + \sqrt{M_2^2 + |\VEC{k}|^2} \\
  & \approx M^S_1 + M^S_2 + \frac{|\VEC{k}|^2}{2\mu_r}  \label{eqn:ENR}
\end{align}
where we have defined the static, kinetic and reduced masses by $M^S$, $M^K$ and $\mu_r = M_1^KM_2^K / (M_1^K + M_2^K)$ respectively. In a finite-volume there will be an additional contribution to Eq.~(\ref{eqn:ENR}) dependent on the infinite-volume scattering phase shift, which will be discussed further below. Eq.~(\ref{eqn:ENR}) also illustrates the density of back-to-back states on our ensembles. As an example, examining the $a=0.09$ fm ensemble, and taking $M_{\eta_b} = 9.399(2)$ GeV from the PDG \cite{PDG:2016}, the smallest allowed ${|\VEC{k}|^2}/{2\mu_r} \approx 20$ MeV or $0.0092$ in lattice units with all other back-to-back states separated by multiples of this value. Consequently, due to the bottomonium mass being large compared to the smallest allowed momentum, adjacent back-to-back states are sufficiently close in energy that fitting the momentum states as a discrete sum would require a vast set of correlators projected onto each separate ${|\VEC{k}|^2}/{2\mu_r}$ (with the methodology used in \cite{Dudek:SDPhase}). Practically, this would be overly computationally expensive and instead, the fact that the states with $\VEC{k}\ne \VEC{0}$ are related by the dispersion relation (and are not independent as the sum would assume) should be included.

This can be achieved by first expanding the non-perturbative coefficient $Z_{X^2}(\VEC{k})$ as a polynomial in $|\VEC{k}|^2 / \mu_r^2$, as dictated by rotational symmetry and by ensuring the Taylor coefficients have the same dimension, then keeping all terms needed to a certain precision. After this the correlator can be written as 
\begin{align}
 & C(t)  = \sum_{X^2}   e^{-(M^S_1+M^S_2)t} \sum_{k}\Big\{ \sum_{i=0}^{\infty} Z_{X^2}^{2l} \frac{|\VEC{k}|^{2l}}{\mu_r^{2l}} \Big\} e^{-\frac{|\VEC{k}|^2}{2\mu_r}t}   \\
  & = \sum_{X^2}   e^{-(M^S_1+M^S_2)t} \int^{\frac{\pi}{a}}_{-\frac{\pi}{a}} \frac{d^3k}{(2\pi)^3} \Big\{ \sum_{i=0}^{\infty} Z_{X^2}^{2l} \frac{|\VEC{k}|^{2l}}{\mu_r^{2l}} \Big\} e^{-\frac{|\VEC{k}|^2}{2\mu_r}t} \label{eqn:2ptFitTwo2}
\end{align}
where going from the first to the second line we have replaced the finite sum by an integral. Taking the limits of the integral to $\pm \infty$ and performing the integrals over $\VEC{k}$ analytically yields the fit function given in Eq.~(\ref{eqn:2ptFitTwo}). Once it is shown that it is possible to replace the finite sum by the indefinite integral within our statistical precision then it is valid to use the above fit function with our data.

\begin{figure}[th]
  \centering
  \includegraphics[width=0.49\textwidth]{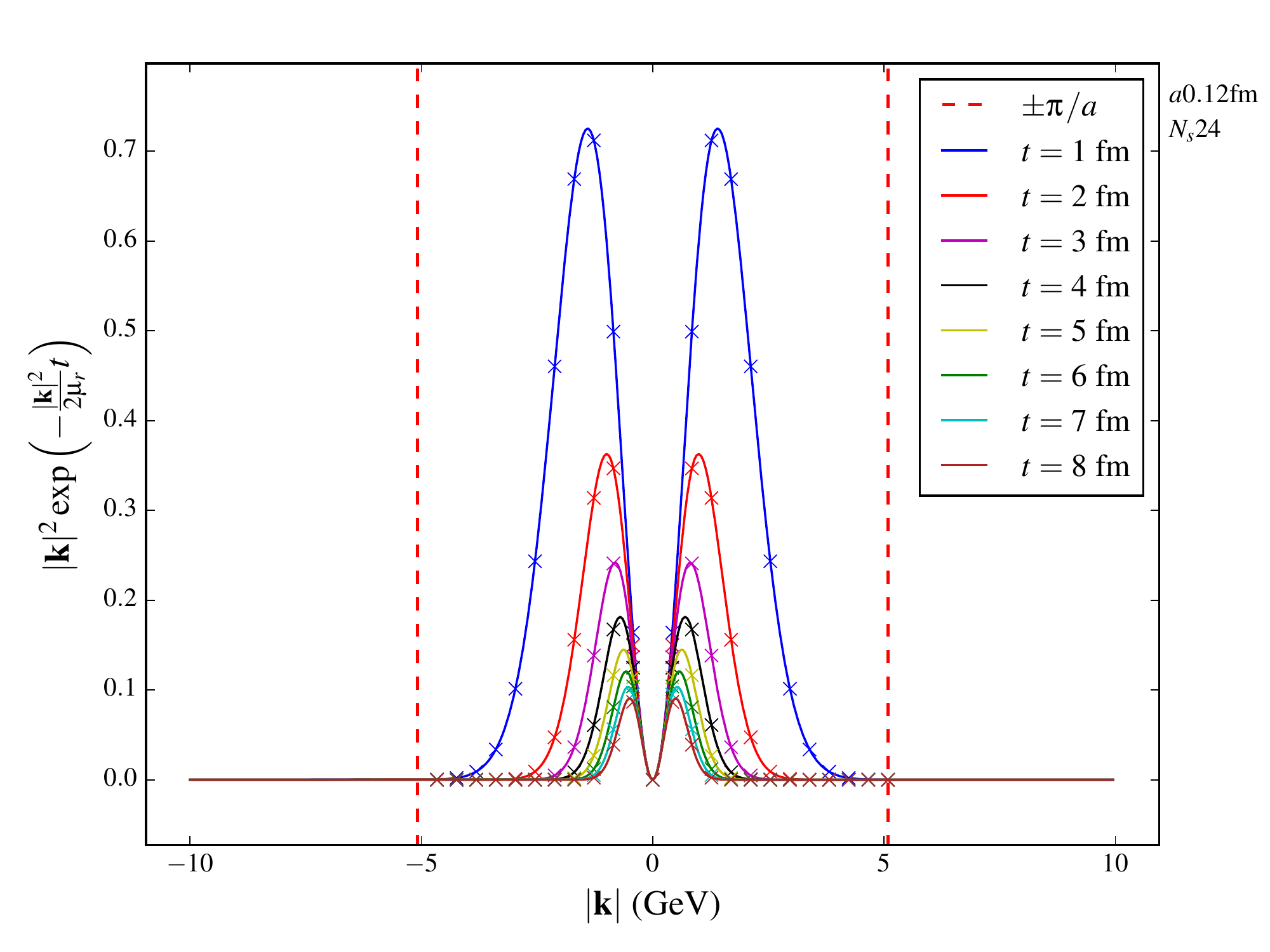}  
  \caption{The integrands of the moments given in Eq.~(\ref{eqn:Dlt}) at multiple times. The crosses represent the discrete finite-volume momentum contributions on the coarse (Set $1$) ensemble as discussed in the text. Due to the Gaussian time dependence, the integrand peak moves towards the origin for larger times. (color online)}
  \label{fig:integrand}
\end{figure}

To do so, using spherical coordinates in Eq.~(\ref{eqn:2ptFitTwo2}), we define the quantities that we need to compare as
\begin{align}
I^{(l)}(t) & = \frac{1}{\mu_r^{2l}}\int^{\infty}_{-\infty} d|\VEC{k}| |\VEC{k}|^{2l+2} e^{-\frac{|\VEC{k}|^2}{2\mu_r}t} \label{eqn:Ilt} \\
D^{(l)}(t) & = \frac{1}{\mu_r^{2l}}\sum_{|\VEC{k}|} |\VEC{k}|^{2l+2} e^{-\frac{|\VEC{k}|^2}{2\mu_r}t}. \label{eqn:Dlt} 
\end{align}
The integrands of both are shown diagrammatically in Figure \ref{fig:integrand}, where it is observed that due to the Gaussian time-dependence the peaks of the integrand move towards the origin with larger $t$. As such, one objective is to choose a large enough $\hat{t}$ such that a sufficient majority of the integrand is within the maximum momentum $\pi / a$.  We can replace the discrete finite-volume fit function with it's infinite-volume counterpart if the relative difference between them is less than our statistical errors. Specifically if
\begin{align}
  &\left| \frac{\sum_{l=0}^{l_{max}}Z^{2,(l)}I^{(l)}(t) - \sum_{l=0}^{\infty}Z^{2,(l)}D^{(l)}(t)}{\sum_{l=0}^{l_{max}}Z^{2,(l)}I^{(l)}(t)}\right| \label{eqn:CompareFit0}\\ 
  \le &\frac{\left| \sum_{l=0}^{l_{max}}Z^{2,(l)}I^{(l)}(t) - \sum_{l=0}^{\infty}Z^{2,(l)}D^{(l)}(t)\right|}{\left|Z^{2,(0)}I^{(0)}\right|} \\
  \le &\sum_{l=0}^{l_{max}}\frac{Z^{2,(l)}}{Z^{2,(0)}}\frac{\left|I^{(l)}(t) - D^{(l)}(t)\right|}{I^{(0)}} + \sum_{l=l_{max}+1}^{\infty}\frac{Z^{2,(l)}}{Z^{2,(0)}} \frac{D^{(l)}(t)}{I^{(0)}} \nonumber \\
  \le &\sum_{l=0}^{l_{max}}\frac{\left|I^{(l)}(t) - D^{(l)}(t)\right|}{I^{(0)}} + \sum_{l=l_{max}+1}^{\infty}\frac{D^{(l)}(t)}{I^{(0)}} \label{eqn:CompareFit}\\
  \le &\frac{\delta C(t)}{C(t)}
\end{align}
where $l_{max}$ is the maximum number of moments to be included in the fit function, the inequality in the second line holds as the moment integrands are positive (shown diagrammatically in Figure \ref{fig:integrand}), in the third line the Cauchy inequality has been used, and in the fourth line it is assumed that the leading moment gives the dominant contribution ($Z^{2,(l)} \le Z^{2,(0)}$). Studying Eq.~(\ref{eqn:CompareFit}) instead of Eq.~(\ref{eqn:CompareFit0}) is a conservative option. 

\begin{figure}[t]
  \centering
  \includegraphics[width=0.49\textwidth]{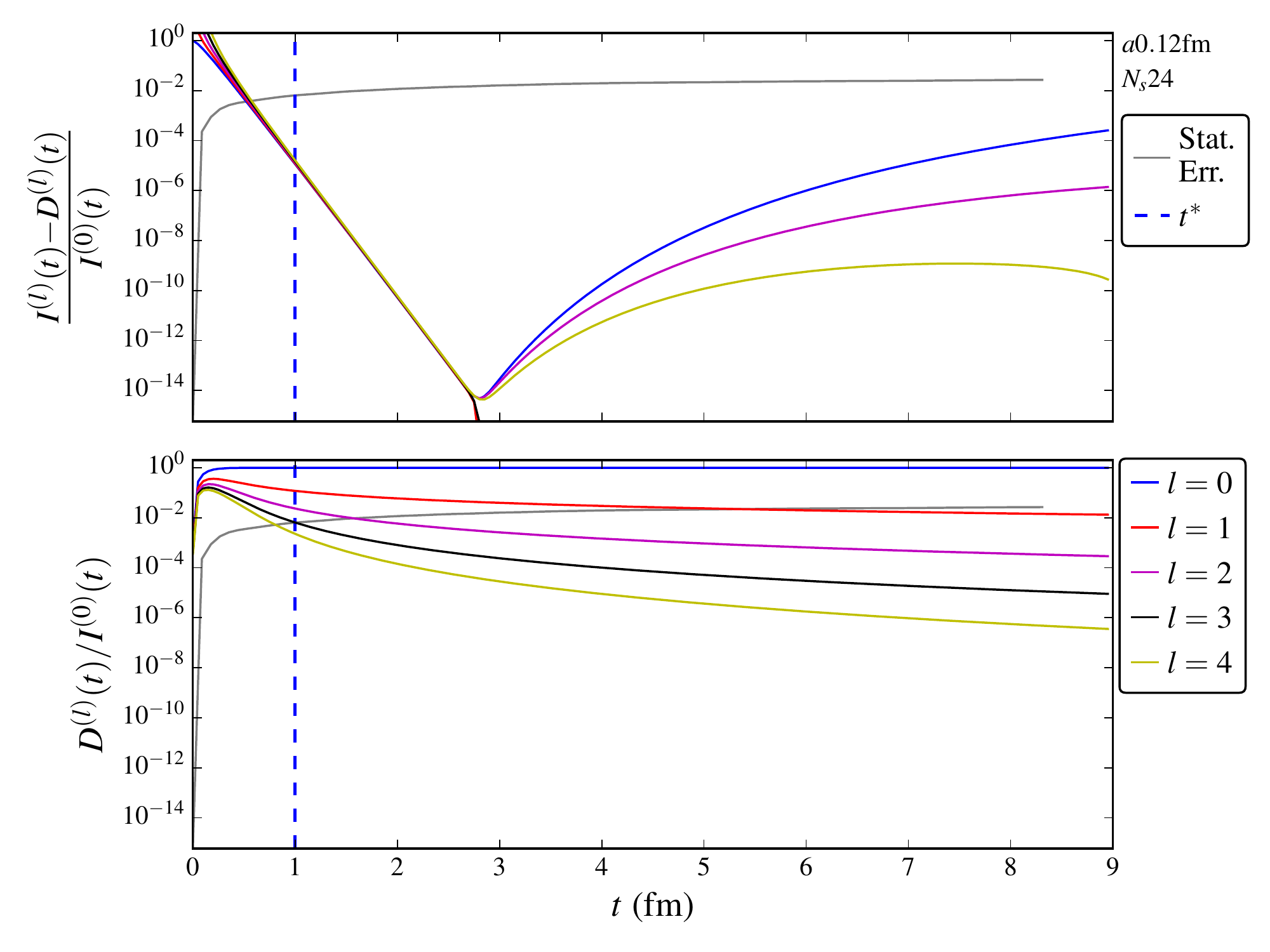}  
  \caption{The difference between the discrete finite-volume and infinite-volume continuum moments (upper) and which moments can be neglected compared to our statistical precision (lower) as discussed in the text. (color online)  }
  \label{fig:moments}
\end{figure}

Each part of the first term in Eq.~(\ref{eqn:CompareFit}) represents how similar $I^{(l)}(t)$ and $D^{(l)}(t)$ need to be in order to be considered equivalent within statistical precision. This is shown in Figure \ref{fig:moments}. For a particular $l_{max}$, the second term represents when the higher moments look like noise within statistical precision, also shown in Figure \ref{fig:moments}. Each Figure was generated with the coarse ensemble parameters (listed as Set $1$ in Table \ref{tab:GluonEnsembles}) as this ensemble has the largest lattice spacing (and hence smallest $\pi / a$ value $-$ the upper limit on the integral of interest) and also the smallest $N_s$ (the number of discrete momenta used in the finite-volume sum). As such, the other ensembles will give better approximations and studying Set $1$ is conservative.  Overlaid on each plot is the smallest relative statistical error from the data on any ensemble. Due to the constant signal-to-noise ratio, the number of configurations and the size of the lattice spacing, the smallest statistical error was the $2\eta_b$ correlator on the fine ensemble. Only examining situations below this curve is the most conservative option for all data generated. As can be observed in Figure \ref{fig:moments}, the discrete finite-volume sums are well represented by the indefinite integrals. Additionally, in order to neglect the higher moments within our statistical precision, a choice of $\hat{t}=1$ fm and $l_{max} = 2$ is sufficient. 

Expanding the finite-volume two-particle energy non-relativistically in Eq.~(\ref{eqn:ENR}) neglected a possible finite-volume energy shift which depends on the infinite-volume scattering amplitude. In the small scattering-length limit, the energy shift is known to be volume suppressed \cite{Parisi:Noise}. The two-particle systems under study are in this limit as the $\eta_b$ and $\Upsilon$ are compact due to the heavy-quark mass, with a size of $0.2-0.3 $fm. As such, the low-momentum energy shifts are not expected to be appreciable given the large volume ensembles we employ. Energy shifts in higher momentum states from Eq.~(\ref{eqn:ENR}) are exponentially suppressed due to the Gaussian integral in Eq.~(\ref{eqn:2ptFitTwo2}).  Consequently, these too are not appreciable and no large influence of finite-volume energy shifts are seen (c.f., the effective mass figures in Sec. \ref{sec:QCD4b}). Quantifying these finite-volume scattering shifts is outside the remit of this study. Regardless, the scattering shifts would be positive and push the finite-volume two-particle energy higher and not contribute to a misidentification of a bound tetraquark below the non-interacting threshold.

\section{Two-Point Correlator Fit Functions With A Harmonic Oscillator}
\label{app:FitHO}

In the non-relativistic limit the free propagator (to leading order) is
\begin{align}
 \Delta(\VEC{x}, t)  =  \int \frac{d^3p}{(2\pi)^3} \exp\left(i\VEC{p} \cdot \VEC{x}\right) \exp\left(-\left\{m + \frac{{{p}}^2}{2m}\right\}t\right). \label{eqn:appprop}
\end{align}
The free two-meson propagator with $\VEC{P}_{tot}=0$,  both starting at common origin $x_0 = (\VEC{0}, 0)$ and ending at time $t$ is given by
\begin{align}
\tilde{\Delta}(t) =   \int d^3{x} ~\Delta_1(\VEC{x},t) \Delta_2(\VEC{x},t). \label{eqn:app2p}
\end{align}
Using Eq.~(\ref{eqn:appprop}) in Eq.~(\ref{eqn:app2p}) produces the large-time behaviour of the free two-meson propagator as 
\begin{align}
\tilde{\Delta}(t) =  \left(\frac{\mu_r}{2\pi t}\right)^{3/2}  e^{-(m_1+m_2)t}.
\label{eq:free}
\end{align}
This agrees with the leading behaviour derived in App.~\ref{app:Fit}. Next, for the harmonic oscillator case, the one dimensional  Hamiltonian is
\begin{align}
\frac{\partial \psi}{\partial t}=  E \psi = -\frac{1}{2m}\frac{\partial^2\psi}{\partial x^2} +\frac{\kappa}{2} x^2 \psi.
\end{align}
Solutions of this system can be related  to the solutions of the Mehler differential equation via 
 \begin{align}
\frac{\partial \phi}{\partial \tilde{t}} = \frac{\partial^2\phi}{\partial \rho^2} - \rho^2 \phi
\end{align}
with the identifications $\omega = \sqrt{\kappa /m}$, $t = 2 \tilde{t}/ \omega $ and $r = (\kappa m)^{-1/4} \rho  $.
The Greens function (propagator) for the Mehler differential equation is given by
\begin{align}
\Delta(\rho_1, \rho_2, \tilde{t}) &=  \frac{1}{\sqrt{2\pi \sinh(2\tilde{t})}} \exp\Big( -\coth{(2\tilde{t})} \frac{(\rho_1^2 + \rho_2^2)}{2} \nonumber \\
 & \hspace{2.7cm} + \csch{(2\tilde{t})}\rho_1\rho_2\Big). \label{eqn:MehlerProp}
\end{align}
To normalize this propagator one can first compare the large $t$ behaviour of this solution with the known behaviour of the harmonic oscillator propagator, $\lim_{t\rightarrow \infty} G(t)  =  |\Psi(\VEC{0})|^2 e^{-\frac{1}{2} \omega t}$, where the wavefunction at the origin is given by $\Psi(\VEC{0}) =  ({m\omega}/{\pi})^{1/4}$.
As such, the harmonic oscillator solution is
\begin{align}
  & \Delta(\VEC{x}, 0, t) =   \frac{\sqrt{m \omega}}{\sqrt{2\pi \sinh(\omega t)}}  \exp\left( - \frac{m\omega\VEC{x}^2}{2} \coth{(\omega t)} \right). \label{eqn:hoprop} 
\end{align}
The three-dimensional solution can then be obtained using the separability of each spatial direction,
so that the zero spatial-momentum single-particle correlator in a harmonic oscillator potential is given by
\begin{align}
\int d^3x  \Delta (\VEC{x}, 0, t)  = \left(\frac{1}{\cosh(\omega t)}\right)^{3/2} e^{-mt}. \label{eq:SHO1}
\end{align}
Finally, the  equal mass two-particle propagator starting at a common origin ${x}_0 = (\VEC{0}, 0)$ and ending at a time $t$ with $\VEC{P}_{tot}=\VEC{0}$, in the presence of an external harmonic oscillator potential, is found from using Eq.~(\ref{eqn:hoprop}) in Eq.~(\ref{eqn:app2p}), to give 
\begin{align}
&\tilde{\Delta}(t) =  \left( \frac{m\omega}{2\pi}\frac{1}{\sinh(2 \omega t)}\right) ^{3/2} e^{-2mt}. \label{eq:SHO2}
\end{align}
By comparing (\ref{eq:SHO2}) to (\ref{eq:free}), and noting that $m = 2\mu_r$,  we see that the free two-meson harmonic oscillator propagator reduces to the non harmonic oscillator case as $\omega \rightarrow 0$.

\bibliographystyle{apsrev4-1}
\bibliography{./CH}

\end{document}